\documentclass[twocolumn]{aastex62}  
\usepackage{xcolor, bm, amsmath, enumitem, mfirstuc}
\usepackage[T1]{fontenc} 
\usepackage{array,multirow}
\usepackage[hang]{footmisc}
\newcommand{\trans}[3]{\capitalisewords{#1}\,\textsc{#2}\,#3\,\AA{}}
\newcommand{\spec}[2]{\capitalisewords{#1}\,\textsc{#2}}
\newcommand{\powr}{\textsc{PoWR}}
\newcommand{\tlusty}{\textsc{tlusty}}
\newcommand{\genec}{\textsc{genec}}
\newcommand{\msun}{$M_\odot$}
\newcommand{\mdot}{$\dot{M}_\star$}
\newcommand{\vsini}{$v \,\mathrm{sin}(i)$}
\newcommand{\vinf}{$v_\infty$}
\newcommand{\vesc}{$v_\mathrm{esc}$}

\newcommand{\hii}{H\,\textsc{ii}}

\newcommand{\hei}{He\,\textsc{i}}
\newcommand{\heii}{He\,\textsc{ii}}
\newcommand{\teff}{$T_\mathrm{eff}$}
\newcommand{\tstar}{$T_\star$}
\newcommand{\logg}{$\log(g)$}

\newcommand{\logl}{$\log(L_\star)$}
\newcommand{\lstar}{$L_\star$}
\newcommand{\mstar}{$M_\star$}
\newcommand{\mspec}{$M_\star^\mathrm{spec}$}
\newcommand{\mevol}{$M_\star^\mathrm{evol}$}
\newcommand{\rstar}{$R_\star$}

\newcommand{\zsun}{$Z_\odot$}

\newcommand{\ebv}{$E(B-V)$}
\newcommand{\ebvh}{$E(B-V)_\mathrm{host}$}

\newcommand{\qh}{$Q$(H)}
\newcommand{\qhe}{$Q$(He)}

\newcommand{\kms}{km\,s$^{-1}$}
\newcommand{\ximin}{$\xi_\mathrm{min}$}

\newcommand{\princeton}{Department of Astrophysical Sciences, Princeton University, 4 Ivy Lane, Princeton, NJ 08544, USA; {\color{xlinkcolor}grace.telford@princeton.edu}} \newcommand{\carnegie}{The Observatories of the Carnegie Institution for Science, 813 Santa Barbara Street, Pasadena, CA 91101, USA}
\newcommand{\rutgers}{Rutgers University, Department of Physics and Astronomy, 136 Frelinghuysen Road, Piscataway, NJ 08854, USA}
\newcommand{\utaustin}{Department of Astronomy, The University of Texas at Austin, 2515 Speedway, Stop C1400, Austin, TX 78712-1205, USA}
\newcommand{\heidelberg}{Zentrum f\"ur Astronomie der Universit\"at Heidelberg, Astronomisches Rechen-Institut, M\"onchhofstr.\ 12-14, 69120 Heidelberg, Germany}
\newcommand{\stsci}{Space Telescope Science Institute, 3700 San Martin Drive, Baltimore, MD 21218, USA}

\shorttitle{Metal-Poor O-Star Winds and Evolution}
\shortauthors{Telford et al.}

\begin{document}

\title{Observations of Extremely Metal-Poor O Stars: Weak Winds and Constraints for Evolution Models}

\author[0000-0003-4122-7749]{O.\ Grace Telford}
\altaffiliation{Carnegie-Princeton Fellow}
\affiliation{\princeton}
\affiliation{\carnegie}
\affiliation{\rutgers}
\author[0000-0002-0302-2577]{John Chisholm}
\affiliation{\utaustin}
\author[0000-0002-2090-9751]{Andreas A.\ C.\ Sander}
\affiliation{\heidelberg}
\author[0000-0001-5205-7808]{Varsha Ramachandran}
\affiliation{\heidelberg}
\author[0000-0001-5538-2614]{Kristen B.\ W.\ McQuinn}
\affiliation{\rutgers}
\affiliation{\stsci}
\author[0000-0002-4153-053X]{Danielle A.\ Berg}
\affiliation{\utaustin}


\begin{abstract}

Metal-poor massive stars drive the evolution of low-mass galaxies, both locally and at high redshift.
However, quantifying the feedback they impart to their local surroundings remains uncertain because models of stellar evolution, mass loss, and ionizing spectra are unconstrained by observations below 20\% solar metallicity (\zsun{}).
We present new Keck Cosmic Web Imager optical spectroscopy of three O stars in the nearby dwarf galaxies Leo~P, Sextans~A, and WLM, which have gas-phase oxygen abundances of 3--14\%\,\zsun{}.
To characterize their fundamental stellar properties and radiation-driven winds, we fit \powr{} atmosphere models to the optical spectra simultaneously with Hubble Space Telescope far-ultraviolet (FUV) spectra and multi-wavelength photometry.
We find that all three stars have effective temperatures consistent with their spectral types and surface gravities typical of main-sequence dwarf stars.
Yet, the combination of those inferred parameters and luminosity for the two lower-$Z$ stars is not reproduced by stellar evolution models, even those that include rotation or binary interactions. 
The scenario of multiple-star systems is difficult to reconcile with all available data, suggesting that these observations pose a challenge to current evolution models.
We highlight the importance of validating the relationship between stellar mass, temperature, and luminosity at very low $Z$ for accurate estimates of ionizing photon production and spectral hardness.
Finally, all three stars' FUV wind profiles reveal low mass-loss rates and terminal wind velocities in tension with expectations from widely adopted radiation-driven wind models.
These results provide empirical benchmarks for future development of mass-loss and evolution models for metal-poor stellar populations.
\end{abstract} 


\section{Introduction\label{sec:intro}}

The recently launched James Webb Space Telescope (JWST) is now successfully detecting large numbers of galaxies at very high redshift ($z$; e.g., \citealt[][]{roberts-borsani22, arrabal-haro23, curtis-lake23}), well into the epoch when the intergalactic medium (IGM) was reionized at $z \gtrsim 6$ \citep[e.g.,][]{becker01, fan06}. 
At these early times (less than 1\,Gyr after the Big Bang), metal deposition by stars and supernovae  had not yet enriched galaxies' star-forming gas to the metallicities ($Z$) observed in galaxies of comparable mass in the present-day universe.
Indeed, JWST studies are now routinely requiring very low stellar $Z$ ($\lesssim$\,10\%\, of the solar metallicity, \zsun{}) to explain both the nebular emission and spectral energy distributions (SEDs) of $z > 6$ galaxies (e.g., \citealt[][]{cameron23, endsley23, topping24}).
The hottest and most massive O-type stars in these metal-poor galaxies dominate the rest-frame UV and optical light that we observe; produce the ionizing radiation that powers their nebular emission; and are the progenitors of supernova explosions that regulate star formation and facilitate the escape of ionizing photons into the IGM \citep[e.g.,][]{byler17, trebitsch17, dayal18}.

To understand the nature of these early galaxies, stellar population synthesis (SPS) models are required to infer their physical properties from observed integrated light \citep[e.g.,][]{tinsley80, conroy13}.
Models of stellar evolution and spectra, both key ingredients in SPS codes, are sensitive to the treatment of various physical processes in massive-star atmospheres, including mixing (convection, rotation), metal opacities, and mass-loss via radiation-driven winds \citep[e.g.,][]{maeder00, eldridge22, vink22}.
In the very low-$Z$ regime ($\lesssim$ 20\%\,\zsun{}), SPS models remain fundamentally limited due to a lack of empirical data that can assess the validity of the theoretical approaches that are typically adopted. 

Both optical and far-ultraviolet (FUV) spectra of individual massive stars are required to measure key stellar and wind properties for comparison to stellar-model predictions.
Large datasets of suitable quality and wavelength coverage have been assembled over decades for O and B stars in the Milky Way (\zsun{}; e.g., \citealt{prinja90, przybilla10, holgado20}), and in the lower-$Z$ environments of the Large and Small Magellanic Clouds (LMC and SMC; 50\% and 20\%\,\zsun{}, respectively; \citealt{hunter07, trundle07}) with the latter recently boosted by the ULLYSES and X-Shooting ULLYSES programs \citep{roman-duval20, vink23}.
Below 20\%\,\zsun{}, however, few spectra of massive stars exist \citep{bresolin06, bresolin07, evans07, evans19, garcia13, lorenzo22, gull22} because the necessary low-metallicity environments are almost exclusively found in more distant dwarf galaxies at the edge of or beyond the Local Group, which are faint and expensive to observe. 
Additional observations in this low-$Z$ regime are required to constrain how the properties of massive stars and their evolutionary pathways change at the metallicities relevant in the early universe.

Massive stars lose mass via radiation-driven winds, which impacts the evolutionary path and endpoint for a star of a given initial stellar mass (\mstar{}) and $Z$, as well as the ionizing photon production of that star over its lifetime \citep{lamers99, kudritzki00, martins21}.
These winds, accelerated to high terminal velocities (\vinf{}), also deposit substantial energetic feedback into their surroundings and contribute to regulating the star-formation process \citep{leitherer92, rosen22}.
As hot-star winds are largely driven by momentum transfer via absorption and scattering of photons in line transitions of ionized metals \citep{castor75,pauldrach86,vink99}, theory predicts that mass-loss rates (\mdot{}) decrease toward lower $Z$.
However, different theoretical treatments make remarkably different predictions of how \mdot{} and the \vinf{} scale with $Z$ and other stellar properties (particularly the bolometric luminosity, \lstar{}, and the effective temperature, \teff{}; e.g., \citealt{vink01, krticka18, bjorklund21}).  

Observational studies are therefore essential to clarify the $Z$ dependence of mass loss from OB stars and avoid incorrect extrapolations from higher to lower metallicity.
\citet{mokiem07} constructed an empirical scaling relation between \mdot{} and \lstar{} at Milky Way, LMC, and SMC metallicities, finding both a normalization and $Z$ dependence consistent with the modeling of \citet{vink01}.
However, those \mdot{} drawn from the literature were inhomogeneous in terms of data quality, wavelength coverage, and analysis technique.
The \citet{vink01} mass-loss prescription has since been found to over-predict \mdot{} for O stars at 20\%\,\zsun{} \citep[e.g.,][]{bouret13}, and multiple observational analyses have found evidence for a steeper scaling between \mdot{} and $Z$ \citep{ramachandran19, marcolino22, rickard22}.

Some efforts have been made to observationally constrain the stellar and wind properties of OB stars in more distant dwarf galaxies ($\sim$0.7--1.5\,Mpc).
\citet{tramper11, tramper14} analyzed optical spectra of several stars in the galaxies IC~1613, NGC~3109, and WLM, finding higher \mdot{} than theoretical predictions based on metallicity estimates from the gas-phase oxygen abundances of those galaxies (9--12\%\,\zsun{}; \citealt{marble10}).
\citet{bouret15} re-analyzed three of those stars by fitting atmosphere models to FUV spectra, and concluded that the \mdot{} measured from H$\alpha$ emission alone were overestimated and that the stars' iron abundances may be SMC-like.
Though the sample size was small, the inferred \mdot{} and stellar parameters did not follow a well-defined power-law scaling.
For eight OB stars in IC~1613, \citet{garcia14} studied the observed \vinf{} via UV spectroscopy, finding no clear differences compared to similar SMC and LMC stars. Contrary to expectations from measured gas-phase and B supergiant oxygen abundances \citep{marble10, bresolin07}, the iron abundance of the young stars in IC~1613 seems to be similar to the SMC value. 
In summary, while these previously studied metal-poor galaxies are interesting sub-solar $\alpha$/Fe environments, few massive stars with sub-SMC iron abundances have been observed.

\begin{table*}[!ht]
\begin{center}
\caption{Basic properties of the three target stars and their host galaxies. \label{tab:target_stars}}
\begin{tabular}{lccccc|lccc}
Star & R.A. & Decl. & $m_V$ & $A_V$ & Spectral & Host & 12+$\log$(O/H) & $Z_\mathrm{gas}/Z_\odot$ & Distance \\
Name & (J2000) & (J2000) & (mag) & (mag) & Type  & Galaxy & (nebular) & (nebular) & (Mpc) \\
\hline
LP26 & 10:21:45.1217 & +18:05:16.93 & 21.51 & 0.04 & O7--8 V & Leo P & $7.17\pm0.04$ & 3\% & $1.62\pm0.15$  \\
S3 & 10:10:58.1866 & $-$04:43:18.45 & 20.80 & 0.09  & O9 V & Sextans A & $7.49\pm0.04$ & 6\% & $1.38\pm0.05$ \\
A15 & 00:02:00.5333 & $-$15:29:52.41 & 20.25 &  0.04 & O7 V((f)) & WLM & $7.83\pm0.06$ & 14\% & $0.98\pm0.04$ \\
\end{tabular}
\end{center}
\vspace{-10pt}
\tablecomments{Following \citet{telford21}, we adopt the star identifiers LP26, S3, and A15 from \citet{evans19}, \citet{garcia19}, and  \citet{bresolin06}, respectively. Right ascension is reported in hours, minutes, seconds, and declination is in degrees, arcminutes, arcseconds. 
Foreground extinction $A_V$  is from \citet{green15}. Visual magnitudes and spectral types are from \citet{mcquinn15}, \citet{garcia19}, and \citet{bresolin06}. Gas-phase oxygen abundances from \citet{skillman89, skillman13} and \citet{lee05} are converted to solar assuming ($12+\log$(O/H))$_\odot = 8.69$ \citep{asplund09}. 
Distances are from \citet{mcquinn15}, \citet{dalcanton09}, and \citet{jacobs09}.}
\end{table*}

\citet{telford21} increased the small dataset that can constrain wind properties below 20\%\,\zsun{} by presenting FUV spectra of three O-dwarf (luminosity class V) stars in nearby, metal-poor dwarf galaxies (with gas-phase oxygen abundances of 3--14\%\,\zsun{}; see Table~\ref{tab:target_stars}) observed with the Cosmic Origins Spectrograph (COS) on the Hubble Space Telescope (HST). 
Those authors characterized the observed photospheric and wind features and estimated the fundamental stellar parameters from the observed stellar SEDs, but lacked the optical spectroscopy necessary for quantitative analysis and \mdot{} inference. 
Here, we build upon that work by presenting new optical spectra of the three stars observed with the Keck Cosmic Web Imager (KCWI), as well as multi-wavelength measurements of the stellar and wind properties obtained from detailed atmosphere modeling of the optical and FUV spectra.

The remainder of the paper is organized as follows.
Section~\ref{sec:data} presents the observations used in our analysis: FUV and optical spectroscopy, as well as stellar photometry measured from archival HST imaging.
Section~\ref{sec:methods} describes our stellar atmosphere modeling technique, and the best-fit models and inferred stellar and wind parameters are presented in Section~\ref{sec:best_models}.
In Section~\ref{sec:model_comparison}, we compare our results to the predictions of widely used stellar evolution models, mass-loss prescriptions, and grids of ionizing spectra predicted by atmosphere models.
Finally, we summarize our conclusions in Section~\ref{sec:conclusions}. 

Throughout, we adopt the \citet{asplund09} solar abundances as our reference, with bulk metal mass fraction $Z_\odot=0.0142$.
Mass fractions of individual elements are denoted as, e.g., $X_\mathrm{C}$ for the carbon mass fraction.
In general, when discussing the metallicities of galaxies, we mean the gas-phase oxygen abundance (12+$\log$(O/H)), while stellar metallicities refer to the bulk metal mass fraction $Z$.
As mentioned above, evidence exists for non-solar O/Fe abundance ratios in metal-poor dwarf galaxies, with important implications for the opacities and \mdot{} of massive stars that are primarily set by the Fe abundance.
Thus, we use the $X_\mathrm{Fe}/X_\mathrm{{Fe,}\odot}$ of observed stars to determine the stellar model $Z$ against which to compare.
Wherever model predictions are discussed, we rescale the bulk metal mass fraction $Z$ from the solar reference adopted by each set of models to express all metallicities relative to the \citet{asplund09} solar abundance scale. 
Finally, we use the terms extremely or very metal-poor to mean $\leq10$\%\,\zsun{}, following \citet{kunth00}.


\section{Observations and Data Reduction\label{sec:data}}


\subsection{Optical Spectroscopy\label{sec:optical_spectra}}

\begin{table*}
\begin{center}
\caption{Details of the Keck/KCWI observations. \label{tab:observations}}
\tabcolsep=0.15cm
\begin{tabular}{lccccccc}
Star & R.A. & Decl. & Position Angle & Exposure Time & Typical Seeing & Continuum SNR & Flux \\
 & (J2000) & (J2000) & ($^{\circ}$ E of N) & (s) & ($''$) & (4500--4700\,\AA{}) & Standard \\
\hline
LP26 & 10:21:44.91 & +18:05:20.6 & 150 & 12000 [$10\times1200$] & 1.9 & 32 & Feige34 \\  
S3 & 10:10:58.42 & $-$04:43:25.8 & 110 & 9600 [$8\times1200$] & 1.3 & 50 & Feige34, GD108 \\ 
A15 & 00:02:00.05 & $-$15:29:52.1 & 90 & 7200 [$6\times1200$] & 1.0 & 46 & LTT949 \\
\end{tabular}
\end{center}
\vspace{-10pt}
\tablecomments{Coordinates are the centers of the KCWI FoV for each observation (see Table~\ref{tab:target_stars} for the target-star coordinates). 
Seeing is the FWHM measured from a 2D Gaussian fit to each star in a white-light image constructed from the coadded KCWI datacube. }
\end{table*}

\begin{figure*}[!htp]
\begin{centering}
  \includegraphics[width=\linewidth]{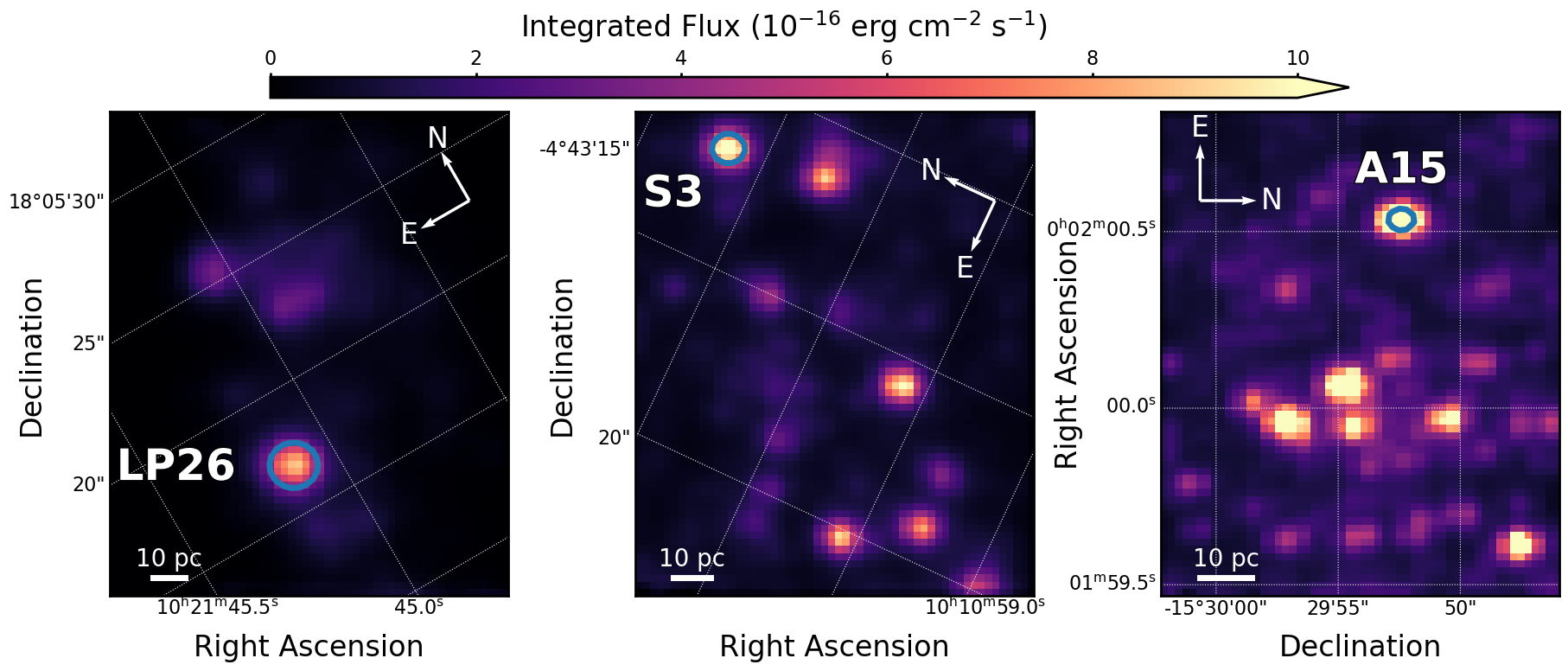}
\caption{\textbf{KCWI fields observed in the three dwarf galaxies.} From left to right, white-light images of Leo~P, Sextans~A, and WLM are shown. Each image is constructed by integrating the flux in the co-added KCWI observations of each galaxy over the full wavelength range. The three stars analyzed here are labeled and encircled in blue. The 1D spectra used in this analysis are the summed spectra in the spaxels falling within those blue contours, which enclose 50\% of the stellar flux. White bars in the bottom-left corners show 10\,pc at the distance of each galaxy for scale.
\label{fig:kcwi_images}}
\end{centering}
\end{figure*}

Spectroscopy at visible wavelengths is required to determine the fundamental properties of O-type stars.
Specifically, the relative strengths of \hei{} and \heii{} photospheric absorption lines is the primary \teff{} diagnostic for mid-late O stars, while the Balmer line profiles encode the surface gravity, (\logg{}; e.g., \citealt{simon-diaz20}).
We therefore observed optical spectra of the three metal-poor O stars in our sample using the Keck Cosmic Web Imager (KCWI) on the 10-m Keck II Telescope.
This integral field unit (IFU) spectrograph is highly sensitive at blue wavelengths, ideal for measuring the key diagnostic absorption features between 4000--5000\,\AA{}.

All three dwarf galaxy targets were observed by NASA Keck program N194\footnote{The KCWI data presented in this paper can be downloaded from the Keck Observatory Archive: \url{https://koa.ipac.caltech.edu/}}. 
Leo~P was observed on 2021 January 13 (2020B semester, PI: J.\ Chisholm), followed by WLM on 2021 August 10 and Sextans~A on 2022 January 5 (2021B semester, PI: O.\ G.\ Telford). 
We used the same instrument configuration for all observations, selecting the medium slicer, BM grating, and a central wavelength (cenwave) of 4500\,\AA{}. 
This setup yields a spectral resolution of $R$$\sim$4000, a $16\farcs5\times20\farcs0$ field of view (FoV), spatial pixels (or spaxels) of angular size $0\farcs29 \times 0\farcs68$, and covers $\sim$3970--4930\,\AA{}.
Table~\ref{tab:observations} summarizes the exposure times and typical conditions during the observations of the three galaxies.
We interleaved our science exposures with observations of off-galaxy fields for both Sextans~A and WLM to measure the sky background. In the case of Leo~P, its bright stellar and nebular emission covers a small enough area that we were able to obtain a background observation by nodding the FoV up to $5\farcs4$ to the northeast (with the position angle fixed), keeping the target star LP26 in the FoV throughout.

Leo~P was also observed by the same program for 30 minutes on 2022 January 5 at a different cenwave of 4700\,\AA{} to cover the [O\,\textsc{iii}]$\lambda\lambda$\,4959,\,5007 nebular lines, and the emission-line spectra from both Leo~P observations were presented in \citet{telford23}. 
However, we do not include the 2022 data in this analysis because the continuum signal-to-noise ratio (SNR) of $\sim$20 is lower than achieved in the earlier observations, and we verified that combining both datasets does not improve the final data quality. 
We prefer to use only the deeper 2021 data at the same cenwave observed for all three stars in our sample to maintain consistent methodology. 

We reduce the KCWI data using the IDL-based KCWI data reduction pipeline\footnote{\url{https://github.com/Keck-DataReductionPipelines/KcwiDRP}} (DRP; version 1.2.1; \citealt{morrissey18}). 
We follow the same process described in Section~2.1 of \citet{telford23} to calibrate the images using bias, dark, and internal/dome flat frames; reject cosmic rays; obtain a wavelength calibration using arc-lamp frames; and transform the data from 2D images to 3D data cubes. 
The offset field observations (or off-galaxy regions of the data cubes in the case of Leo~P) were used to subtract the sky background.
We use the interactive tool in the KCWI DRP to derive inverse sensitivity curves from the flux standard spectra obtained during all observing runs (Table~\ref{tab:observations}) and transform units of counts to flux. 
Residuals in the fits to the standard-star continua were $\lesssim$\,5\%. 

We combine the reduced and flux-calibrated individual exposures using CWITools \citep{osullivan20}. 
A WCS solution is obtained for each exposure, given the known coordinates of the target star, from a Gaussian fit to the star's light profile.
The exposures are then aligned and interpolated onto a common spatial and wavelength grid with square pixels 0\farcs29 on a side and 0.5-\AA{} wavelength sampling (or $\sim$37.5 \kms{}).
Finally, the data cubes constructed from individual exposures are averaged, weighted by exposure time.
Figure~\ref{fig:kcwi_images} shows white-light images constructed by integrating each spaxel in the coadded data cubes over all wavelengths.
The target star in each image is encircled in blue and labeled, where the circle size corresponds to the inferred seeing (i.e., the full width at half maximum, or FWHM, of a 2D Gaussian fit to the star's light profile; see next paragraph). 
Stars observed under better conditions appear elongated along the horizontal direction in the coadded images because the KCWI detector pixels are rectangular, and have been resampled to a uniform grid of square pixels.
Each target star is nicely separated from any comparably bright OB stars in high-resolution HST imaging (0.05$''$ pixels) of the same fields. 

\begin{figure*}[!ht]
\begin{centering}
  \includegraphics[width=\linewidth]{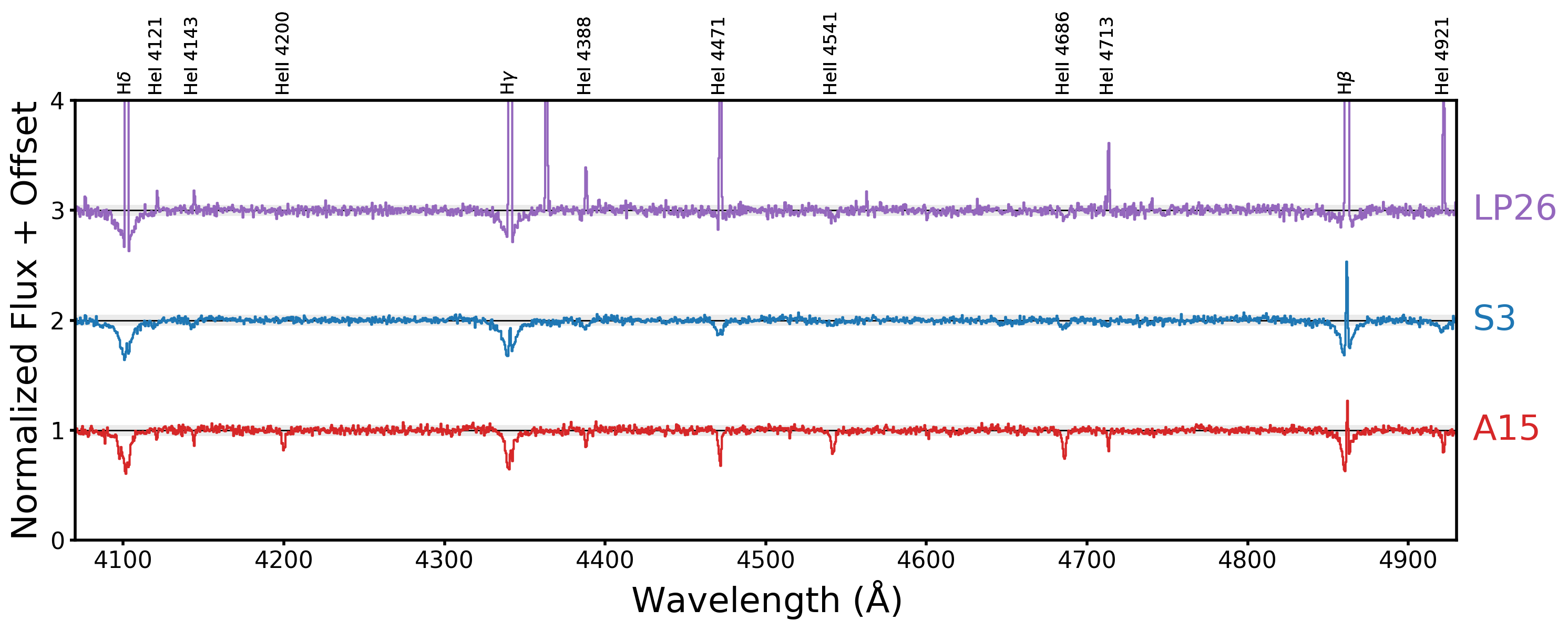}
\caption{\textbf{KCWI optical spectra of the three O stars.} Continuum-normalized KCWI spectra of LP26 (top, purple), S3 (middle, blue), and A15 (bottom, red) plotted as a function of rest-frame wavelength. 
An integer offset is added for clarity, and the continuum level for each star is indicated by a horizontal black line with gray shading to show $\pm$\,5\%. 
The key diagnostics of O-star \teff{} (\hei{} and \heii{} lines) and \logg{} (Balmer lines) are labeled at the top of the plot.
All three spectra show narrow nebular emission filling in the centers of the Balmer lines, and the \hei{} transitions are also affected for S3 (primarily the 4471\,\AA{} line) and LP26 (all lines). 
These normalized spectra are used in our \powr{} modeling analysis in Sections~\ref{sec:procedure} and \ref{sec:best_models} below.
\label{fig:kcwi_spectra}}
\end{centering}
\end{figure*}

From these coadded data cubes, we extract the 1D spectrum for each star by fitting a 2D Gaussian model to its spatial profile in the white-light image and summing the spectra in all spaxels within the FWHM, which captures 50\% of the total flux from each star. 
Adopting a larger radius (out to two times the FWHM, which encloses 94\% of the flux) does not increase the final SNR substantially, but does increase the contribution of nebular emission to the final spectra.
We allow for different FWHM in the two spatial dimensions to account for the rectangular spaxels in the raw KCWI images.
In the case of LP26, the 2D Gaussian was fit to the profile of the \hii{} region in which the star is embedded, which is larger than the star itself.
We tested instead using the FWHM of the flux-standard star to define the aperture within which the 1D spectrum of LP26 is constructed, but found that the FWHM of the standard was similar to that of the \hii{} region. 

Because observing conditions inevitably vary throughout the night, we considered extracting a 1D spectrum within the FWHM of the center of each target star's light profile in the individual exposures, then combining those 1D spectra.
However, we found that this does not improve the SNR of the final 1D stellar spectrum, and more importantly, does not change the equivalent width or shape of any of the stellar lines used in our atmosphere modeling.
Thus, we prefer to use the CWITools coadds for consistency with our previous analysis of KCWI observations of Leo~P \citep{telford23}.
We also tried excluding the exposures taken under the poorest seeing during each observing run, but found that including all available data maximizes the continuum SNR, and this choice does not change the stellar absorption line shapes.

For comparison to the optical spectra predicted by stellar atmosphere models (see Sections~\ref{sec:procedure} and \ref{sec:best_models} below), observed optical spectra are typically normalized to a continuum level of 1. 
For each star, we model the continuum as a fifth-degree polynomial, masking all absorption and emission lines, and divide the observed flux by that model continuum to obtain the normalized spectrum used in this analysis.
All three spectra contained a residual sinusoidal feature over $\sim$4170--4300\,\AA{} with amplitude $\lesssim$\,5\%, which was also visible in the standard-star spectra and is thus an instrumental artifact.
To remove this unphysical feature, we divide out a spline fit to the continuum in that region only, masking the \trans{he}{ii}{4200} absorption line, but this choice does not affect the conclusions drawn from our modeling below.
The SNR reported in Table~\ref{tab:observations} is calculated from the normalized continuum rms between 4500--4700\,\AA{}, with the absorption lines in that range masked.
Finally, we correct the observed wavelengths to the heliocentric reference frame, with small velocity shifts of $\sim$20\,km\,s$^{-1}$ (i.e., less than one pixel at this spectral resolution). 

The final, normalized spectra used in our analysis below are shown in Figure~\ref{fig:kcwi_spectra}, vertically offset for clarity, in order of increasing host-galaxy metallicity from top to bottom: LP26 (purple), S3 (blue), and A15 (red). 
Black lines indicate the continuum level for each star, with gray shading to demonstrate that the normalized data fall within $\pm$\,5\% of the flat continuum level.
All three spectra are affected by narrow nebular emission, but this is cleanly separable from the Balmer-line wings that are critical to constrain \logg{}.
The \hei{} photospheric absorption lines are completely filled in by nebular emission for LP26, which is surrounded by a bright \hii{} region, but are clearly detected in the spectra of S3 and A15.
All three spectra contain clear \heii{} absorption, confirming their classification as O-type stars.
Based on the ratio of the equivalent widths of the observed \trans{he}{ii}{4540} and \trans{he}{i}{4471} lines \citep{martins18}, we confirm the spectral types of S3 and A15 as O9 and O7, respectively. 
We cannot quantitatively assign a spectral type to LP26 due to the infilling of all optical \spec{he}{i} lines by nebular emission.
All three stars have broad Balmer wings, consistent with luminosity class V. 


\subsection{Far-Ultraviolet Spectroscopy\label{sec:fuv_spectra}}

Most important diagnostics of stellar mass loss and metal abundances for hot stars are highly ionized metal transitions found in the FUV part of the spectrum. 
Thus, we use the HST/COS FUV spectra of the three target stars presented in \cite{telford21} in this analysis. 
A comprehensive discussion of the data and reduction can be found in Section~2.3 of that paper, but we briefly summarize key aspects here.

The COS spectra were obtained using the moderate-resolution G130M and G160M gratings during April--June of 2020 (GO-15967; PI: J.\ Chisholm). 
For each star, the CalCOS pipeline-reduced spectra\footnote{The HST/COS data analyzed in this paper can be downloaded from MAST: \dataset[10.17909/hs6f-a370]{http://dx.doi.org/10.17909/hs6f-a370}} from the two gratings were downloaded from the Mikulski Archive for Space Telescopes (MAST), resampled onto a common wavelength grid at the native sampling of the G160M grating, and coadded. 
Based on the velocity FWHM of Milky Way interstellar medium (ISM) absorption features in the spectra, the spectral resolution is better than 50\,km\,s$^{-1}$. 
We bin the spectra by 12 pixels to maximize SNR while ensuring 1.5--2 samples per resolution element. 
Given our goal of modeling the weak photospheric absorption features in these metal-poor stars, achieving the highest SNR possible is essential.

The FUV data will be compared to model spectra in Section~\ref{sec:best_models} below, providing both line diagnostics and a constraint on the overall SED shape ($\pm5$\% flux-calibration uncertainties\footnote{\url{https://hst-docs.stsci.edu/cosihb}}).
The model spectra account for dust extinction, both in Milky Way and internal to the host dwarf galaxies, so we do not apply any extinction correction to the observed spectra. 
For the star LP26 in Leo~P only, we correct the FUV observations for the contribution of nebular continuum emission from its surrounding \hii{} region, which is not included in the model spectra.
\citet{telford21} found that a substantial contribution to the FUV flux from nebular continuum (up to a maximum of 19\% at the reddest wavelengths in the G160M data) was required to explain the shape of LP26's SED.
Thus, we divide the observed spectrum by the best-fit nebular continuum in that SED modeling to ensure a fair comparison between the observed COS spectrum of LP26 and the \powr{} model spectra.
No significant nebular continuum contribution was detected for S3 or A15, consistent with their environments outside of bright \hii{} regions.


\subsection{Near-Ultraviolet, Optical, and Near-Infrared Photometry from Archival Imaging\label{sec:photometry}}

Finally, the stellar SED constructed from photometry across a wide wavelength baseline is required to constrain \lstar{}, \mstar{}, and reddening \ebv{} due to dust along the line of sight to the star.
We employ the high-quality stellar photometry measured from archival near-ultraviolet (NUV), optical, and near-infrared (NIR) HST imaging presented in Table~3 of \citet{telford21}.
Details on the process of measuring point-spread function (PSF) photometry with \textsc{dolphot} \citep{dolphin00, dolphin16}, including parameter choices and quality metrics, are given in Section~2.2 of that paper. 
These measurements typically have quite high SNR of $>$\,100 in the NUV and optical, and $\sim$10--30 in the NIR. 
We use the observed magnitudes and fluxes, uncorrected for Milky Way foreground dust along the line of sight to the targets, and include dust extinction in the atmosphere models to which we compare below.

Accurate distances to the host galaxies are essential to translate observed into absolute magnitudes.
Thus, we use distances determined from the tip of the red-giant branch and, for Leo~P, the horizontal branch in optical color-magnitude diagrams (Table~\ref{tab:target_stars}).
These are among the most precise distance-measurement techniques for nearby galaxies \citep[e.g.,][]{beaton18}, and we account for distance uncertainties in the \lstar{} measurements reported below.


\section{\powr{} Stellar Atmosphere Models\label{sec:methods}}


\subsection{The \powr{} Code\label{sec:powr}}

To analyze the observations, we employ the Potsdam Wolf-Rayet \powr{} stellar-atmosphere code \citep{grafener02,hamann03,sander15}. 
\powr{} is a 1D state-of-the-art code solving the statistical equations together with the radiative transfer in an expanding, stationary outflow outside of local thermodynamic equilibrium (i.e., in a non-LTE environment).
The radiative transfer is performed in the co-moving frame to properly account for the stellar wind and the quasi-hydrostatic photosphere.
After a model atmosphere has converged, the emergent spectrum is obtained by computing the formal integral along rays in the observer's frame. 
Fundamental input parameters are \tstar{}, \lstar{}, \mstar{} (or alternatively, \logg{}), and the chemical composition. Formally, \tstar{} denotes the effective temperature corresponding to the inner boundary of the model, at a Rosseland optical depth ($\tau_\mathrm{Ross}$) of 20. For a better comparison, we will list the more common \teff{} defined at $\tau_\mathrm{Ross} = 2/3$ throughout this work, which is an output parameter of \powr{}. All listed radii and surface gravity values are also referring to the same optical depth.

The velocity field in the supersonic wind is prescribed by a $\beta$-type velocity law:
\begin{equation}
  v(r) = v_\infty \left( 1 - \frac{R_\star}{r} \right)^\beta,
\end{equation}
where we adopt $\beta=0.8$ and adjust the free parameter $v_\infty$ as necessary to reproduce the wind features in the UV spectra. 
The velocity and density stratification in the subsonic part of the atmosphere is obtained from a consistent solution of the hydrostatic equation, accounting for the full radiative force from continua and lines \citep{sander15}, and is smoothly connected to the $\beta$-law supersonic wind regime. 

The \mdot{} of the wind is a free parameter that is adjusted to match to the spectral appearance of wind-sensitive features, particularly metal resonance lines in the FUV spectra. 
Wind inhomogenieties are treated in the microclumping approach: the wind is assumed to be composed of small clumps with density $D\rho(r)$, where $D$ is the clumping factor and $\rho(r)$ is the density profile of a homogenous wind with the same \mdot{} \citep{hamann98}.
The interclump space is void, so the wind volume-filling factor is $D^{-1}$.
We adopt a depth-dependent $D$, increasing from unity to $D_\infty = 10$ between the sonic point and $10\,R_\star$.
This is a commonly adopted value (e.g., \citealt{hainich19}) and is supported by a recent analysis of similar O~V stars in the 20\%\,\zsun{} SMC by \citet{rickard22}, who found $5 \leq D_\infty \leq 20$.
Because \mdot{}$\sqrt{D}$ is constant for microclumping, we account for this factor of 2 uncertainty in $D$ in our \mdot{} measurements reported below.

The \powr{} code offers the option to include X-rays in the wind \citep{baum92}, but our analysis revealed that there was no need to include X-rays to match the wind-sensitive metal resonance lines in the observed FUV spectra.
While X-ray emission from the stellar wind appears ubiquitous for O stars in the Milky Way and LMC \citep[e.g.,][]{rauw15,crowther22}, no observations exist that can constrain the X-ray fluxes of our target stars (or even of massive stars in the SMC). 
We thus  use the least complex model that can explain the observations to avoid introducing additional free parameters and do not include X-rays in our presented models (following, e.g., \citealt{rickard22}).

For the microtubulent velocity ($\xi$) in the formal integral, we adopt a depth-dependent approach \citep{shenar15} where $\xi(r) = 0.3\,v(r)$, with a minimum value of \ximin{}. 
The resulting element-dependent Doppler velocity $v_\text{Dop}$ used in the observer's frame calculation is then calculated via: 
\begin{equation}
 v_\text{Dop}(r) = \sqrt{ v_\text{th}(r)^2 + \xi(r)^2},
\end{equation}
where $v_\text{th}$ is the thermal velocity of the considered element of the stellar atmosphere.

Detailed model atoms of H, He, C, N, O, Mg, Si, P, and S are included in the non-LTE calculations, while the vastly more numerous and complex transitions of Fe-group elements (atomic numbers 21--28) are treated in a superlevel approach (described in \citealt{grafener02}).
Though our data cannot directly constrain the abundances of all considered elements, they are important opacity sources in the stellar atmospheres that must be accounted for to enable accurate inference of fundamental stellar properties from their observed spectra \citep[e.g.,][]{sander17}.


\subsection{Choice of Initial Parameters for \powr{} Models\label{sec:initial_guesses}}

For each of the three low-$Z$ stars in this analysis, we aim to find a \powr{} atmosphere model that simultaneously matches the HST/COS FUV spectrum, Keck/KCWI optical spectrum, and HST photometry (Section~\ref{sec:data}).
Initial guesses for each parameter of interest in our \powr{} models are chosen based on existing information about the target stars to minimize the number of computationally expensive model iterations needed to find a good fit to the observations.
We adjust these initial parameters, following the procedure in Section~\ref{sec:procedure} below, to find the parameter set that best matches all available data.
Each iteration requires human intervention and is not automated, so the resulting best-fit models are not sensitive to the initial-guess parameters.
We explain our choice of initial parameters here, and provide the values in Table~\ref{tab:initial_guesses} in Appendix~\ref{app:initial_guesses}.

We initially adopt the stellar properties (\teff{}, \logg{}, and \lstar{}) inferred by \citet{telford21} from model SEDs fit to the FUV spectra and HST photometry (see their Table~7). 
For the wind properties, we pick the nearest point in the publicly available SMC-metallicity \powr{} grid\footnote{\url{http://www.astro.physik.uni-potsdam.de/~wrh/PoWR/SMC-OB-III/}} (the OB-III grid with ``low'' mass-loss and wind strength parameter $\log(Q)=-14.0$; \citealt{hainich19}) and scale their \mdot{} and \vinf{}
to the presumed iron abundance ($X_\mathrm{Fe}$) for each of our O-star targets. For this, we assume the power-law scalings $\dot{M}_\star \propto Z^{0.69}$ \citep{vink01} and $v_\infty \propto Z^{0.13}$ \citep{leitherer92}.

Initial mass fractions of most metals are scaled to 3\%, 6\%, and 20\% of the solar values for models of LP26 in Leo P, S3 in Sextans A, and A15 in WLM, respectively, assuming \citet{asplund09} solar abundances (except Fe-group elements, for which we adopt the updated solar abundances in \citealt{scott15}). 
These choices are informed by the gas-phase oxygen abundances measured in the host galaxies and by the best-fit SED models from \citet{telford21}. 
The higher $Z$ of A15 compared to WLM's gas-phase oxygen abundance (14\%\,\zsun{}; \citealt{lee05}) is also consistent with the atmosphere modeling of an O-supergiant star in the same galaxy by \citet{bouret15}. 
The initial He mass fraction ($Y$) is interpolated between primordial ($Y_\mathrm{p}=0.2453$; \citealt{aver22}) and solar ($Y_\odot=0.2485$) values assuming a linear scaling with the total metal mass fraction $Z$. 

We adopt initial mass fractions that deviate from the solar pattern for the CNO elements, as galaxies' gas-phase C/O and N/O ratios are known to correlate with O/H (though, with substantial scatter). 
The typical observed abundance ratios (by number) in metal-poor dwarf galaxies of $\log\mathrm{(C/O)} = -0.7$ and $\log\mathrm{(N/O)} = -1.4$, respectively \citep{berg19}, are substantially lower than the solar values of $\log\mathrm{(C/O)}_\odot = -0.26$ and $\log\mathrm{(N/O)}_\odot = -0.86$ \citep{asplund09}.
We adopt in our initial models the observed gas-phase abundances of N and O in the three host galaxies.
The observed $\log\mathrm{(N/O)}$ in Leo~P, Sextans~A, and WLM are $-1.35$, $-1.54$, and $-1.49$, respectively \citep[][]{skillman13, kniazev05, lee05}, which are all close to the typical value for metal-poor galaxies quoted above.
Since we do not have C abundance measurements for the three host galaxies, we adopt initial C mass fractions in our \powr{} models such that $\log\mathrm{(C/O)} = -0.7$.
For the model of LP26 only, we also adopt the abundances of S and He measured in its surrounding \hii{} region \citep{skillman13, aver22}.

We convolve each model spectrum to the instrumental resolution (of COS in the FUV and of KCWI in the visible part of the spectrum) and apply a rotational broadening kernel, taking as an initial guess for \vsini{} the values reported in \citet{telford21} from modeling the \trans{c}{iii}{1176} feature in the COS spectra. 
The resolution and SNR of our spectra preclude using a Fourier-transform technique \citep[e.g., \texttt{iacob-broad};][]{simon-diaz14} to measure \vsini{}, so we instead estimate this quantity from the observed profiles of the photospheric He and metal lines.
We adopt $\xi_\text{min} = 10\,\mathrm{km}\,\mathrm{s}^{-1}$, a typical value for SMC O dwarfs \citep[e.g.,][]{bouret03}, as an initial guess for the minimum microturbulent velocity. 

Finally, we apply dust extinction to each model.
We model the extinction due to Milky-Way foreground dust toward each star with a fixed reddening \ebv{} from the \citet{green15} dust map (see Table~\ref{tab:target_stars}) and a \citet{fitzpatrick99} extinction law.
We also include the extinction internal to each host galaxy with \ebvh{} calculated from the best-fit $A_V^\mathrm{host}$ from the SED fitting in \citet{telford21} as an initial guess, and adopting the preferred extinction law from that modeling (\citealt{fitzpatrick99} for LP26 and S3, and \citealt{gordon03} for A15).

\begin{table*}
\begin{center}
\caption{Stellar and wind parameters in the best-fit \powr{} models.\label{tab:params}}
\tabcolsep=0.75cm
\begin{tabular}{lccc}
Parameter & LP26 & S3 & A15 \\
\hline 
\teff{} (kK) & $34 \pm 1$ & $31 \pm 1$ & $36 \pm 1$ \\
$\log(g/\mathrm{cm\,s}^{-2})$ & $4.15 \pm 0.1$ & $3.9 \pm 0.1$ & $4.0 \pm 0.05$ \\  
$\log(L_\star/L_\odot)$ & $5.0 \pm 0.09$ & $5.15 \pm 0.05$ & $5.15 \pm 0.06$ \\
\mspec{} (\msun{}) & $43^{+32}_{-18}$ & $49^{+30}_{-19}$ & $34^{+15}_{-10}$ \\
$\log(\dot{M}_\star/M_\odot\,\mathrm{yr}^{-1})$ & $-$$8.8 \pm 0.35$ & $-$$9.1 \pm 0.35$ & $-$$8.4 \pm 0.35$ \\
\vinf{} (km\,s$^{-1}$) & $500 \pm 200$ & $300 \pm 200$ & $1100 \pm 300$ \\ 
\vsini{} (km\,s$^{-1}$) & $200 \pm 80$ & $200 \pm 50$ & $45 \pm 20$ \\
$\xi_\mathrm{min}$ (km\,s$^{-1}$) & $5 \pm 5$ & $5 \pm 5$ & $5 \pm 5$ \\
$X_\mathrm{He}$ & ($0.25 \pm 0.1$) & ($0.25 \pm 0.05$) & $0.35 \pm 0.05$ \\
$\log(X_\mathrm{C})$ & ($-4.58 \pm 0.3$) & ($-4.29 \pm 0.3$) & ($-3.92 \pm 0.3$) \\
$\log(X_\mathrm{N})$ & ($-5.17 \pm 0.7$) & ($-5.06 \pm 0.7$) & $-3.64 \pm 0.3$ \\
$\log(X_\mathrm{O})$ & ($-3.76 \pm 0.7$) & ($-3.46 \pm 0.7$) & ($-3.09 \pm 0.3$) \\
$\log(X_\mathrm{Fe})$ & ($-4.44 \pm 0.3$) & ($-4.14 \pm 0.3$) & ($-3.62 \pm 0.3$) \\
\ebv{}$_\mathrm{host}$ (mag) & $0.000 \pm 0.01$ & $0.056 \pm 0.01$ & $0.022 \pm 0.01$\\
\end{tabular}
\end{center}
\vspace{-10pt}
\tablecomments{Adopted best-fit values in our \powr{} models and estimated uncertainties.
Abundances are reported as mass fractions, not as number fractions relative to hydrogen.
Abundances that did not need to be adjusted from the initial guesses to match the observations are reported in parentheses. 
}
\end{table*}


\subsection{Procedure for Fitting \powr{} Models to Observations of the Metal-Poor O Stars\label{sec:procedure}}

Starting from the initial model for each star described above, we assess by eye the agreement of the calculated model spectrum with the observations, and adjust the model parameters iteratively until the model spectrum acceptably reproduces diagnostic features in the data.
We first find the \teff{} and \logg{} that produce a good match to the \spec{he}{i} and \spec{he}{ii} line strengths and Balmer line wings in the optical spectrum.
Metal and \spec{he}{ii} lines in the FUV spectra provide secondary diagnostics of these fundamental stellar parameters, especially \teff{}. 
Because these lines are also sensitive to He and metal abundances, we simultaneously adjust mass fractions of individual elements as required to match all observed photospheric transitions of that element.
If there are no detected lines of a given element, or there is no clear need to adjust the abundance of an element to be consistent with the data, we adopt our initial guess for its abundance in our final model.
Given the lines detected in our observations, we ultimately only vary the abundances of He, C, N, O, and Fe in our \powr{} modeling described in Section~\ref{sec:procedure} below.

We also adjust \vsini{} to reproduce the observed profiles of the photospheric He and metal lines.
Rotational broadening strongly affects these line shapes and the depths of the Balmer absorption lines, but importantly, does not affect the wings of the Balmer lines that we use to determine \logg{}. 
We then adjust $\xi_\text{min}$ to match the observed profiles of optical \spec{he}{i} lines (only in the case of A15, which has low \vsini{} resulting in line profiles narrow enough to constrain microturbulence) and observed strengths of metal absorption lines in the FUV (for S3 and LP26; discussed further in Section~\ref{sec:best_models} below).

Next, we adjust two parameters that affect the shape and normalization of the model: \ebvh{} and \lstar{}.
The observed FUV spectrum is normalized by the emergent model continuum. The high density of metal absorption lines in the FUV makes it challenging to determine the continuum level directly from the data.
The quality of the FUV normalization and its agreement with the normalized model spectrum is therefore a sensitive diagnostic of \lstar{} and the total (foreground + host galaxy) dust extinction. 
We adjust \ebvh{} and \lstar{} iteratively to achieve the best possible match to the observed FUV continuum shape and HST NUV, optical, and NIR photometry.

Finally, we find the wind parameters required to match the observed metal resonance line profiles in the FUV. The terminal velocity \vinf{} determines the blueward extent of the broad absorption due to the wind, while \mdot{} sets the strength of the absorption or (weak, unsaturated) P-Cygni profiles.
The \trans{n}{iii}{4634--4640} emission, which is only detected in the spectrum of A15, could potentially provide an additional constraint for our models, but this emission is highly sensitive to the details of line blanketing and non-LTE effects \citep{rivero-gonzalez11} rather than providing a direct wind diagnostic.

Once the set of model parameters that best matches the full suite of observations for each star is identified, we run additional \powr{} models to estimate uncertainties.
We adjust one model parameter at a time, changing each higher and lower from the preferred value until the resulting model is inconsistent with the observations.
Though this method does not capture the degeneracies among the various parameters, it is sufficient to approximate the range of allowed stellar properties given the fundamental assumption that the stars do not have companions and the choices made in the \powr{} modeling.


\section{Results: \powr{} Models of the Metal-Poor O Stars\label{sec:best_models}}

Here we present the \powr{} models that best match the observed FUV and optical spectra and HST photometry of the three target stars.
We discuss each model in turn and report the corresponding best-fit stellar and wind parameters in Table~\ref{tab:params}. 
Figures~\ref{fig:lp26}, \ref{fig:s3}, and \ref{fig:a15} compare our preferred \powr{} models (red lines) for LP26, S3, and A15, respectively, to the observations (black points and lines) of each star. 
In all three figures, the FUV spectrum is shown in the three panels of the left column; the stellar SED is shown in the top-right panel; and the optical spectrum is shown in the middle-right and bottom-right panels. 
The FUV spectra are included in the stellar SED plots because they extend the wavelength range sampled by the HST photometry.


\subsection{LP26 in Leo~P\label{sec:results_lp26}}

\begin{figure*}[!htp]
\begin{centering} 
  \includegraphics[width=\linewidth]{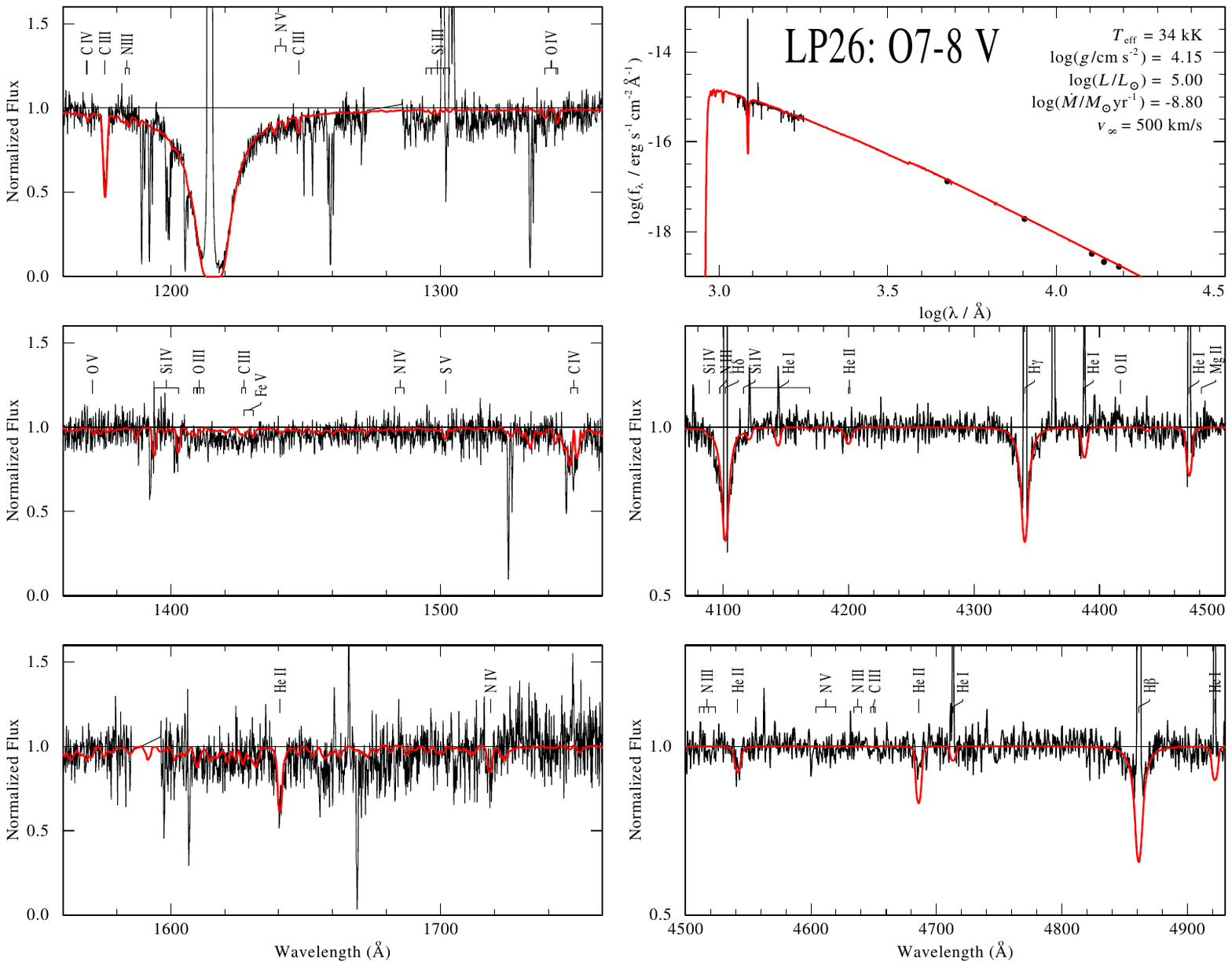}
\caption{\textbf{\powr{} model of LP26 in Leo~P.} 
The best \powr{} model (red) compared to the observations of LP26 (black). 
\uline{Left column}: 
The FUV spectra are plotted with the observed wavelength range divided across the three panels, normalized by the emergent model continuum.
Prominent ISM absorption, nebular emission, and airglow (1215, 1302--1306\,\AA{}) features are apparent, but only Ly$\alpha$ absorption is included in the \powr{} model.
\uline{Right  column}: 
The top panel shows the absolute flux of the stellar SED plotted as a function of wavelength (both log-scaled), including both the observed FUV spectrum (black line) and HST photometry (black circles). 
Error bars on the photometry are smaller than the point size.
The middle and bottom panels show the normalized optical spectra, with the observed wavelength range divided across the two panels. 
Again, narrow nebular emission is obvious in the Balmer lines, and for this star only, fills in all stellar \hei{} absorption lines.
\label{fig:lp26}}
\end{centering}
\end{figure*}

LP26 in Leo~P is the most distant and faintest of the three stars; resides in the lowest-$Z$ host galaxy; and powers a bright \hii{} region.
The continuum SNR of $\sim$10 in the FUV spectrum and $\sim$30 in the optical spectrum combined with high \vsini{} prohibit detection of many weak diagnostic lines in this very metal-poor star, and strong nebular emission lines unfortunately fill in most of the \hei{} transitions in the optical.
Despite these challenges, we are able to constrain many key stellar properties thanks to clearly detected FUV absorption features (especially \trans{c}{iii}{1176}, \trans{c}{iv}{1550}, and \trans{he}{ii}{1640}) and well-resolved Balmer line wings and \heii{} absorption in the KCWI data (Figure~\ref{fig:lp26}).

The optical and FUV \heii{} lines favor a \teff{} of 34\,kK, within the uncertainties of the $37.5^{+5.9}_{-5.5}$\,kK previously estimated from SED fitting by \citet{telford21}. 
The lower best-fit \teff{} reported here may be more consistent with the hardness of the stellar ionizing spectrum implied by the relative strengths of nebular H$\beta$ and \trans{he}{i}{4471} emission from the surrounding \hii{} region \citep{telford23}; we return to this point in Section~\ref{sec:lp26_hiiregion} below.
The observed \trans{he}{ii}{4686} absorption is substantially weaker than predicted by our best \powr{} model and has an asymmetric profile, both of which are puzzling. 
While this is the most susceptible of the optical \heii{} transitions to infilling by nebular and/or stellar wind emission, such effects should also impact the FUV \trans{he}{ii}{1640} line, which is well matched by the model.
It is also likely that nebular continuum emission from the \hii{} region contributes to the optical spectrum (on the order of 10\% of the observed continuum level), so dividing out the continuum level could create the appearance of weaker optical \spec{he}{ii} absorption than the lines actually formed in the stellar photosphere.
Yet, tests with model spectra showed that this effect cannot account for the combination of well-matched \trans{he}{ii}{4541} and weak observed \trans{he}{ii}{4686} absorption.

The Balmer line profiles favor a relatively high \logg{} of 4.15, typical for dwarf stars and consistent with the position of LP26 on the main sequence in the HST color-magnitude diagram of Leo~P \citep{mcquinn15}. 
This value is consistent within the uncertainties with the $\log(g) = 4.0\pm 0.2$ reported by \citet{telford21}.
The H$\beta$ stellar absorption feature is more dramatically filled in by nebular emission than H$\delta$, but the wings of the three Balmer lines are all well matched by a single \logg{}. 
We find that the best overall match to the shapes of the FUV and optical line profiles is given by a \vsini{} of 200\,km\,s$^{-1}$, lower than the $370\pm90$\,km\,s$^{-1}$ measured from the FUV \trans{c}{iii}{1176} profile \citep{telford21}, but the shapes of the Balmer wings in the \powr{} models are not substantially different for those two different \vsini{} values.
We attribute the need for a lower \vsini{} here to a combination of a difference in intrinsic line broadening between the \powr{} model and the \tlusty{} models used in \citet{telford21}, and the additional information provided by the \hei{} and \heii{} line profiles in the KCWI data included in this analysis. 
We recommend adopting the updated \vsini{} reported here because it is informed by additional data compared to the \citet{telford21} value, but emphasize that (1) the estimated uncertainty of $\pm80$\,\kms{} is large, due to low SNR and broad lines, and (2) the key conclusion that this star is a fast rotator remains unchanged.

\begin{figure*}[!htp]
\begin{centering} 
  \includegraphics[width=\linewidth]{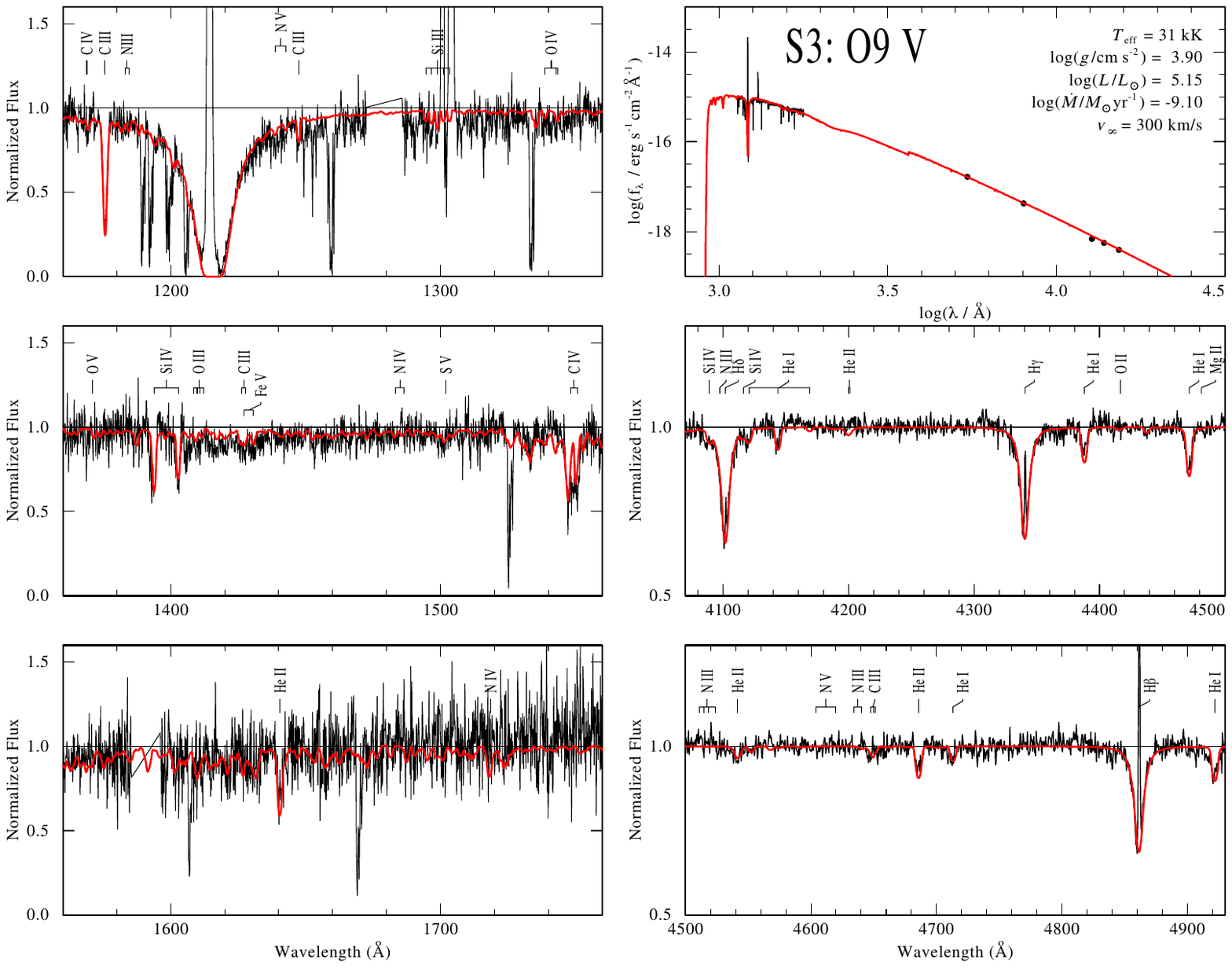}
\caption{\textbf{\powr{} model of S3 in Sextans~A.} 
As in Figure~\ref{fig:lp26}, the best model (red) compared to observations of S3 (black).
\label{fig:s3}}
\end{centering}
\end{figure*}

\begin{figure*}[!htp]
\begin{centering} 
  \includegraphics[width=\linewidth]{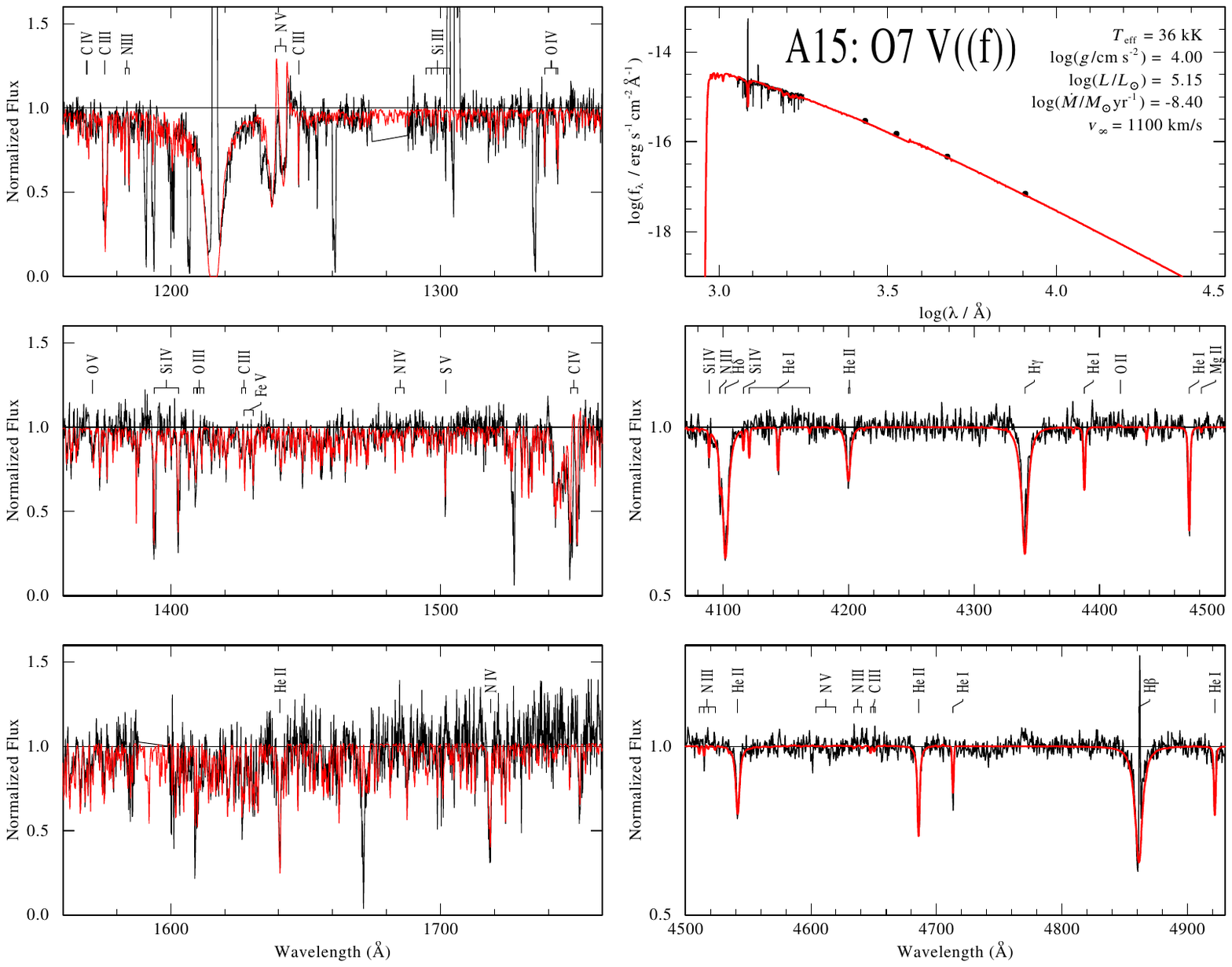}
\caption{\textbf{\powr{} model of A15 in WLM.} 
As in Figure~\ref{fig:lp26}, the best model (red) compared to observations of A15 (black). The narrow absorption lines in the spectra indicate a lower \vsini{}, making the metal lines in the FUV spectrum more clearly discernible than in the spectra of the other two stars at the $\mathrm{SNR}\sim10$ of the HST/COS data.
\label{fig:a15}}
\end{centering}
\end{figure*}

We find that $\log(L_\star)=5.0$ with only foreground Milky Way dust extinction \citep{green15}, and no contribution from dust in the host galaxy, produces the best match to the shape of LP26's SED and FUV spectrum. 
This is consistent with the $\log (L_\star/L_\odot) = 5.1\pm0.2$ and host-galaxy extinction $A_V^\mathrm{host}=0.01^{+0.06}_{-0.01}$ inferred from modeling the stellar SED \citep{telford21}, and with the small dust content implied by the Balmer decrement of the \hii{} region emission \citep{skillman13, telford23}.

Few metal lines are clearly detected in the COS and KCWI spectra of LP26, preventing robust constraints on its metal abundances. 
For our initially adopted abundance set, we do not find strong discrepancies between the observed and modeled metal lines for any ions except for \spec{c}{iii}. 
Given the best-fit \teff{} and \logg{}, the \powr{} model predicts a substantially stronger \trans{c}{iii}{1176} feature than is observed, but we found that even decreasing the C abundance to the lowest $\log \mathrm{(C/O)} \simeq -1.1$ observed in metal-poor dwarf galaxies did not resolve the discrepancy.
Instead, decreasing \ximin{} from 10\,km\,s$^{-1}$ to 5\,km\,s$^{-1}$ while keeping the initial $\log \mathrm{(C/O)} = -0.7$ was able to roughly reproduce the observed strength of the \trans{c}{iii}{1176} line, and did not affect the appearance of any other spectral features in the model.
This result is consistent with low microturbulent velocities previously measured for metal-poor O~V stars in the Magellanic Bridge \citep{ramachandran21}.

The metal resonance lines in the FUV spectrum of LP26 indicate a weak stellar wind.
The \trans{n}{v}{1238,\,1242} and \trans{si}{iv}{1393,\,1402} doublets appear to be purely photospheric, though the \spec{si}{iv} profile is affected by either narrow (non-wind) emission or a high noise level.
The \trans{c}{iv}{1548,\,1550} profile is challenging to interpret due to the superimposed narrow Milky Way ISM absorption lines, but it appears to be broadened due to a stellar wind, with a blueward extent several pixels beyond the bluest narrow ISM absorption line.
We find that a model with $\log(\dot{M}_\star) = -8.8$ and a quite low $v_\infty = 500$\,km\,s$^{-1}$ produces the best match to the observed \trans{c}{iv}{1548,\,1550} profile.
This best-fit \vinf{} is substantially lower than expected from extrapolating observations of the winds of more metal-rich stars to 3\%\,\zsun{} (the initial guess for \vinf{} was 1720\,km\,s$^{-1}$ for LP26), as we discuss in Section~\ref{sec:windspeed} below.


\subsection{S3 in Sextans~A\label{sec:results_s3}}

The FUV spectrum of the other very metal-poor star in our sample, S3 in Sextans~A, has the lowest SNR of the three, but the strongest photospheric and wind lines are still clearly detected. 
On the other hand, the KCWI spectrum of S3 is the highest-quality of the three, reaching SNR of $\sim$50 in the continuum. 
S3 does not reside inside a bright \hii{} region, but the Balmer lines clearly show nebular emission from diffuse ionized gas, and slight nebular infilling is discernible in some \hei{} lines (particularly 4471\,Å{}). 
The optical \hei{} and \heii{} lines imply a \teff{} of 31\,kK for this star, consistent with the $32.5^{+2.9}_{-3.0}$\,kK inferred from its observed SED \citep{telford21}. 
This agrees nicely with the prominent \spec{si}{iii} absorption at 1294--1300\,\AA{} (additional, redder transitions are obscured by geocoronal emission lines) and multiple strong \spec{c}{iii} lines in the FUV spectrum.
Not all optical He lines are exactly matched by the model, but we checked that changing neither \teff{} nor the He abundance can improve the model's agreement with the data.
We attribute this to infilling by nebular \hei{} emission, or possibly wind emission for the \trans{he}{ii}{4686} line.

The Balmer line wings are best matched by a \logg{} of 3.9, indicative of an unevolved star still on the main sequence and significantly higher than the $\log(g)=3.5^{+0.1}_{-0.2}$ estimated from modeling the stellar SED.
This higher \logg{} is further supported by the clearly detected FUV \spec{si}{iii} absorption at 1294--1300\,\AA{}, which becomes much weaker in lower-\logg{} models. 
As for LP26, we find that adopting \vsini{} of 200\,km\,s$^{-1}$ gives the best match the observed profiles of optical and FUV photospheric absorption lines. 
This value is lower than the $290\pm90$\,km\,s$^{-1}$ inferred from the \trans{c}{iii}{1176} profile \citep{telford21}, but consistent within the uncertainty.
The choice of \vsini{} affects the depths of the Balmer lines more than the shapes of the wings, thus, adopting the higher \vsini{} would not change our best-fit \logg{}. 
We find that the SED shape is best matched by $\log(L_\star)=5.15$ and total $E(B-V)=0.085$ (0.029 Milky Way foreground; \citealt{green15}), assuming a \citet{fitzpatrick99} extinction law in Sextans~A, both consistent with previous SED fitting results \citep{telford21}.
However, this \lstar{} is unexpected in combination with the lower \teff{} and higher \logg{} constrained by the KCWI spectrum, as we discuss in Section~\ref{sec:evolution} below.

We did not find any need to adjust the metal abundances from the initial guesses to match the observed metal lines in the FUV spectrum and the He lines in the optical for S3.
Similar to our findings for LP26 (Section~\ref{sec:results_lp26}), adopting a lower $\xi = 5$\,km\,s$^{-1}$ provides a good match to the observed \spec{c}{iii} FUV lines given the adopted $\log \mathrm{(C/O)} = -0.7$ and best-fit \teff{} of 31\,kK, and is more physically motivated than significantly decreasing the C mass fraction.
Again, the FUV \spec{n}{v} and \spec{si}{iv} lines appear purely photospheric, but the \trans{c}{iv}{1548,\,1550} line is clearly affected by the stellar wind, and is best matched by a low $v_\infty=300$\,km\,s$^{-1}$ and $\log(\dot{M}_\star) = -9.1$.


\subsection{A15 in WLM\label{sec:results_a15}}

A15 in WLM is the closest and brightest of the three stars. It is also the only one with low \vsini{} and therefore narrow lines which are more easily detectable above the noise level in the FUV and optical spectra. 
The best-fit \teff{} and \logg{} of 36\,kK and 4.0, respectively, agree nicely with the values reported in \citet{telford21} from modeling the stellar SED ($37.5^{+7.2}_{-2.7}$\,kK and $4.0^{+0.2}_{-0.1}$), though we find that a slightly lower \lstar{} of 10$^{5.15}\,L_\odot$ better matches the data than their reported $\log (L_\star/L_\odot) = 5.3^{+0.2}_{-0.1}$. 
A \vsini{} of 45\,km\,s$^{-1}$ was required to match the observed FUV and optical line profiles, which again is consistent with, but lower than, the $80\pm45$\,km\,s$^{-1}$ measured from the \trans{c}{iii}{1176} profile by \citet{telford21}.
The low \vsini{} of this star enabled us to constrain $\xi =5\,\mathrm{km\,s}^{-1}$ based on the narrow observed profiles of the optical He lines and FUV metal lines (\trans{c}{iii}{1247}, \trans{S}{v}{1502}).
This is the same lower $\xi$ value that was required to match the strength of the \trans{c}{iii}{1176} profiles in the COS spectra of LP26 and S3, but determined by a different method, suggesting that low $\xi$ is common among metal-poor O dwarfs. 

This star's lower \vsini{} and higher $Z$ compared to the other two enabled tighter abundance constraints.
We found that the N and He abundances both had to be increased to match simultaneously the observed strengths of all the FUV and optical transitions of those elements.
The enhanced N and He abundances are consistent with CNO-cycle products exposed at the stellar surface. 
This suggests that A15 is either evolved or may be mixed by strong rotation. 
The latter is only possible if A15 is rotating substantially faster than the low observed \vsini{} (i.e., if we are viewing the star at a low inclination). 
However, we find no evidence of C depletion, as would be expected for CNO-cycle products (C is quickly depleted by the CN cycle, but O destruction occurs more slowly; e.g., \citealt{maeder14}).
The factor of 2 uncertainty in our inferred C and N abundances implies a minimum $X_\mathrm{C}/X_\mathrm{N} \geq 0.13$, well above the CN-equilibrium value (0.025; \citealt{maeder87}).
Thus, while we do not observe abundances reflecting CNO-cycle equilibrium in A15, its high $X_\mathrm{N}$ fits in with the observed trend of stronger N enhancement in O stars at lower $Z$ \citep[e.g.,][]{martins24}.
The finding that unevolved O dwarfs in the SMC, similar to A15, are already N-enriched is consistent with the expectation of more efficient chemical transport in compact, metal-poor stars \citep{meynet02, brott11}.

\mdot{} and \vinf{} were constrained from the \trans{n}{v}{1238,\,1242} and \trans{c}{iv}{1548,\,1550} wind profiles in the COS spectrum of A15 (\trans{s}{iv}{1393,\,1402} again appears photospheric). 
We prioritized matching the blueward extent and depth of the wind lines, but no combination of \mdot{} and \vinf{} could also reproduce the peaks of the \spec{n}{v} emission, given the fixed parameters adopted in our models. 
The best-fit wind parameters are $v_\infty=1100$\,km\,s$^{-1}$ and $\log(\dot{M}_\star) = -8.4$, both higher than found for the two more metal-poor stars in our sample, as expected for line-driven winds.
The best-fit stellar and wind parameters do not reproduce the observed weak \trans{n}{iii}{4634--4640} emission in the optical, but this is not concerning because the these lines are highly sensitive to non-LTE effects and are therefore not a reliable wind diagnostic.


\section{Discussion: Comparison to Theoretical Models at Low Metallicity\label{sec:model_comparison}}

Next, we place the inferred stellar and wind properties of the three stars into the context of theoretical expectations at very low $Z$.
We compare the best-fit parameters of the \powr{} model for each star presented in Section~\ref{sec:best_models} above to stellar evolution models, \mdot{} and \vinf{} scalings, and predicted ionizing spectra.


\begin{figure*}[!htp]
\begin{centering} 
  \includegraphics[width=\linewidth]{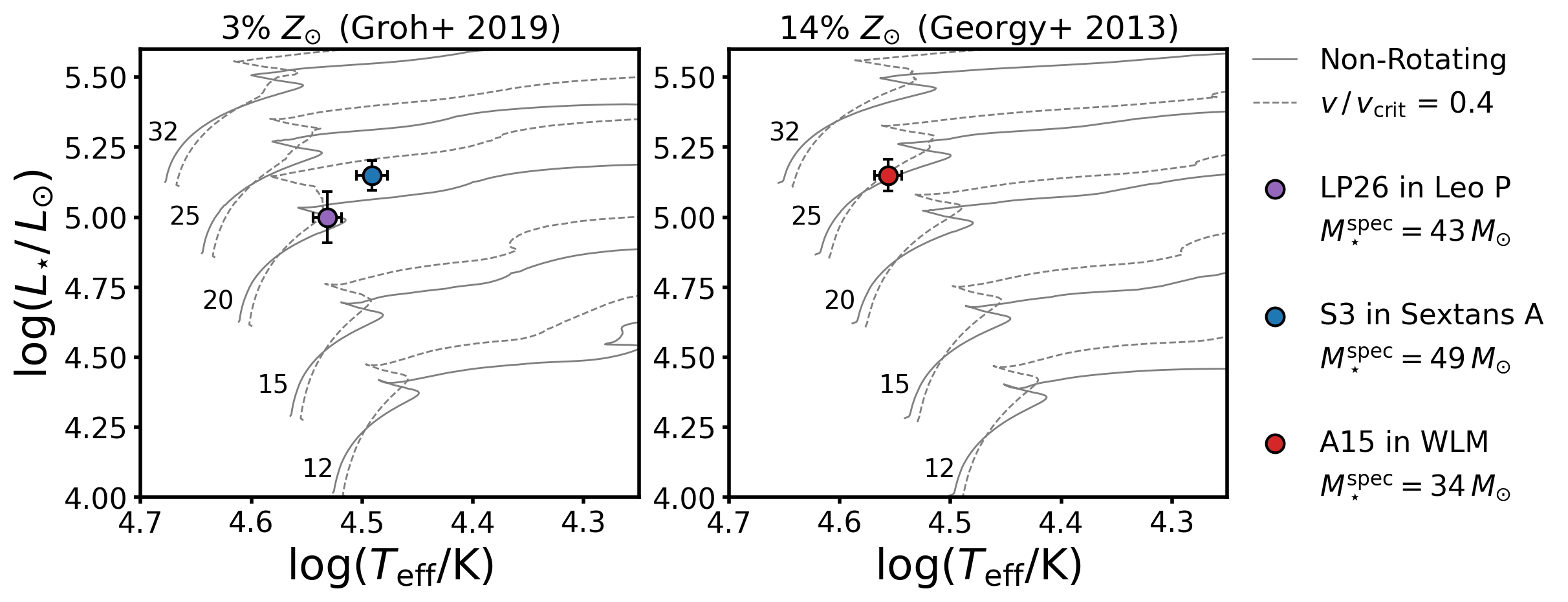}
\caption{\textbf{Comparison to \genec{} stellar evolution models.} \underline{Left}: \lstar{} plotted as a function of \teff{} (both on a logarithmic scale).
Stellar evolution model tracks at 3\%\,\zsun{} from \citet{groh19} are shown as gray lines for initial stellar masses of 12, 15, 20, 25, and 32\,\msun{}.
Solid lines are non-rotating models, while dashed lines are models initially rotating at $0.4\,v_\mathrm{crit}$.
The two stars in our sample with metallicities closer to 3\%\,\zsun{}, LP26 and S3, are shown as purple and blue circles, respectively.
Error bars indicate the uncertainties on our best-fit parameters, estimated as the range of values that produces \powr{} models consistent with the observations of each star (see Section~\ref{sec:procedure} and Table~\ref{tab:params}).
\underline{Right}: Same as the left panel, except that stellar evolution models at 14\%\,\zsun{} from \citet{georgy13} are shown. These higher-$Z$ models are shifted $\sim$0.03 dex toward cooler \teff{} compared to the models in the left panel. 
A15 in WLM is indicated by the red circle with error bars. 
\label{fig:geneva}}
\end{centering}
\end{figure*}

\subsection{Stellar Evolution Models\label{sec:evolution}}

In Figure~\ref{fig:geneva}, we compare the properties inferred from our \powr{} modeling of the three low-$Z$ stars to stellar models calculated with the Geneva Stellar Evolution Code (\genec{}) at 3\% (left panel; \citealt{groh19}) and 14\%\,\zsun{} (right panel; \citealt{georgy13}).
The Geneva models are widely used in SPS codes (e.g., Starburst99; \citealt{leitherer99, leitherer14}) and have been computed for both rotating and non-rotating stars.
In the rotating models, stars are assumed to initially rotate at $v = 0.4 v_\mathrm{crit}$, where $v_\mathrm{crit}$ is the critical velocity at which the star breaks apart.
This yields rotation speeds between 110--220\,km\,s$^{-1}$ on the main sequence \citep{ekstrom12}, comparable to the \vsini{} inferred for the two fast-rotating stars S3 and LP26.

Model tracks of single stars in the Hertzprung-Russell diagram (HRD; i.e., \lstar{} vs.\ \teff{}) are shown as gray lines in Figure~\ref{fig:geneva}, with solid and dashed lines indicating non-rotating and rotating models, respectively.
The initial mass of each model is labeled at the left of each panel, near the zero-age main sequence.
Stars evolve toward higher \lstar{} (upward) and lower \teff{} (to the right), exiting the main sequence when they are no longer burning H in their cores (sharp rightward turns in the model tracks).
The two very low-$Z$ stars in our sample, LP26 and S3, are shown as purple and blue circles, respectively, in the left panel compared to the 3\%\,\zsun{} models of \citet{groh19}.
A15 is shown as a red circle in the right panel for comparison to the models of \citet{georgy13} at 14\%\,\zsun{} (their adopted SMC metallicity).

The locations of the three observed stars in the HRD lie near the evolutionary tracks with initial masses of $\sim$20\,\msun{} (for LP26 and S3) and $\sim$25\,\msun{} (for A15).
This is consistent with the \mstar{} reported by \citet{telford21} from modeling the observed stellar SEDs, constrained by the \textsc{parsec} evolution models \citep{bressan12}: 22$^{+7}_{-5}$, 21$^{+4}_{-2}$, and 29$^{+13}_{-3}$\,\msun{} for LP26, S3, and A15. 
Yet, only A15 is hot enough to be on the main sequence of the considered evolutionary tracks in that mass range, as would be expected from the relatively high \logg{} of all three stars (3.9--4.15; Table~\ref{tab:params}).
Both LP26 and S3 appear to have evolved off the main sequence based on their positions in the HRD alone, but both stars have broad Balmer wings that are inconsistent with the lower $\log(g)\lesssim3.5$ expected for post-main-sequence stars. 
For stars with more efficient mixing processes, the main sequence can extend down to $\sim$$25\,$kK \citep{brott11,higgins19}. 
We also compared our results to the \textsc{parsec} \citep{bressan12, tang14} and \textsc{BoOST} \citep{brott11, szecsi22} stellar evolution models.
For both, the end of the main sequence is indeed shifted to lower \teff{} such that LP26 lies near its end, but S3 still appears more evolved.
Yet, even late-main-sequence stars would not necessarily appear as dwarfs, though this would be in line with recent predictions by \citet{martins21} assuming lower \mdot{} than the \genec{} models.

More striking is the discrepancy between the evolutionary (\mevol{}) and spectroscopic (\mspec{}) stellar masses, where 
\begin{equation}
M_\star^\mathrm{spec}=L_\star \left( \frac{5780}{T_\mathrm{eff}} \right)^4 10^{\log (g) - 4.44},
\end{equation}
with \teff{} in K, $g$ in cm\,s$^{-2}$, and \mspec{}, \lstar{} in solar units \citep{heap06}.
We find \mspec{} of 43$^{+32}_{-18}$, 49$^{+30}_{-19}$, and 34$^{+15}_{-10}$\,\msun{} for LP26, S3, and A15.
These values are higher than the \mevol{} suggested by comparison to the \genec{} models for all three stars, and the uncertainties in stellar parameters estimated from our \powr{} modeling allow for minimum \mspec{} of 25, 30, and 24\,\msun{} for LP26, S3, and A15, respectively. 
The large uncertainties on \mspec{} are consistent with the implied \mevol{} for A15, but there is substantial tension in the two mass estimates for S3 and LP26.
This ``mass discrepancy'' is well known for massive stars in the Milky Way and Magellanic Clouds \citep[e.g.,][]{herrero92, serenelli21}, but it is more common to find \mevol{}$\,>\,$\mspec{}, in the opposite sense of the result here.
Yet, some studies have found \mevol{}$\,<\,$\mspec{} \citep[e.g.,][]{ramachandran18}, especially for high-mass stars $\gtrsim 35$\,\msun{} \citep{markova18, bestenlehner20}.
The cause of the mass discrepancy remains unknown. 

The combinations of \teff{}, \lstar{}, and \mspec{} we infer are challenging to explain, especially for the lower-$Z$ stars LP26 and S3. 
Their \teff{} are lower than typical expectations for main-sequence stars of their \mspec{}, suggesting that they could have an unusual evolutionary history.
At the same time, the \textsc{genec} evolution models for stars above 40\,\msun{} would predict far higher \lstar{} than we observe ($\log(L_\star/L_\odot) \gtrsim 5.5$).
While stars that are observed to be over-luminous can be explained by advanced evolutionary status, or perhaps recent interaction with a companion star, it is harder to identify a physical mechanism that can cause a star to appear under-luminous. 
Therefore, the discrepancy we find for these low-$Z$ stars may be pointing to an incomplete treatment of relevant physics in stellar evolution and/or atmosphere models.
For example, lower-$Z$ stars are expected to be more compact due to lower metal opacities in their atmospheres, but such model predictions have never been empirically calibrated against detailed observations of very metal-poor OB stars.
Thus, it is possible that the physics implemented in current stellar evolution codes result in an incorrect relationship between stellar size and other fundamental properties in the metal-poor regime.
Alternatively, the Balmer-line wings could be impacted by some unaccounted-for physics that impacts their utility as a \logg{} diagnostic at very low $Z$.

Another possibility, particularly at the large distances of these target stars, is that they are unresolved multiple-star systems.
Recent years have seen an accumulation of evidence that the majority of massive stars form in multiple systems \citep[e.g.,][]{sana12, offner23}.
However, detecting a companion star in single-epoch spectroscopy with modest SNR and resolution can be challenging. 
If a system contains two stars with similar \teff{} and brightness, then there may be nothing unusual about the line profiles or strengths of \teff{}-sensitive lines, depending on the orbital phase when the system was observed.
In this case, the only hint of a multiple system would be a higher \lstar{} than expected from single-star evolution models or the measured properties of similar spectral-type stars \citep[e.g.,][]{bouret13, rickard22}.

At first, this seems like a natural solution to the puzzling properties of LP26 and S3.
But inspection of Figure~\ref{fig:geneva} shows that to bring LP26 and S3 onto the main sequence of lower-\mstar{} tracks would require a reduction in \lstar{} of more than a factor of two ($>$\,0.3\,dex), implying multiple similar-mass companions (i.e., triple systems).
Moreover, it is unlikely to find a binary system with two identical spectral types (let alone two such systems), so we would expect to see spectroscopic signatures of different-\teff{} stars if these systems were unresolved binaries, but no such discrepancies are observed. 

We also checked for changes in the stars' radial velocities or absorption line profiles across the individual COS visits and KCWI exposures, and also checked coadds of the first and second half of the KCWI observations for each star (to increase the SNR over individual 20-minute exposures). 
But we found no changes in stellar absorption-line widths or velocities over time suggestive of companions. 
S3 is also included in a recent catalog of Sextans~A OB stars \citep{lorenzo22}. 
Those authors report that the star (s029 in their naming convention) is a single-lined spectroscopic binary, but state that significant radial velocity shifts were only detected in the very weak \trans{he}{ii}{4541} line and do not show the multi-epoch spectroscopy. 
Thus, S3 could be in a multiple-star system, but we do not have clear observational evidence in hand. 
Altogether, while it is tempting to invoke multiplicity to explain the unusual combination of stellar properties that we infer, this scenario does not easily explain all of the observations of LP26 and S3.

Many massive-star binaries will interact during their lifetimes \citep[e.g.,][]{sana12, eldridge22}, raising the possibility that we are observing the products of recent mass-transfer events. 
To assess whether this can explain the properties of LP26 and S3, we compared their locations in the HRD to the predictions of BPASS binary evolution models \citep{eldridge17}, similar to the comparison to single-star models shown in Figure~\ref{fig:geneva}.
The BPASS models follow the evolution of the primary (donor) star following mass transfer, but not the secondary (accretor).
We selected models with initial $M_\star^\mathrm{primary}$ between 15--25\,\msun{}, comparable to the \mevol{} of LP26 and S3 implied by our comparison to \genec{} single-star models, and a representative range of mass ratios (i.e., $M_\star^\mathrm{secondary}/M_\star^\mathrm{primary}$) and orbital periods.
This exercise revealed no BPASS tracks that could simultaneously match the positions of LP26 and S3 in the HRD and their high \logg{}, ruling out the possibility that they are the inflated, evolved primary stars in post-interaction systems.

Both LP26 and S3 are fast rotators, consistent with expectations for secondary stars that were spun up via mass gain from an evolved companion \citep{de-mink13}. 
Elevated N abundances are not required to explain the observations (though the uncertainties allow for up to a factor of 5 higher $X_\mathrm{N}$ than the host galaxies' gas-phase abundances; Table~\ref{tab:params}), so recent accretion from a companion star's unevolved envelope is plausible \citep{maeder09}.
Though relatively few models following the evolution of secondary stars following mass transfer exist in the literature, \citet{renzo21} showed that the internal structure of accretor stars can differ markedly from that predicted by single-star models.
Given that the impacts of accretion on stellar structure and evolution remain poorly understood, further theoretical work in this area will be crucial to constrain possible formation pathways of stars with the observed properties of LP26 and S3.

The primary star in such a binary system could potentially have lost enough mass that it is now a hot, compact stripped star that would appear faint in the FUV and optical compared to a companion O-type star, and therefore be challenging to detect in our data \citep[e.g.,][]{gotberg17}. 
It could also have exploded as a supernova (or directly collapsed to a black hole; e.g., \citealt{sukhbold16}).
While there are no known supernova remnants (SNRs) near LP26 or S3, it is entirely possible that supernovae did explode nearby fairly recently but have now faded from view (given SNR lifetimes of $\sim10^5$\,yr). 
Yet, \citet{telford21} found no large offsets between the stellar radial velocities and the systemic velocities of the host galaxies that would suggest that the stars were ejected from the natal system by a supernova kick.
We also lack evidence for a disk of accreted material around either star (e.g., double-peaked Balmer emission). 

Alternatively, the stars could be the products of mergers, which are expected to occur in $\sim$30\% of massive-star binaries \citep{sana12}. 
After a merger, stars are predicted to undergo a luminous phase during which they rapidly rotate before shedding mass and angular momentum and contracting back to the main sequence as slowly rotating magnetic stars \citep{schneider19}.
LP26 or S3 could potentially be observed in the over-luminous and fast-rotating post-merger phase, but this cannot account for their high \logg{}. 
Moreover, it would be very unlikely to find two stars in such a short-lived evolutionary phase.

To conclude, we have presented a surprising set of inferred properties for both very metal-poor O stars in our sample, though the star A15 appears consistent with single-star evolution models within the uncertainties. 
It is not clear that either effectively single or interacting binary companions can explain all of the available observations of LP26 and S3, though we cannot definitively rule out that possibility. 
Additional observations of a larger sample of metal-poor O stars are required to assess whether these two stars are outliers or typical at very low $Z$.
These results should inform future development of massive-star evolution models below 20\%\,\zsun{}.

 
 \subsection{Stellar Mass-Loss Rates\label{sec:massloss}}
 
Stellar evolution codes must account for massive stars' mass loss over their lifetimes \citep[e.g.,][]{brott11, bressan12, ekstrom12}.
These models typically adopt prescriptions for \mdot{} as a function of stellar properties, particularly \lstar{}, \teff{}, and $Z$, calibrated by detailed theoretical calculations of line driving in hot-star atmospheres.
Empirical \mdot{} measured from optical and FUV spectra of stars in the Milky Way and Magellanic Clouds have been used to test these models \citep[e.g.,][]{mokiem07, ramachandran19, marcolino22}, but few \mdot{} measurements for massive stars in galaxies below the 20\%\,\zsun{} of the SMC have been reported \citep[e.g.,][]{tramper14, bouret15}. 
Moreover, the analysis of three stars in IC~1613 and WLM by \citet{bouret15} suggests that those galaxies may have SMC-like Fe abundances, consistent with our results for the star A15 in WLM.
Thus, substantial uncertainty remains in the dependence of \mdot{} on $Z$, especially in the very low-$Z$ regime.

 \begin{figure}[!t]
\begin{centering} 
  \includegraphics[width=\linewidth]{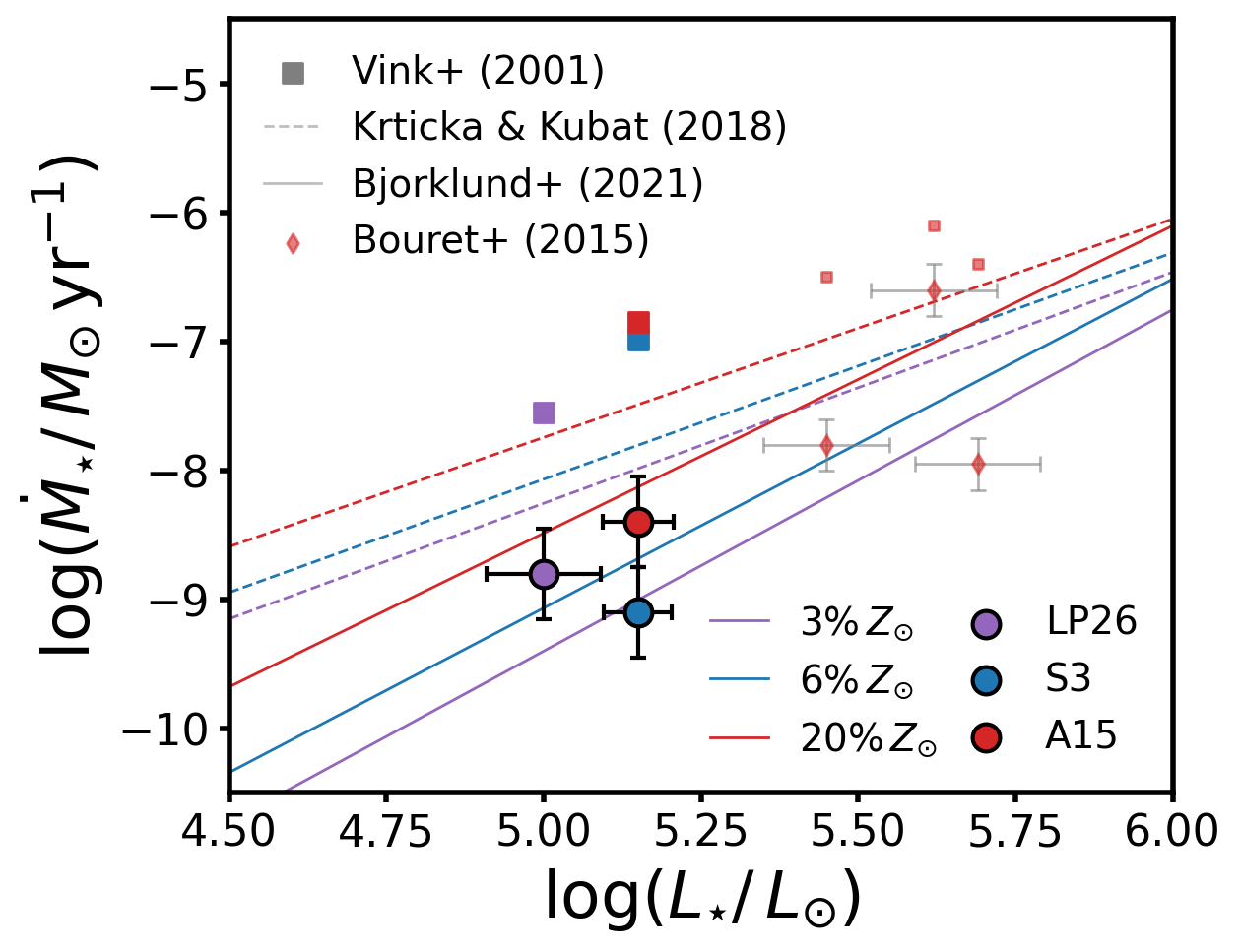}
\caption{\textbf{Comparison to theoretical mass-loss rates.} The power-law relationships between \mdot{}, \lstar{}, and $Z$ reported by \citet{krticka18} and \citet{bjorklund21} are shown as solid and dashed lines, respectively. The \citet{vink01} \mdot{} prescription depends upon additional parameters (\teff{}, \mspec{}, \vinf{}/$v_\mathrm{esc}$), so we calculated that model's predictions for the inferred stellar parameters for each of the three stars in our sample, shown as squares. Red, blue, and purple correspond to metallicities of 20\%, 6\%, and 3\%\,\zsun{} for both model predictions and for the three stars analyzed here (circles with error bars). 
For comparison, we also show the measurements of \citet{bouret15} for three O stars in WLM and IC~1613 as light red diamonds (all three had inferred $X_\mathrm{Fe}$ of 20\% solar).
The inferred \mdot{} for our low-$Z$ stars lie at the low-\mdot{} end of the range of model predictions.
\label{fig:massloss}}
\end{centering}
\end{figure}

The stellar and wind properties inferred from our \powr{} models enable a first comparison to the predictions of theoretical mass-loss recipes extending to substantially sub-SMC metallicities. 
In Figure~\ref{fig:massloss}, we plot \mdot{} as a function of \lstar{} for the three stars in our observational sample as colored circles, where purple, blue, and red correspond to LP26, S3, and A15, respectively (same as in Figure~\ref{fig:geneva}).
For comparison, we show predicted power-law relationships between \mdot{} and \lstar{} at 3\%, 6\%, and 20\%\,\zsun{} (purple, blue, and red, corresponding to our adopted Fe abundances for the three observed stars) from \citet{bjorklund21} and \citet{krticka18} as solid and dashed lines, respectively. 
We also calculate the \mdot{} predicted by the widely used \citet{vink01} prescription given the best-fit \lstar{}, \teff{}, \mspec{}, and the ratio of \vinf{} to the surface escape velocity (\vesc{}) for each star, shown as squares with colors encoding $Z$ as for the other two model predictions.
If we adopt the approximate \mevol{} of the three stars instead, the \citet{vink01} \mdot{} predictions increase by $\sim$0.2-0.5\,dex.

Of the three models to which we compare, the empirical \mdot{} we obtain for these three metal-poor O stars agree best with the \citet{bjorklund21} mass-loss recipe. 
The best-fit \mdot{} for all three stars lie well below the predictions across 3--20\%\,\zsun{} of both the \citet{vink01} and \citet{krticka18}, except perhaps for LP26, which lies between the \citealt{bjorklund21} and \citealt{krticka18} models. 
Importantly, this result holds despite uncertainties due to the possibility of artificially high \lstar{} due to companion stars: if the stars' true \lstar{} are lower than our inferred values by $\sim$0.2--0.6\,dex (see Section~\ref{sec:evolution}), then the inferred $\log(\dot{M}_\star)$ would also shift lower by $\Delta\log(\dot{M}_\star)=0.75\Delta\log(L_\star)$ to reproduce the observed FUV wind line profiles. 
Even if all three stars shifted toward lower \lstar{} (to the left in Figure~\ref{fig:massloss}) in that scenario, the inferred \mdot{} values would remain lower than predicted by either the \citet{vink01} or \citet{krticka18} models.
Thus, uncertainty in these parameters cannot bring our results into agreement with the higher-\mdot{} models.

 \begin{figure*}[!t]
\begin{centering} 
  \includegraphics[width=\linewidth]{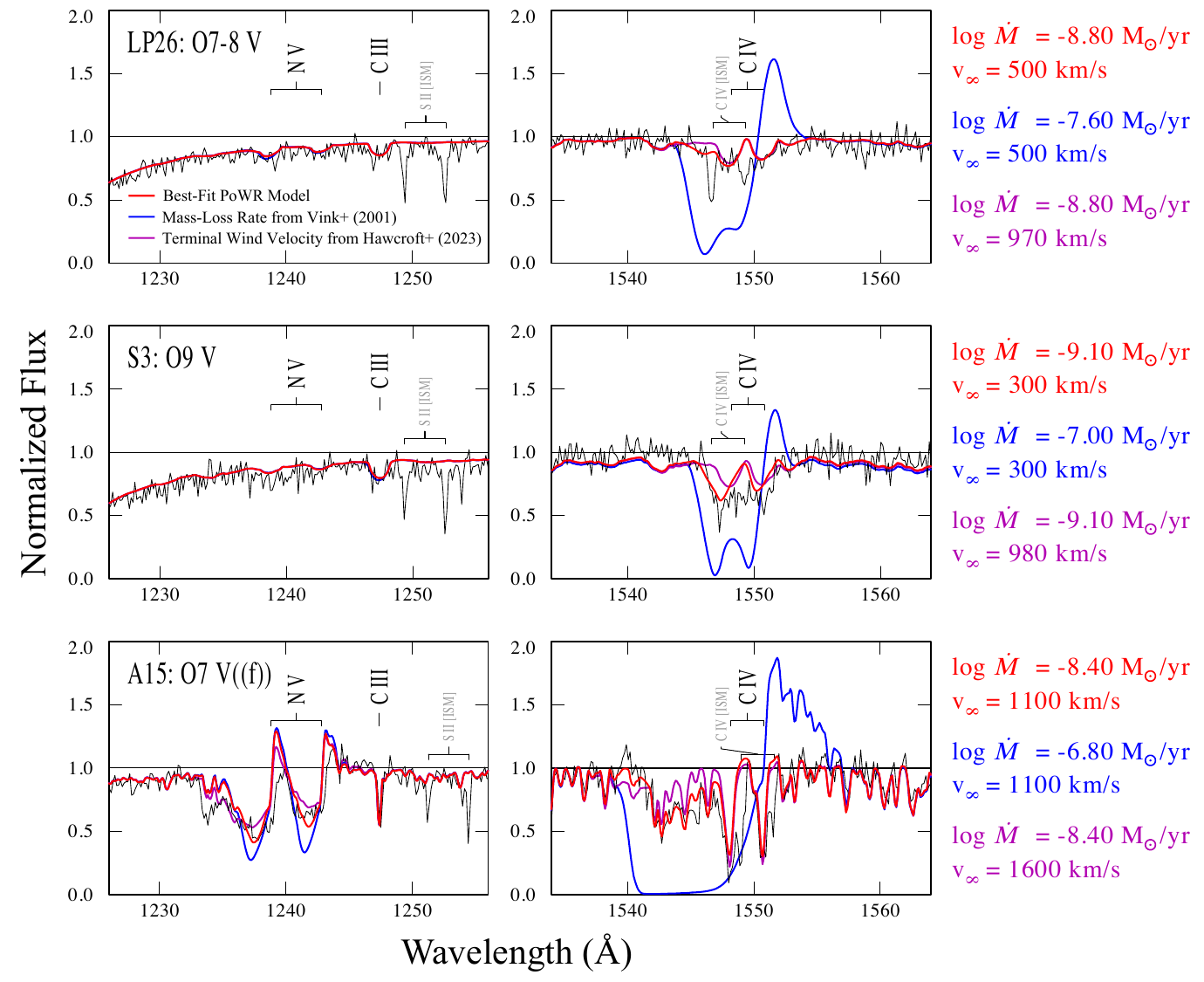}
\caption{\textbf{Impact of $\bm{\dot{M}_\star}$ and $\bm{v_\infty}$ on FUV wind profiles.} 
Normalized flux plotted as a function of rest-frame wavelength for the regions surrounding the two main FUV wind diagnostic features: \trans{N}{v}{1238,\,1242} (left column) and \trans{C}{iv}{1548,\,1550} (right column).
The observations for the three stars are shown in black for LP26, S3, and A15 from top to bottom. 
Three models are shown in each panel: in red, our best-fit \powr{} model for that star (presented in Section~\ref{sec:best_models}); in blue, a model with the same parameters except for \mdot{} set equal to the prediction from the \citet{vink01} mass-loss recipe (Section~\ref{sec:massloss}); and in purple, a model with the same parameters as the best fit except for \vinf{} set equal to the expectation from extrapolating the empirical \vinf{} vs.\ \teff{} scaling of \citet{hawcroft23} to lower $Z$ (Section~\ref{sec:windspeed}).
The \citet{vink01} \mdot{} yield FUV wind profiles that are obviously inconsistent with the observations.
Similarly, the \vinf{} expected from the \citet{hawcroft23} relation causes the \spec{c}{iv} profiles to extend too far to the blue.
The purple models produce nearly photospheric \spec{c}{iv} absorption for S3 and LP26 that are clearly too narrow compared to the observed profiles, demonstrating that low, but non-negligible, \mdot{} are required.
\label{fig:windlines}}
\end{centering}
\end{figure*}

The need for low \mdot{} to explain the observations of these stars is emphasized by Figure~\ref{fig:windlines}, which shows the \spec{n}{v} (left column) and \spec{c}{iv} (right column) wind profiles for the three stars.
As in Figures~\ref{fig:lp26}--\ref{fig:a15}, the COS spectra are shown in black and our best-fit \powr{} models are shown in red, with the best \mdot{} and \vinf{} values reported in red text. 
For comparison, the blue lines show \powr{} models computed with the \mdot{} predicted by the \citet{vink01} mass-loss recipe, reported in blue text (and shown as squares in Figure~\ref{fig:massloss}).
All other parameters are held fixed, and we checked that setting $D_\infty=1$ to ensure a fair comparison to the \citet{vink01} models produced negligible changes in the resultant wind profiles. 
The \spec{c}{iv} wind profiles in the blue models with high \mdot{} are clearly ruled out by the observations. 

The purple lines in Figure~\ref{fig:windlines} show \powr{} models with higher \vinf{}.
We discuss these models in Section~\ref{sec:windspeed} below, but here remark that for both LP26 and S3, this change causes the \spec{c}{iv} profiles to appear nearly photospheric (i.e., not showing any signature of a stellar wind). 
The photospheric \spec{c}{iv} lines in the purple models do not reproduce the observed \spec{c}{iv} profiles, which have more extended blueshifted absorption. 
Thus, low but non-negligible \mdot{} are required to explain the FUV observations of these two very low-$Z$ stars.

The analysis presented here provides empirical evidence that winds of low-\lstar{} O stars below 20\%\,\zsun{} are best described by theoretical models with lower \mdot{} values.
The only other reported \mdot{} from FUV spectra of O stars in galaxies more metal-poor than the SMC are shown for comparison in Figure~\ref{fig:massloss} \citep{bouret15}.
Their three higher-\lstar{} targets in WLM and IC~1613 are shown as light red diamonds, along with their expected \mdot{} from the \citet{vink01} prescription.
While those authors inferred SMC-like $X_\mathrm{Fe}$ of 20\% solar for all three of their targets, they too found lower \mdot{} than predicted by the \citet{vink01} model.
Interestingly, there is substantial scatter in their \mdot{} measurements, such that the highest-\mdot{} star is consistent with the \citet{krticka18} model -- but this is at the high-\lstar{} extreme where differences among the models are smaller. 
Taken together, the observations of all six O stars in low-$Z$ galaxies agree best with the \citet{bjorklund21} models and favor lower \mdot{} than typically assumed in the literature.

Our low \mdot{} are also consistent with recent studies of OB stars in the Milky Way and Magellanic Clouds that found evidence for a steeper decline in \mdot{} with $Z$ \citep{ramachandran19} and \lstar{} \citep{rickard22} than expected from either mass-loss models, or from earlier empirical studies \citep[e.g.,][]{mokiem07}.
\citet{bjorklund21} also show that their predicted \mdot{} scaling yields a better match than that of \citet{vink01} to empirical \mdot{} inferred for SMC O stars by \citet{bouret13}.
Both theoretical and empirical work have also proposed a $Z$-dependent slope of the \mdot{}--\lstar{} relationship \citep[e.g.,][]{bjorklund21, marcolino22}, but its strength (and even sense) remains unsettled.
Our sample is too small and diverse in $Z$ to characterize the dependence of \mdot{} on both $Z$ and \lstar{} in the very metal-poor regime, but this open question motivates future analysis of FUV stellar wind features for a larger sample of low-$Z$ O stars spanning a wider range of \lstar{}.

The reported uncertainties on \mdot{} reflect the range of values that produce wind profiles consistent with the observations, given the best-fit values of all other free parameters in our \powr{} models, plus an additional 0.15\,dex to account for the factor of 2 uncertainty in $D_\infty$ (see Section~\ref{sec:powr}).
Yet, particularly at low \lstar{}, some wind material could be shocked to high enough temperatures to emit X-rays \citep[e.g.,][]{lucy80, lagae21}.
It is therefore possible that the total \mdot{} across all phases of the wind could be higher than the \mdot{} only in the phase traced by the wind profiles in the FUV observations. 
So far, no insights about the strength of wind-intrinsic X-rays exist at very low $Z$, so the FUV-based analysis presented here represents the state-of-the-art constraint on their wind properties.

In principle, X-rays from a shocked wind could also impact the modeled FUV wind profiles.
For the temperature regimes of our targets, atmosphere models use additional X-rays mainly to increase the strength of the \trans{n}{v}{1238,\,1242} line. 
Yet, this line is already well reproduced in our models, even without including additional X-rays. 
Moreover, independent analyses of the same observed stars with and without including X-rays in atmosphere models have yielded similar wind parameters \citep[e.g.,][]{bouret13, rickard22}.
We conclude that while treating X-rays might be able to improve the \powr{} models' ability to match simultaneously the \spec{n}{v} and \spec{c}{iv} wind profiles, such modifications are unlikely to change our inferred \mdot{} enough to reconcile our results with the order-of-magnitude higher \mdot{} predictions of some models seen in Figure~\ref{fig:massloss}.


\subsection{Terminal Wind Velocities\label{sec:windspeed}}

 \begin{figure}[!t]
\begin{centering} 
  \includegraphics[width=\linewidth]{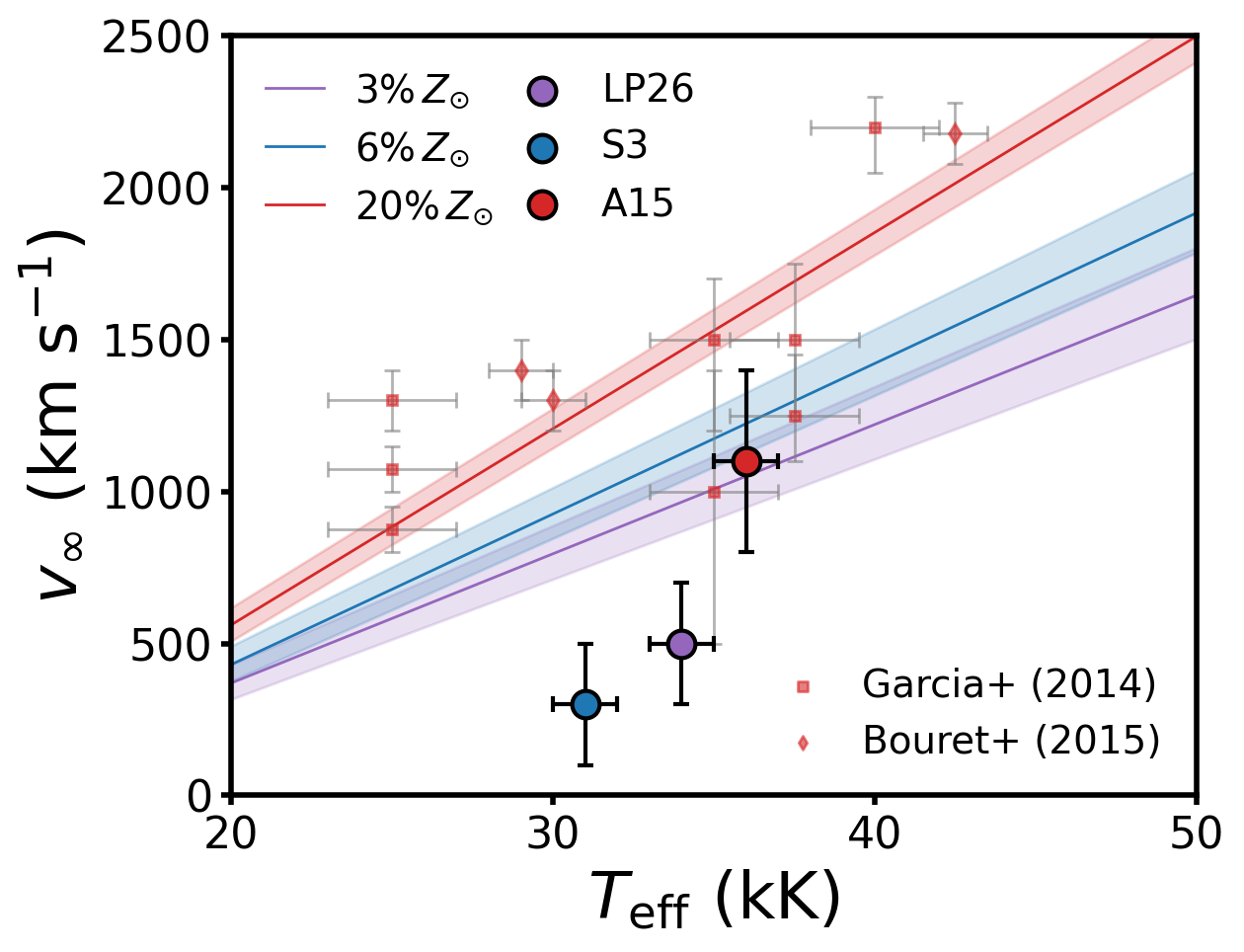}
\caption{\textbf{Comparison to extrapolation of empirical terminal wind velocities at higher $\bm{Z}$.} Solid lines show the scaling relation between \vinf{}, \teff{}, and $Z$ fit to observations in the Milky Way, LMC, and SMC by \citet{hawcroft23}, extrapolated to lower $Z$ (color-coded as in Figure~\ref{fig:massloss}). 
Circles with error bars show the measured \vinf{} as a function of \teff{} for the three stars analyzed here.
For comparison, we also show measurements for O stars in WLM and IC~1613 (all with inferred SMC-like $X_\mathrm{Fe}$) drawn from \citet{garcia14} and \citet{bouret15} as light red squares and diamonds, respectively.
We find lower \vinf{} for all three metal-poor stars studied here than expected from the best-fit relation to observations at 20--100\%\,\zsun{}, though our result for A15 in WLM lies within the scatter of published \vinf{} of individual stars at similar $Z$ and \teff{} in metal-poor galaxies beyond the SMC. 
\label{fig:vfinal}}
\end{centering}
\end{figure}

A complete physical understanding of radiation-driven winds, particularly their momenta and mechanical feedback deposited into the surrounding ISM, depends critically on \vinf{} \citep[e.g.,][]{castor75, weaver77}.
For Milky Way OB stars, \vinf{} has long been known to correlate with both \vesc{} $\left(\propto \sqrt{M_\star/R_\star}\right)$ and \teff{} \citep[][]{kudritzki00}. 
Its predicted weak $Z$ dependence, $v_\infty \propto Z^{0.13}$ \citep{leitherer92}, is generally found to be consistent with observations. 
For example, recent empirical studies of \vinf{} measured from the blueward extent of FUV wind-line profiles of OB stars in the Milky Way, LMC, and SMC reported similar $v_\infty \propto Z^{0.08-0.22}$ \citep{marcolino22, hawcroft23}.
Yet, like for \mdot{}, few \vinf{} measured from FUV wind profiles have been reported for OB stars in lower-$Z$ galaxies than the SMC, leaving open the question of whether the same behavior holds in the extremely metal-poor regime.

Figure~\ref{fig:vfinal} shows the best-fit \vinf{} as a function of \teff{} for our three metal-poor stars as circles (colored by adopted Fe abundance as in Figures~\ref{fig:geneva} and \ref{fig:massloss}).
For comparison, we show the \vinf{} vs. \teff{} relationships of \citet{hawcroft23} at the adopted $Z$ of each star as solid lines, extrapolated from the empirical relationship between \vinf, $Z$, and \teff{} fit to observed OB stars at 20--100\%\,\zsun{}.
Shaded regions indicate the uncertainty in their model parameters, not the typical scatter in individual \vinf{} measurements ($\pm$500\,km\,s$^{-1}$), which is larger than the shaded regions.
For all three metal-poor stars, the inferred \vinf{} lies well below the expectation for their \teff{} and $Z$ from the \citet{hawcroft23} relationship.

Figure~\ref{fig:windlines} illustrates that the \vinf{} values expected from the empirical scaling for Milky Way, LMC, and SMC O stars is incompatible with the observations of our metal-poor stars.
The purple lines show \powr{} models with the same parameters as our best-fit models (red lines), except that \vinf{} has been set to the value calculated from the \citet{hawcroft23} relation for the star's \teff{} and $Z$. 
As discussed in Section~\ref{sec:massloss}, the higher \vinf{} at fixed \mdot{} changes the \spec{c}{iv} profiles in the LP26 and S3 models such that they are nearly photospheric.
Yet, for S3 (middle row), the purple line is discernibly below the red line just blueward of the \spec{c}{iv} blue edge in the best model. 
If \mdot{} is increased, that blueward absorption becomes more pronounced, producing a \spec{c}{iv} profile that is clearly affected by the wind, and also inconsistent with the data (black lines). 
The same is true for the LP26 model; both  extremely low-$Z$ stars require low \vinf{} to reproduce the blueward extent of their observed \spec{c}{iv} profiles.
For A15, the \citet{hawcroft23} \vinf{} produces broader \spec{n}{v} and \spec{c}{iv} wind profiles; the former is a comparably good match to the data as the best-fit model, but the \spec{c}{iv} absorption extends beyond the observed blue edge.
Like for the other two stars, \mdot{} would need to be increased to compensate for the higher \vinf{}, producing an even more discrepant \spec{c}{iv} profile, so the higher \vinf{} is ruled out.
Still, our best-fit \vinf{} of 1100\,\kms{} for A15 is consistent within the scatter of measurements for individual SMC stars in \citet{hawcroft23}.

To place our results in context, Figure~\ref{fig:vfinal} also shows published \vinf{} of 11 OB stars in the metal-poor galaxies WLM and IC~1613 inferred from FUV spectroscopy.
Measurements from \citet{garcia14} and \citet{bouret15} are shown as light red squares and diamonds, respectively.
Both papers report that all of their targets have SMC-like $X_\mathrm{Fe}$ and sample luminosity classes from dwarfs to supergiants.
As expected for that Fe abundance, the empirical \vinf{} from the literature scatter about the best-fit \vinf{}--\teff{}--$Z$ relationship of \citet{hawcroft23} at 20\%\,\zsun{}.
Our best-fit \vinf{} for the star A15 in WLM, with the same inferred $X_\mathrm{Fe}$, lies below the empirical \vinf{} vs. \teff{} for the SMC, but interestingly, lies within a group of stars in IC~1613 that scatter to lower \vinf{} than the overall trend.
Though the sample size is small, this cluster of lower-\vinf{} stars around $\sim$35--38\,kK is striking.
Future observations of a larger sample of OB stars in galaxies with gas-phase O abundances below 20\%\,\zsun{} would help to clarify whether this decrease in typical \vinf{} is real, or if the scatter across the full \teff{} range is larger and extends to lower \vinf{} than suggested by the currently available data.

Radiation-driven wind theory predicts a linear scaling between \vinf{} and \vesc{} set by the force multiplier parameter, $\alpha$ \citep[e.g.,][]{kudritzki89, lamers99}. 
Empirical studies of Galactic hot stars ($\geq 21$\,kK) have found a ratio of \vinf{}/\vesc{}$\,\simeq\,$2.65, with lower ratios observed for cooler stars with winds driven by different ions \citep{lamers95, kudritzki00}. 
However, it is unclear whether a constant \vinf{}/\vesc{} ratio holds at lower $Z$ \citep{garcia14, hawcroft23}.
Given the \vesc{} in our \powr{} models of the three metal-poor stars (1300, 1230, and 1100 \kms{} for LP26, S3, and A15, respectively), we find \vinf{}/\vesc{} of 0.38, 0.24, and 1.0.
These values are well below the canonical ratio of 2.65, and lower for the two very metal-poor stars.
This may hint at an extension of the trend for lower \vinf{}/\vesc{} for stars in the SMC compared to the LMC reported by \citet{hawcroft23}.

Echoing the earlier discussion in Section~\ref{sec:massloss}, these \vinf{} measurements are for the phase of the wind that can be probed by FUV spectroscopy.
It is possible that these stars have higher \mdot{} and/or \vinf{} if super-ionization due to shocks is indeed important in weak winds.
Unfortunately, we do not have observational diagnostics of this scenario (e.g., X-rays, or high-ionization UV lines like \trans{o}{vi}{1032,\,1038}; \citealt{puls20}).
The lack of strong \spec{n}{v} wind features in the two stars with lower $Z$ and \mdot{} suggests that super-ionization is not important in their winds, but examples of X-ray-emitting Galactic O stars that also have weak \spec{n}{v} features are known (e.g., \citealt{martins05, huenemoerder12}).
\vinf{} inferred from unsaturated P-Cygni profiles are often regarded as lower limits \citep{lamers95, crowther16}.
Yet, encouragingly, \citet{hawcroft23} showed that even weak-wind stars in their sample follow the same \vinf{} scaling relationships as the stars with saturated wind profiles.
Moreover, the stars with \vinf{} drawn from the literature shown in Figure~\ref{fig:vfinal} show stronger P-Cygni profiles than the three metal-poor stars analyzed here, and still are consistent with our results.
Given the available data, our analysis provides the best constraints to date on the strength of stellar winds in the very low-$Z$ regime.


\subsection{Ionizing Spectra\label{sec:ionizing_spectra}}

 \begin{figure*}[!t]
\begin{centering} 
  \includegraphics[width=\linewidth]{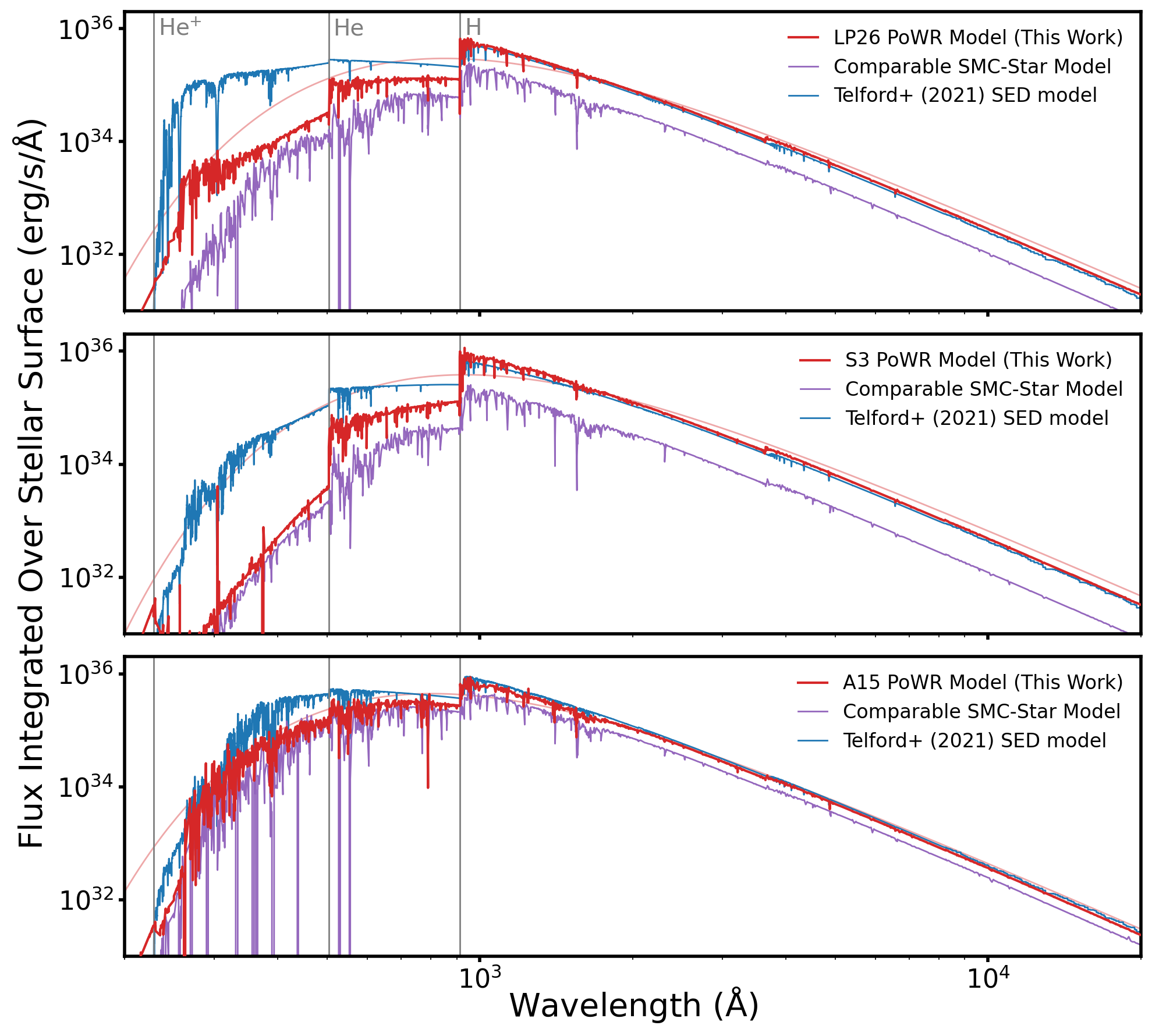}
\caption{\textbf{Comparison of model ionizing spectra.} Total flux emitted from the stellar surface plotted as a function of wavelength from 20\,\AA{} to 2\,$\mu$m, both on logarithmic scales, for LP26, S3, and A15, ordered from top to bottom.
Vertical gray lines indicate the wavelengths below which photons are energetic enough to ionize H, He, and He$^+$.
The dark red line in each panel shows the best-fit \powr{} model presented in Section~\ref{sec:best_models}.
For reference, a blackbody spectrum for the same \teff{} and \rstar{} as the best-fit \powr{} model is plotted as the light red line, highlighting that even at very low $Z$, a blackbody is not an appropriate approximation of an O star's ionizing spectrum.
The best-fit models to the observed stellar SEDs reported by \citet{telford21} are shown in blue; these ionizing spectra are all harder than the \powr{} models reported here due to their higher \teff{} (Table~\ref{tab:model_comparison}; Section~\ref{sec:seds}). 
Finally, the purple models show the theoretical expectations for stars with similar \teff{} and \logg{} to each of the three observed stars at SMC metallicity, drawn from the \citet{hainich19} grid of \powr{} models.
The lower \lstar{} of the theoretical SMC models results in lower ionizing flux compared to the best models of our three metal-poor O stars.
\label{fig:ionizing_seds}}
\end{centering}
\end{figure*}

An important application of stellar evolution and spectral models at very low $Z$ is estimating the ionizing photon production rate of massive stellar populations in low-mass galaxies.
This impacts our understanding of the role played by early, chemically unevolved galaxies in the process of cosmic reionization, as well as feedback and nebular emission in low-$Z$ galaxies at all cosmic epochs. 
Yet, because we lack observational constraints on \mdot{} prescriptions, stellar evolution, and the ionizing spectra of O stars below 20\%\,\zsun{}, substantial uncertainty remains in the shape and normalization of massive stellar populations' ionizing spectra in the very low-$Z$ regime.
The atmosphere models tailored to FUV and optical spectra of very low-$Z$ O stars presented here enable a first comparison to theoretical models representative of those adopted in SPS codes to interpret the light observed from metal-poor galaxies.


\subsubsection{Comparison to Theoretical Atmosphere Models at SMC Metallicity}

\begin{table}
\begin{center}
\caption{Comparison to other model SEDs.\label{tab:model_comparison}}
\tabcolsep=0.25cm
\begin{tabular}{lccc} 
Parameter & LP26 & S3 & A15 \\
\hline 
\multicolumn{4}{c}{\textbf{\powr{} Models (This Work)}} \\
\teff{} (kK) & 34 & 31 & 36 \\
$\log(g/\mathrm{cm\,s}^{-2})$ & 4.15 & 3.9 & 4.0 \\
$\log(L_\star/L_\odot)$ & 5.0 & 5.15 & 5.15 \\
\mstar{} (\msun{}) & 43 & 49 & 34 \\
$\log(Q(\mathrm{H)/s}^{-1})$ & 48.26 & 48.13 & 48.63 \\
$\log(Q(\mathrm{He)/s}^{-1})$ & 46.78 & 45.69 & 47.51 \\
\qhe{}/\qh{} & 0.034 & 0.003 & 0.075 \\
\hline
\multicolumn{4}{c}{\textbf{SMC OB-III \powr{} Models}} \\
\teff{} (kK) & 34 & 31 & 36 \\
$\log(g/\mathrm{cm\,s}^{-2})$ & 4.2 & 4.0 & 4.0 \\
$\log(L_\star/L_\odot)$ & 4.55 & 4.54 & 4.97 \\
\mstar{} (\msun{}) & 17 & 15 & 23 \\
$\log(Q(\mathrm{H)/s}^{-1})$ & 47.87 & 47.61 & 48.49 \\
$\log(Q(\mathrm{He)/s}^{-1})$ & 46.38 & 45.43 & 47.39 \\
\qhe{}/\qh{} & 0.033 & 0.007 & 0.080 \\
\hline
\multicolumn{4}{c}{\textbf{\citet{telford21} SED Models}} \\
\teff{} (kK) & 37.5 & 32.5 & 37.5 \\
$\log(g/\mathrm{cm\,s}^{-2})$ & 4.0 & 3.5 & 4.0 \\
$\log(L_\star/L_\odot)$ & 5.1 & 5.1 & 5.3 \\
\mstar{} (\msun{}) & 22 & 21 & 29 \\
$Z$/\zsun{}, Grid & 2\%, W & 3\%, V & 20\%, S \\
$\log(Q(\mathrm{H)/s}^{-1})$ & 48.63 & 48.55 & 48.84 \\
$\log(Q(\mathrm{He)/s}^{-1})$ & 47.89 & 47.30 & 48.00 \\
\qhe{}/\qh{} & 0.184 & 0.056 & 0.146 \\
\end{tabular}
\end{center}
\vspace{-10pt}
\tablecomments{\underline{Top}: key properties of the \powr{} models related to the normalization and shape of the stellar SED. \lstar{} and \mstar{} are reproduced from Table~\ref{tab:params}. \underline{Middle}: comparison to the most similar \powr{} models in a theoretical grid at SMC metallicity \citep{hainich19}. \underline{Bottom}: comparison to the best-fit models to the stellar SEDs constructed from the \textsc{ostar2002} grid of \tlusty{} models \citep{lanz03} and \textsc{parsec} evolution models \citep{bressan12}. }
\end{table}

The top section of Table~\ref{tab:model_comparison} summarizes the key properties that impact ionizing photon production for our best-fit \powr{} models to the three metal-poor O stars.
\teff{}, \logg{}, \logl{}, and \mstar{}, all of which set the shape and normalization of the ionizing SED, are reproduced from Table~\ref{tab:params}.
We also report the production rate of photons capable of ionizing H (\qh{}; $\geq 13.6$\,eV) and He (\qhe{}; $\geq 24.6$\,eV) in the models.
These models predict $\log(Q(\mathrm{H)/s}^{-1})$ ranging from 48.13--48.63, broadly consistent with previous atmosphere modeling of observed O7--9 stars in the SMC \citep{ramachandran19, rickard22}.

First, we consider how our results compare to expectations from a purely theoretical grid of model SEDs at low $Z$. 
The middle section of Table~\ref{tab:model_comparison} summarizes the properties of models with similar \teff{} and \logg{} to our best-fit models to the three metal-poor stars drawn from the \citet{hainich19} grid of \powr{} models at SMC metallicity.
Figure~\ref{fig:ionizing_seds} also presents a graphical comparison of the emergent flux integrated over the stellar surface as a function of wavelength for LP26, S3, and A15, ordered from top to bottom.
Our best-fit \powr{} models (the same models shown in the top-left panels of Figures~\ref{fig:lp26}--\ref{fig:a15}) are shown as red lines, and the \citet{hainich19} models with comparable stellar parameters at SMC metallicity are shown as purple lines.
For reference, radiation to the left of the gray vertical lines is capable of ionizing H, He, and He$^+$ (from right to left). 

The most obvious difference between our \powr{} models and the SMC-star models in Figure~\ref{fig:ionizing_seds} is the normalization, particularly for LP26 and S3.
The \powr{} model grid adopts \lstar{} for a given \teff{} and \logg{} from single-star evolution models \citep{brott11}, and these theoretical \lstar{} are lower than our best-fit values (similar to the discussion in Section~\ref{sec:evolution}).
\lstar{} strongly affects the normalization of the stellar SED, and therefore the absolute \qh{} and \qhe{}: both are higher in our \powr{} models than in the comparable SMC-star models.
The high \lstar{} inferred for our three metal-poor O stars thus imply higher ionizing photon production rates than would be predicted from purely theoretical models of stellar evolution and ionizing spectra.

On the other hand, we find good agreement in the hardness of the ionizing spectra.
While small changes in the shapes of the red and purple models are visually apparent in Figure~\ref{fig:ionizing_seds}, particularly for He-ionizing photons, the y-axes span 5 orders of magnitude, so subtle differences where the flux is low do not substantially change \qhe{}/\qh{}.
Interestingly, this comparison suggests little evolution in the hardness of O stars' ionizing spectra across a factor of several in $Z$, from 3--20\%\,\zsun{}.


\subsubsection{Comparison to SED-Fitting Results\label{sec:seds}}

Next, we compare to the ionizing spectra predicted from our earlier modeling of these stars' SEDs, published before optical spectra were available to constrain their \teff{} and \logg{}.
Stellar photometry is readily available for many metal-poor galaxies in the local universe, whereas optical spectroscopy of the necessary resolution and SNR is expensive and therefore rare.
Thus, it is important to understand how SED-based constraints on \qh{} and the hardness of ionizing spectra compare to the quantitative spectroscopic analysis presented here.

The bottom section of Table~\ref{tab:model_comparison} reproduces the \teff{}, \logg{}, \logl{}, and \mstar{} estimated  for the three metal-poor O stars by \citet{telford21} from fitting their HST/COS FUV spectra and HST NUV-through-NIR photometry.
We also report \qh{}, \qhe{}, and \qhe{}/\qh{} for those best-fit SED models, drawn from \citet{telford23} for LP26 and newly calculated for S3 and A15 here by integrating the model fluxes above the minimum photon energy required to ionize H and He.
The updated \teff{} and \logl{} reported here for our best-fit \powr{} models are consistent within the (large) uncertainties on the SED-based quantities. 
However, the SED-based \teff{} are also higher for all three stars than found in our models fit to the Keck/KCWI optical spectra in particular (and informed by all available observations), which enabled the \citet{telford21} models to account for the high \lstar{} required to match the normalization of the observed SEDs (because $L_\star \propto T_\mathrm{eff}^4$) within the constraints of the \textsc{parsec} single-star evolution models adopted in that analysis. 

The impact of systematically higher \teff{} in the SED-based models is apparent in Figure~\ref{fig:ionizing_seds}.
The \citet{telford21} models are shown as blue lines, all of which have higher \qh{} and \qhe{} compared to the best-fit \powr{} models (red lines), though both sets of models reproduce the observed FUV-through-NIR parts of the SEDs. 
The SED-based models also predict harder ionizing spectra (i.e., higher \qhe{}/\qh{}) for all three stars than our best-fit \powr{} models.
This tendency toward higher \teff{} and \qhe{}/\qh{} to explain high \lstar{} (as would be observed for unresolved binaries) should be kept in mind for SED-fitting analyses of resolved massive-star populations, and underscores the importance of optical spectroscopy to constrain \teff{} for O stars.


\subsubsection{The Special Case of LP26\label{sec:lp26_hiiregion}}

Finally, we consider LP26, the one star in our sample for which additional constraints on the ionizing spectrum are available from its surrounding \hii{} region.
Analysis of the nebular emission lines in the Keck/KCWI data constrained $\log(Q(\mathrm{H)/s}^{-1}) = 48.57^{+0.07}_{-0.09}$ and $Q(\mathrm{He})/Q(\mathrm{H})=0.12\pm0.03$ \citep{telford23}.
Interestingly, while the SED-based \qh{} is consistent with the constraint from the \hii{} region emission, the \qh{} predicted by our best-fit \powr{} model is lower (Table~\ref{tab:model_comparison}). 
The \hii{}-region analysis adopted an escape fraction $f_\mathrm{esc}=0$, so if ionizing photons are escaping the \hii{} region, then the true \qh{} of the ionizing O star would need to be even higher, and thus in stronger tension with the results presented here.
Moreover, the \powr{} model of LP26 predicts a softer ionizing spectrum than allowed by the observed \hii{} region emission.
These differences between the best-fit \powr{} model (red line) and \tlusty{} model fit to the stellar SED (blue line) are readily apparent in the top panel of Figure~\ref{fig:ionizing_seds}.

Whereas the estimated stellar properties from modeling LP26's observed SED were consistent with its \hii{}-region emission, we now find that the additional constraints on the stellar \teff{} and \logg{} from the stellar optical spectrum complicate our interpretation.
The optical \hei{} and \heii{} photospheric absorption lines do not allow for a high enough \teff{} to explain the hardness of the ionizing spectrum powering the surrounding \hii{} region.
The most plausible solution to this puzzle is likely a multiple-star system with a history of interaction.
For example, a compact stripped-star companion would be challenging to detect in FUV/optical light, but could contribute significant ionizing flux \citep{gotberg17}.
Past mass transfer from that companion star could also have spun up LP26 to its high \vsini{}.
However, this scenario does not readily explain the combination of high \lstar{} and high \logg{}; this may require another luminous star to be part of the system, or some other mechanism that is not yet understood.
Long-term monitoring of LP26 would be useful to constrain the multiplicity of the system and rule out some possible explanations for its peculiar properties.


\subsection{Implications for Modeling Metal-Poor Massive Stellar Populations\label{sec:implications}}

Here, we connect our analysis of these three very low-$Z$ O stars to the broader landscape of massive stellar population modeling in metal-poor galaxies.
Models of stellar evolution, spectra, and SEDs are essential tools in various areas of astrophysics: they are used to interpret observations of evolved stars, supernovae, and gravitational-wave events; predict feedback and nucleosynthesis from massive stars and supernovae; and are fundamental ingredients in the SPS codes required to infer galaxies' star-formation histories and ionizing photon production.
Massive-star models informed by empirical results are thus highly important, but the dearth of observations has been the limiting factor below 20\%\,\zsun{}.

This analysis has revealed some interesting tensions between our observations of very metal-poor O stars and model expectations.
The fundamental stellar properties that we infer for the two stars in galaxies below 10\%\,\zsun{} are not simultaneously matched by any publicly available stellar evolution models, even those that treat rotation or binary interactions (Section~\ref{sec:evolution}). 
Either the stars are under-luminous for their \mspec{}, which is challenging to explain, or they are over-luminous for their apparent \mevol{}. 
In the latter case, multiple-star systems may be able to alleviate the tension, but the fine-tuned scenarios that could explain all observational constraints are unlikely.
We also highlight the large impact of \lstar{} on stellar ionizing photon production (Section~\ref{sec:ionizing_spectra}), emphasizing the need to understand the relationship between \mstar{} and \lstar{} at very low $Z$ to accurately predict metal-poor galaxies' ionizing fluxes.
Altogether, the inferred properties of LP26 and S3 pose a challenge to current stellar evolution models in the still-unexplored extremely low-$Z$ regime.

Our \mdot{} and \vinf{} results (Section~\ref{sec:massloss}--\ref{sec:windspeed}) paint a picture of weaker winds than expected at and below 20\%\,\zsun{}. 
That we infer \mdot{} more than an order of magnitude lower than the \citet{vink01} recipe, which is often adopted in stellar evolution calculations, is particularly striking, and implies that metal-poor massive stars' evolutionary pathways and end products are less affected by mass loss than current models predict.
However, a physical understanding of these weak stellar winds remains elusive because no theoretical models match all of the available constraints from observations.
For example, while the models of \citet{bjorklund21} broadly agree with our low inferred \mdot{} at low $Z$, they also predict that \vinf{} increases strongly with decreasing \mdot{}, resulting in $\gtrsim 3000$\,\kms{} for their SMC models. 
This is in tension with the low \vinf{} we infer for our three metal-poor stars, as well as empirical results at 20\%\,\zsun{} and higher \citep[e.g.,][]{garcia14, bouret15, hawcroft23}, suggesting that further theoretical work is needed to understand the driving of weak stellar winds at low $Z$.
Improved models of radiation-driven winds, informed by FUV spectra of metal-poor OB stars, are critical to ensure accurate interpretation of metal-poor galaxies' evolution, particularly now that observations of chemically unevolved galaxies both locally and at high $z$ are becoming increasingly routine.


\section{Conclusions\label{sec:conclusions}}

We have presented new Keck/KCWI optical spectra of three very metal-poor O stars, LP26, S3, and A15, in nearby dwarf galaxies (Section~\ref{sec:optical_spectra}, Figures~\ref{fig:kcwi_images}--\ref{fig:kcwi_spectra}).
To determine their stellar and wind properties, we fit \powr{} atmosphere models to these observations combined with published HST/COS FUV spectra and HST NUV, optical, and NIR photometry.
Our key findings are:

\begin{enumerate}

\item We refine fundamental stellar parameter measurements over previous estimates from modeling the stellar SEDs \citep{telford21}. 
The \teff{}, \logg{}, and \lstar{} determined from our \powr{} models are all consistent within the large uncertainties on the SED-based estimates, except that we require a significantly higher \logg{} to match S3's broad Balmer wings ($3.9\pm0.1$, vs.\ previous $\log(g)=3.5^{+0.1}_{-0.2}$).
Yet, we also measure \teff{} systematically lower by 1.5--3.5\,kK than the best-fit values from SED fitting, underscoring the importance of optical spectroscopy for accurate stellar parameter inference (Section~\ref{sec:best_models}; Figures~\ref{fig:lp26}--\ref{fig:a15}; Tables~\ref{tab:params}--\ref{tab:model_comparison}).

\item The star A15 has strong FUV and optical He and N lines indicating enhanced abundances of both elements, but the data rule out the low C/N ratio expected for pure CNO-cycle products.
This moderate N enrichment in a dwarf star is consistent with efficient mixing at low $Z$. 
While abundances of the two lower-$Z$ stars are more uncertain due to their relatively high best-fit $v \,\mathrm{sin}(i)=200$\,\kms{} and lower-SNR FUV spectra, neither shows clear evidence of elevated N. 
For all three stars, a low $\xi_\mathrm{min}=5$\,\kms{} is required to match the observed \trans{c}{iii}{1176} (for LP26 and S3) and optical \spec{he}{i} lines (for A15), suggesting that microturbulent velocities may typically be small for O-dwarf stars at very low $Z$ (Section~\ref{sec:best_models}; Table~\ref{tab:params}).

\item The inferred \teff{}, \logg{}, and \lstar{} for all three stars yield higher \mspec{} than the \mevol{} implied by single-star evolution models, though \mspec{} and \mevol{} are consistent within the uncertainties for A15.
No evolution tracks, even those including binary interactions, are able to simultaneously match all inferred properties of the extremely metal-poor stars LP26 and S3.
Though the possibility of binaries could alleviate some of the tension, it is challenging to identify a plausible scenario that can account for both stars' apparent \teff{}, \logg{}, and \lstar{}, as well as the lack of evidence for companions in the data.
Our observations of these low-$Z$ stars pose an interesting challenge for current evolution models (Section~\ref{sec:evolution}; Figure~\ref{fig:geneva}).

\item We compare our inferred \mdot{} from the observed FUV metal resonance line profiles to the predictions of radiation-driven wind models, and find the best overall agreement with the \citet{bjorklund21} scaling at the low-\mdot{} end of the range. 
For all three stars, the data constrain \mdot{} to more than an order of magnitude lower than that calculated from the \citet{vink01} mass-loss recipe, which is widely adopted in stellar evolution codes.
Our findings agree with previous empirical results for OB stars in the SMC, and for higher-\lstar{} O stars in more metal-poor dwarf galaxies. 
This emerging empirical support for lower \mdot{} at low $Z$ should inform the assumptions in stellar evolution modeling, as \mdot{} strongly affects massive stars' evolution and end products, which propagate to SPS model predictions 
(Sections~\ref{sec:massloss} and \ref{sec:implications}; Figures~\ref{fig:massloss}--\ref{fig:windlines}).

\item We also compare \vinf{} inferred from the observed wind profiles, particularly the \trans{c}{iv}{1550} feature, to the empirical relationship between \vinf{}, \teff{}, and $Z$ at 20--100\%\,\zsun{} reported by \citet{hawcroft23}, extrapolated to lower $Z$.
The best-fit \vinf{} of all three stars are lower than expected, but A15 lies within the scatter of published \vinf{} for 11 other OB stars with similar abundance patterns in galaxies beyond the SMC. 
The low empirical \vinf{} in the SMC and more metal-poor galaxies, including our results here, are in strong tension with the very high \vinf{} predicted by the \citet{bjorklund21} models that best reproduce the observed \mdot{} for the same metal-poor stars.
Improved modeling of radiation-driven winds is needed in the low-$Z$ regime (Sections~\ref{sec:windspeed} and \ref{sec:implications}; Figures~\ref{fig:windlines}--\ref{fig:vfinal}).

\item We compare the ionizing photon production rates, \qh{}, and hardness of the ionizing spectra in our best-fit \powr{} models to expectations from similar models at SMC metallicity, and to the \citet{telford21} best-fit models to these stars' SEDs. 
The theoretical models for SMC stars, constrained by single-star evolution model expectations, predicts lower \qh{} than our best-fit \powr{} models, emphasizing the need to clarify the relationship between \lstar{} and \mstar{} at very low $Z$ for accurate predictions ionizing photon production by metal-poor galaxies.
We also show that SED fitting is subject to biases in \teff{}, emphasizing the need for optical spectra to constrain \qh{} and ionizing spectral hardness (Section~\ref{sec:ionizing_spectra}--\ref{sec:implications}, Figure~\ref{fig:ionizing_seds},  Table~\ref{tab:model_comparison}).

\item Finally, we highlight that our inferred properties of LP26 predict \qh{} and \qhe{}/\qh{} that are inconsistent with previous constraints on the ionizing source of its surrounding \hii{} region from the observed nebular emission \citep{telford23}. 
Possible explanations include one or more companion stars, past interactions, and/or exotic stellar physics at very low $Z$ (Section~\ref{sec:lp26_hiiregion}).
\end{enumerate} 

The combined Keck/KCWI and HST/COS spectra analyzed here have given us a first look at the stellar and wind properties of very metal-poor O stars in dwarf galaxies near the edge of the Local Group.
Yet, the present sample size is too small for empirical characterization of stellar wind recipes below 20\%\,\zsun{}.
The initial results from our \powr{} modeling emphasize that we have much to learn about low-$Z$ stellar astrophysics from these spectra, despite the limited resolution and SNR that can be achieved for faint and distant O stars.
Additional data are particularly needed to clarify the implications of our findings for the calculation of star-formation rates and ionizing photon production in reionization-era galaxies.
In future work, we will employ similar-quality spectroscopy from new COS programs (HST-GO-16767, HST-GO-17491; PI: O.\ G.\ Telford) targeting larger samples of extremely metal-poor O stars to more rigorously test and calibrate stellar models in this low-$Z$ regime.


\acknowledgements

OGT thanks Jenny Greene, D.\ John Hillier, Calum Hawcroft, Jon Sundqvist, Alex de Koter, Ciar\'an Furey, Frank Backs, Jan Eldridge, and Carles Badenes for helpful discussions; Wolf-Rainer Hamann and Helge Todt for assistance with the \powr{} modeling; Rosalie McGurk and Luca Rizzi for support with the Keck/KCWI observations; and the referee for helpful suggestions that improved the quality of this paper. 

JC thanks Joseph Cassinelli whose care and guidance sparked and inspired the initial drive to understand these massive star winds.

Based on observations with the NASA/ESA Hubble Space Telescope obtained at the Space Telescope Science Institute, which is operated by the Association of Universities for Research in Astronomy, Incorporated, under NASA contract NAS5-26555. Support for Program number GO-15967 was provided through a grant from the STScI under NASA contract NAS5-26555.

This work was supported by a NASA Keck PI Data Award, administered by the NASA Exoplanet Science Institute. Data presented herein were obtained at the W. M. Keck Observatory from telescope time allocated to the National Aeronautics and Space Administration through the agency's scientific partnership with the California Institute of Technology and the University of California. The Observatory was made possible by the generous financial support of the W. M. Keck Foundation.
The authors wish to recognize and acknowledge the very significant cultural role and reverence that the summit of Maunakea has always had within the indigenous Hawaiian community. We are fortunate to have the opportunity to conduct observations from this mountain.
 
OGT acknowledges generous support from a Carnegie-Princeton Fellowship, through Princeton University and the Carnegie Observatories, and from the Momental Foundation.
This work was performed in part at the Aspen Center for Physics, which is supported by National Science Foundation grant PHY-2210452, and was partially supported by a grant from the Simons Foundation.
This research used NASA's Astrophysics Data System, adstex\footnote{\url{https://github.com/yymao/adstex}}, and the arXiv preprint server. 

AACS and VR acknowledge support by the Deutsche Forschungsgemeinschaft (DFG, German Research Foundation) in the form of an Emmy Noether Research Group -- Project-ID 445674056 (SA4064/1-1, PI Sander). AACS further acknowledges support by the Deutsche Forschungsgemeinschaft (DFG, German Research Foundation) – Project-ID 138713538 -- SFB 881 (“The Milky Way System”, subproject P04). This work was partially supported by funding provided by the Federal Ministry of Education and Research (BMBF) and the Baden-Württemberg Ministry of Science as part of the Excellence Strategy of the German Federal and State Governments. This publication has benefited from discussions in a team meeting (PI: Oskinova) sponsored by the International Space Science Institute (ISSI) at Bern, Switzerland. Several figures in this work were created with \textsc{WRplot}, developed by W.-R.\ Hamann.

\facilities{Keck:II (KCWI), HST (COS, ACS, WFC3, WFPC2)}

\software{\powr{} \citep{grafener02,hamann03,sander15}, Astropy \citep{astropy, astropy2, astropy3}, CWITools \citep{osullivan20}, KCWI DRP \citep{morrissey18}, iPython \citep{ipython}, Matplotlib \citep{matplotlib}, NumPy \citep{numpy, numpy2}, SAOImageDS9 \citep{joye03}, SciPy \citep{scipy2}}


\appendix
\section{Parameters in the Initial \powr{} Models}\label{app:initial_guesses}
\setcounter{table}{0}
\renewcommand{\thetable}{A\arabic{table}}

Table~\ref{tab:initial_guesses} reports the values of all \powr{} model parameters that were adjusted in our analysis that we adopted in the starting model for each star. 
The choice of initial guesses was informed by previous analysis of these stars; observations of their host galaxies; and theoretical wind-parameter scaling relationships, as detailed in Section~\ref{sec:initial_guesses}.
Because we judged the goodness of fit by eye and adjusted parameters by hand, the specific initial guesses have little impact on the final result, but starting parameters as close as possible to the best values are beneficial to reduce computation time. 
We report our initial model parameters here to facilitate reproducibility of this analysis.

\begin{table*}
\begin{center}
\caption{Initial guesses for parameters in the \powr{} models.\label{tab:initial_guesses}}
\tabcolsep=0.75cm
\begin{tabular}{lccc}
Parameter & LP26 & S3 & A15 \\
\hline 
\teff{} (kK) & 37 & 32 & 37 \\
$\log(g/\mathrm{cm\,s}^{-2})$ & 4.0 & 3.5 & 4.0 \\
$\log(L_\star/L_\odot)$ & 5.1 & 5.1 & 5.3 \\
$\log(\dot{M}_\star/M_\odot\,\mathrm{yr}^{-1})$ & $-$8.7 & $-$8.3 & $-$8.1\\
\vinf{} (km\,s$^{-1}$) & 1720 & 1500 & 2160 \\ 
\vsini{} (km\,s$^{-1}$) & 370 & 290 & 80 \\
$\xi_\mathrm{min}$ (km\,s$^{-1}$) & 10 & 10 & 10 \\
$X_\mathrm{He}$ & 0.25 & 0.25 & 0.25 \\
$\log(X_\mathrm{C})$ & $-4.58$ & $-4.29$ & $-3.92$ \\
$\log(X_\mathrm{N})$ & $-5.17$ & $-5.06$ & $-4.64$ \\
$\log(X_\mathrm{O})$ & $-3.76$ & $-3.46$ & $-3.09$ \\
$\log(X_\mathrm{Fe})$ & $-4.44$ & $-4.14$ & $-3.62$ \\
\ebv{}$_\mathrm{host}$ (mag) & 0.003 & 0.048 & 0.018 \\
\end{tabular}
\end{center}
\vspace{-10pt}
\tablecomments{Parameters adopted in the first \powr{} models run for the three observed stars, based on observations or theoretical expectations as described in Section~\ref{sec:initial_guesses}. Quantities are reported as in Table~\ref{tab:params}.}
\end{table*}



\begin{thebibliography}{}
\expandafter\ifx\csname natexlab\endcsname\relax\def\natexlab#1{#1}\fi
\providecommand{\url}[1]{\href{#1}{#1}}
\providecommand{\dodoi}[1]{doi:~\href{http://doi.org/#1}{\nolinkurl{#1}}}
\providecommand{\doeprint}[1]{\href{http://ascl.net/#1}{\nolinkurl{http://ascl.net/#1}}}
\providecommand{\doarXiv}[1]{\href{https://arxiv.org/abs/#1}{\nolinkurl{https://arxiv.org/abs/#1}}}

\bibitem[{{Arrabal Haro} {et~al.}(2023){Arrabal Haro}, {Dickinson},
  {Finkelstein}, {Kartaltepe}, {Donnan}, {Burgarella}, {Carnall}, {Cullen},
  {Dunlop}, {Fern{\'a}ndez}, {Fujimoto}, {Jung}, {Krips}, {Larson}, {Papovich},
  {P{\'e}rez-Gonz{\'a}lez}, {Amor{\'\i}n}, {Bagley}, {Buat}, {Casey},
  {Chworowsky}, {Cohen}, {Ferguson}, {Giavalisco}, {Huertas-Company},
  {Hutchison}, {Kocevski}, {Koekemoer}, {Lucas}, {McLeod}, {McLure}, {Pirzkal},
  {Seill{\'e}}, {Trump}, {Weiner}, {Wilkins}, \& {Zavala}}]{arrabal-haro23}
{Arrabal Haro}, P., {Dickinson}, M., {Finkelstein}, S.~L., {et~al.} 2023, \nat,
  622, 707, \dodoi{10.1038/s41586-023-06521-7}

\bibitem[{{Asplund} {et~al.}(2009){Asplund}, {Grevesse}, {Sauval}, \&
  {Scott}}]{asplund09}
{Asplund}, M., {Grevesse}, N., {Sauval}, A.~J., \& {Scott}, P. 2009, \araa, 47,
  481, \dodoi{10.1146/annurev.astro.46.060407.145222}

\bibitem[{{Astropy Collaboration} {et~al.}(2013){Astropy Collaboration},
  {Robitaille}, {Tollerud}, {Greenfield}, {Droettboom}, {Bray}, {Aldcroft},
  {Davis}, {Ginsburg}, {Price-Whelan}, {Kerzendorf}, {Conley}, {Crighton},
  {Barbary}, {Muna}, {Ferguson}, {Grollier}, {Parikh}, {Nair}, {Unther},
  {Deil}, {Woillez}, {Conseil}, {Kramer}, {Turner}, {Singer}, {Fox}, {Weaver},
  {Zabalza}, {Edwards}, {Azalee Bostroem}, {Burke}, {Casey}, {Crawford},
  {Dencheva}, {Ely}, {Jenness}, {Labrie}, {Lim}, {Pierfederici}, {Pontzen},
  {Ptak}, {Refsdal}, {Servillat}, \& {Streicher}}]{astropy}
{Astropy Collaboration}, {Robitaille}, T.~P., {Tollerud}, E.~J., {et~al.} 2013,
  \aap, 558, A33, \dodoi{10.1051/0004-6361/201322068}

\bibitem[{{Astropy Collaboration} {et~al.}(2018){Astropy Collaboration},
  {Price-Whelan}, {Sip{\H{o}}cz}, {G{\"u}nther}, {Lim}, {Crawford}, {Conseil},
  {Shupe}, {Craig}, {Dencheva}, {Ginsburg}, {VanderPlas}, {Bradley},
  {P{\'e}rez-Su{\'a}rez}, {de Val-Borro}, {Aldcroft}, {Cruz}, {Robitaille},
  {Tollerud}, {Ardelean}, {Babej}, {Bach}, {Bachetti}, {Bakanov}, {Bamford},
  {Barentsen}, {Barmby}, {Baumbach}, {Berry}, {Biscani}, {Boquien}, {Bostroem},
  {Bouma}, {Brammer}, {Bray}, {Breytenbach}, {Buddelmeijer}, {Burke},
  {Calderone}, {Cano Rodr{\'\i}guez}, {Cara}, {Cardoso}, {Cheedella}, {Copin},
  {Corrales}, {Crichton}, {D'Avella}, {Deil}, {Depagne}, {Dietrich}, {Donath},
  {Droettboom}, {Earl}, {Erben}, {Fabbro}, {Ferreira}, {Finethy}, {Fox},
  {Garrison}, {Gibbons}, {Goldstein}, {Gommers}, {Greco}, {Greenfield},
  {Groener}, {Grollier}, {Hagen}, {Hirst}, {Homeier}, {Horton}, {Hosseinzadeh},
  {Hu}, {Hunkeler}, {Ivezi{\'c}}, {Jain}, {Jenness}, {Kanarek}, {Kendrew},
  {Kern}, {Kerzendorf}, {Khvalko}, {King}, {Kirkby}, {Kulkarni}, {Kumar},
  {Lee}, {Lenz}, {Littlefair}, {Ma}, {Macleod}, {Mastropietro}, {McCully},
  {Montagnac}, {Morris}, {Mueller}, {Mumford}, {Muna}, {Murphy}, {Nelson},
  {Nguyen}, {Ninan}, {N{\"o}the}, {Ogaz}, {Oh}, {Parejko}, {Parley}, {Pascual},
  {Patil}, {Patil}, {Plunkett}, {Prochaska}, {Rastogi}, {Reddy Janga},
  {Sabater}, {Sakurikar}, {Seifert}, {Sherbert}, {Sherwood-Taylor}, {Shih},
  {Sick}, {Silbiger}, {Singanamalla}, {Singer}, {Sladen}, {Sooley},
  {Sornarajah}, {Streicher}, {Teuben}, {Thomas}, {Tremblay}, {Turner},
  {Terr{\'o}n}, {van Kerkwijk}, {de la Vega}, {Watkins}, {Weaver}, {Whitmore},
  {Woillez}, {Zabalza}, \& {Astropy Contributors}}]{astropy2}
{Astropy Collaboration}, {Price-Whelan}, A.~M., {Sip{\H{o}}cz}, B.~M., {et~al.}
  2018, \aj, 156, 123, \dodoi{10.3847/1538-3881/aabc4f}

\bibitem[{{Astropy Collaboration} {et~al.}(2022){Astropy Collaboration},
  {Price-Whelan}, {Lim}, {Earl}, {Starkman}, {Bradley}, {Shupe}, {Patil},
  {Corrales}, {Brasseur}, {N{\"o}the}, {Donath}, {Tollerud}, {Morris},
  {Ginsburg}, {Vaher}, {Weaver}, {Tocknell}, {Jamieson}, {van Kerkwijk},
  {Robitaille}, {Merry}, {Bachetti}, {G{\"u}nther}, {Aldcroft},
  {Alvarado-Montes}, {Archibald}, {B{\'o}di}, {Bapat}, {Barentsen},
  {Baz{\'a}n}, {Biswas}, {Boquien}, {Burke}, {Cara}, {Cara}, {Conroy},
  {Conseil}, {Craig}, {Cross}, {Cruz}, {D'Eugenio}, {Dencheva}, {Devillepoix},
  {Dietrich}, {Eigenbrot}, {Erben}, {Ferreira}, {Foreman-Mackey}, {Fox},
  {Freij}, {Garg}, {Geda}, {Glattly}, {Gondhalekar}, {Gordon}, {Grant},
  {Greenfield}, {Groener}, {Guest}, {Gurovich}, {Handberg}, {Hart},
  {Hatfield-Dodds}, {Homeier}, {Hosseinzadeh}, {Jenness}, {Jones}, {Joseph},
  {Kalmbach}, {Karamehmetoglu}, {Ka{\l}uszy{\'n}ski}, {Kelley}, {Kern},
  {Kerzendorf}, {Koch}, {Kulumani}, {Lee}, {Ly}, {Ma}, {MacBride}, {Maljaars},
  {Muna}, {Murphy}, {Norman}, {O'Steen}, {Oman}, {Pacifici}, {Pascual},
  {Pascual-Granado}, {Patil}, {Perren}, {Pickering}, {Rastogi}, {Roulston},
  {Ryan}, {Rykoff}, {Sabater}, {Sakurikar}, {Salgado}, {Sanghi}, {Saunders},
  {Savchenko}, {Schwardt}, {Seifert-Eckert}, {Shih}, {Jain}, {Shukla}, {Sick},
  {Simpson}, {Singanamalla}, {Singer}, {Singhal}, {Sinha}, {Sip{\H{o}}cz},
  {Spitler}, {Stansby}, {Streicher}, {{\v{S}}umak}, {Swinbank}, {Taranu},
  {Tewary}, {Tremblay}, {Val-Borro}, {Van Kooten}, {Vasovi{\'c}}, {Verma}, {de
  Miranda Cardoso}, {Williams}, {Wilson}, {Winkel}, {Wood-Vasey}, {Xue},
  {Yoachim}, {Zhang}, {Zonca}, \& {Astropy Project Contributors}}]{astropy3}
{Astropy Collaboration}, {Price-Whelan}, A.~M., {Lim}, P.~L., {et~al.} 2022,
  \apj, 935, 167, \dodoi{10.3847/1538-4357/ac7c74}

\bibitem[{{Aver} {et~al.}(2022){Aver}, {Berg}, {Hirschauer}, {Olive}, {Pogge},
  {Rogers}, {Salzer}, \& {Skillman}}]{aver22}
{Aver}, E., {Berg}, D.~A., {Hirschauer}, A.~S., {et~al.} 2022, \mnras, 510,
  373, \dodoi{10.1093/mnras/stab3226}

\bibitem[{{Baum} {et~al.}(1992){Baum}, {Hamann}, {Koesterke}, \&
  {Wessolowski}}]{baum92}
{Baum}, E., {Hamann}, W.~R., {Koesterke}, L., \& {Wessolowski}, U. 1992, \aap,
  266, 402

\bibitem[{{Beaton} {et~al.}(2018){Beaton}, {Bono}, {Braga}, {Dall'Ora},
  {Fiorentino}, {Jang}, {Mart{\'\i}nez-V{\'a}zquez}, {Matsunaga}, {Monelli},
  {Neeley}, \& {Salaris}}]{beaton18}
{Beaton}, R.~L., {Bono}, G., {Braga}, V.~F., {et~al.} 2018, \ssr, 214, 113,
  \dodoi{10.1007/s11214-018-0542-1}

\bibitem[{{Becker} {et~al.}(2001){Becker}, {Fan}, {White}, {Strauss},
  {Narayanan}, {Lupton}, {Gunn}, {Annis}, {Bahcall}, {Brinkmann}, {Connolly},
  {Csabai}, {Czarapata}, {Doi}, {Heckman}, {Hennessy}, {Ivezi{\'c}}, {Knapp},
  {Lamb}, {McKay}, {Munn}, {Nash}, {Nichol}, {Pier}, {Richards}, {Schneider},
  {Stoughton}, {Szalay}, {Thakar}, \& {York}}]{becker01}
{Becker}, R.~H., {Fan}, X., {White}, R.~L., {et~al.} 2001, \aj, 122, 2850,
  \dodoi{10.1086/324231}

\bibitem[{{Berg} {et~al.}(2019){Berg}, {Erb}, {Henry}, {Skillman}, \&
  {McQuinn}}]{berg19}
{Berg}, D.~A., {Erb}, D.~K., {Henry}, R. B.~C., {Skillman}, E.~D., \&
  {McQuinn}, K. B.~W. 2019, \apj, 874, 93, \dodoi{10.3847/1538-4357/ab020a}

\bibitem[{{Bestenlehner} {et~al.}(2020){Bestenlehner}, {Crowther},
  {Caballero-Nieves}, {Schneider}, {Sim{\'o}n-D{\'\i}az}, {Brands}, {de Koter},
  {Gr{\"a}fener}, {Herrero}, {Langer}, {Lennon}, {Ma{\'\i}z Apell{\'a}niz},
  {Puls}, \& {Vink}}]{bestenlehner20}
{Bestenlehner}, J.~M., {Crowther}, P.~A., {Caballero-Nieves}, S.~M., {et~al.}
  2020, \mnras, 499, 1918, \dodoi{10.1093/mnras/staa2801}

\bibitem[{{Bj{\"o}rklund} {et~al.}(2021){Bj{\"o}rklund}, {Sundqvist}, {Puls},
  \& {Najarro}}]{bjorklund21}
{Bj{\"o}rklund}, R., {Sundqvist}, J.~O., {Puls}, J., \& {Najarro}, F. 2021,
  \aap, 648, A36, \dodoi{10.1051/0004-6361/202038384}

\bibitem[{{Bouret} {et~al.}(2003){Bouret}, {Lanz}, {Hillier}, {Heap}, {Hubeny},
  {Lennon}, {Smith}, \& {Evans}}]{bouret03}
{Bouret}, J.~C., {Lanz}, T., {Hillier}, D.~J., {et~al.} 2003, \apj, 595, 1182,
  \dodoi{10.1086/377368}

\bibitem[{{Bouret} {et~al.}(2015){Bouret}, {Lanz}, {Hillier}, {Martins},
  {Marcolino}, \& {Depagne}}]{bouret15}
---. 2015, \mnras, 449, 1545, \dodoi{10.1093/mnras/stv379}

\bibitem[{{Bouret} {et~al.}(2013){Bouret}, {Lanz}, {Martins}, {Marcolino},
  {Hillier}, {Depagne}, \& {Hubeny}}]{bouret13}
{Bouret}, J.~C., {Lanz}, T., {Martins}, F., {et~al.} 2013, \aap, 555, A1,
  \dodoi{10.1051/0004-6361/201220798}

\bibitem[{{Bresolin} {et~al.}(2006){Bresolin}, {Pietrzy{\'n}ski}, {Urbaneja},
  {Gieren}, {Kudritzki}, \& {Venn}}]{bresolin06}
{Bresolin}, F., {Pietrzy{\'n}ski}, G., {Urbaneja}, M.~A., {et~al.} 2006, \apj,
  648, 1007, \dodoi{10.1086/506200}

\bibitem[{{Bresolin} {et~al.}(2007){Bresolin}, {Urbaneja}, {Gieren},
  {Pietrzy{\'n}ski}, \& {Kudritzki}}]{bresolin07}
{Bresolin}, F., {Urbaneja}, M.~A., {Gieren}, W., {Pietrzy{\'n}ski}, G., \&
  {Kudritzki}, R.-P. 2007, \apj, 671, 2028, \dodoi{10.1086/522571}

\bibitem[{{Bressan} {et~al.}(2012){Bressan}, {Marigo}, {Girardi}, {Salasnich},
  {Dal Cero}, {Rubele}, \& {Nanni}}]{bressan12}
{Bressan}, A., {Marigo}, P., {Girardi}, L., {et~al.} 2012, \mnras, 427, 127,
  \dodoi{10.1111/j.1365-2966.2012.21948.x}

\bibitem[{{Brott} {et~al.}(2011){Brott}, {de Mink}, {Cantiello}, {Langer}, {de
  Koter}, {Evans}, {Hunter}, {Trundle}, \& {Vink}}]{brott11}
{Brott}, I., {de Mink}, S.~E., {Cantiello}, M., {et~al.} 2011, \aap, 530, A115,
  \dodoi{10.1051/0004-6361/201016113}

\bibitem[{{Byler} {et~al.}(2017){Byler}, {Dalcanton}, {Conroy}, \&
  {Johnson}}]{byler17}
{Byler}, N., {Dalcanton}, J.~J., {Conroy}, C., \& {Johnson}, B.~D. 2017, \apj,
  840, 44, \dodoi{10.3847/1538-4357/aa6c66}

\bibitem[{{Cameron} {et~al.}(2023){Cameron}, {Saxena}, {Bunker}, {D'Eugenio},
  {Carniani}, {Maiolino}, {Curtis-Lake}, {Ferruit}, {Jakobsen}, {Arribas},
  {Bonaventura}, {Charlot}, {Chevallard}, {Curti}, {Looser}, {Maseda}, {Rawle},
  {Rodr{\'\i}guez Del Pino}, {Smit}, {{\"U}bler}, {Willott}, {Witstok},
  {Egami}, {Eisenstein}, {Johnson}, {Hainline}, {Rieke}, {Robertson}, {Stark},
  {Tacchella}, {Williams}, {Willmer}, {Bhatawdekar}, {Bowler}, {Boyett},
  {Circosta}, {Helton}, {Jones}, {Kumari}, {Ji}, {Nelson}, {Parlanti},
  {Sandles}, {Scholtz}, \& {Sun}}]{cameron23}
{Cameron}, A.~J., {Saxena}, A., {Bunker}, A.~J., {et~al.} 2023, \aap, 677,
  A115, \dodoi{10.1051/0004-6361/202346107}

\bibitem[{{Castor} {et~al.}(1975){Castor}, {Abbott}, \& {Klein}}]{castor75}
{Castor}, J.~I., {Abbott}, D.~C., \& {Klein}, R.~I. 1975, \apj, 195, 157,
  \dodoi{10.1086/153315}

\bibitem[{{Conroy}(2013)}]{conroy13}
{Conroy}, C. 2013, \araa, 51, 393, \dodoi{10.1146/annurev-astro-082812-141017}

\bibitem[{{Crowther} {et~al.}(2022){Crowther}, {Broos}, {Townsley}, {Pollock},
  {Tehrani}, \& {Gagn{\'e}}}]{crowther22}
{Crowther}, P.~A., {Broos}, P.~S., {Townsley}, L.~K., {et~al.} 2022, \mnras,
  515, 4130, \dodoi{10.1093/mnras/stac1952}

\bibitem[{{Crowther} {et~al.}(2016){Crowther}, {Caballero-Nieves}, {Bostroem},
  {Ma{\'\i}z Apell{\'a}niz}, {Schneider}, {Walborn}, {Angus}, {Brott},
  {Bonanos}, {de Koter}, {de Mink}, {Evans}, {Gr{\"a}fener}, {Herrero},
  {Howarth}, {Langer}, {Lennon}, {Puls}, {Sana}, \& {Vink}}]{crowther16}
{Crowther}, P.~A., {Caballero-Nieves}, S.~M., {Bostroem}, K.~A., {et~al.} 2016,
  \mnras, 458, 624, \dodoi{10.1093/mnras/stw273}

\bibitem[{{Curtis-Lake} {et~al.}(2023){Curtis-Lake}, {Carniani}, {Cameron},
  {Charlot}, {Jakobsen}, {Maiolino}, {Bunker}, {Witstok}, {Smit}, {Chevallard},
  {Willott}, {Ferruit}, {Arribas}, {Bonaventura}, {Curti}, {D'Eugenio},
  {Franx}, {Giardino}, {Looser}, {L{\"u}tzgendorf}, {Maseda}, {Rawle}, {Rix},
  {Rodr{\'\i}guez del Pino}, {{\"U}bler}, {Sirianni}, {Dressler}, {Egami},
  {Eisenstein}, {Endsley}, {Hainline}, {Hausen}, {Johnson}, {Rieke},
  {Robertson}, {Shivaei}, {Stark}, {Tacchella}, {Williams}, {Willmer},
  {Bhatawdekar}, {Bowler}, {Boyett}, {Chen}, {de Graaff}, {Helton}, {Hviding},
  {Jones}, {Kumari}, {Lyu}, {Nelson}, {Perna}, {Sandles}, {Saxena}, {Suess},
  {Sun}, {Topping}, {Wallace}, \& {Whitler}}]{curtis-lake23}
{Curtis-Lake}, E., {Carniani}, S., {Cameron}, A., {et~al.} 2023, Nature
  Astronomy, 7, 622, \dodoi{10.1038/s41550-023-01918-w}

\bibitem[{{Dalcanton} {et~al.}(2009){Dalcanton}, {Williams}, {Seth}, {Dolphin},
  {Holtzman}, {Rosema}, {Skillman}, {Cole}, {Girardi}, {Gogarten},
  {Karachentsev}, {Olsen}, {Weisz}, {Christensen}, {Freeman}, {Gilbert},
  {Gallart}, {Harris}, {Hodge}, {de Jong}, {Karachentseva}, {Mateo}, {Stetson},
  {Tavarez}, {Zaritsky}, {Governato}, \& {Quinn}}]{dalcanton09}
{Dalcanton}, J.~J., {Williams}, B.~F., {Seth}, A.~C., {et~al.} 2009, \apjs,
  183, 67, \dodoi{10.1088/0067-0049/183/1/67}

\bibitem[{{Dayal} \& {Ferrara}(2018)}]{dayal18}
{Dayal}, P., \& {Ferrara}, A. 2018, \physrep, 780, 1,
  \dodoi{10.1016/j.physrep.2018.10.002}

\bibitem[{{de Mink} {et~al.}(2013){de Mink}, {Langer}, {Izzard}, {Sana}, \& {de
  Koter}}]{de-mink13}
{de Mink}, S.~E., {Langer}, N., {Izzard}, R.~G., {Sana}, H., \& {de Koter}, A.
  2013, \apj, 764, 166, \dodoi{10.1088/0004-637X/764/2/166}

\bibitem[{{Dolphin}(2016)}]{dolphin16}
{Dolphin}, A. 2016, {DOLPHOT: Stellar photometry}, Astrophysics Source Code
  Library, record ascl:1608.013.
\newblock \doeprint{1608.013}

\bibitem[{{Dolphin}(2000)}]{dolphin00}
{Dolphin}, A.~E. 2000, \pasp, 112, 1383, \dodoi{10.1086/316630}

\bibitem[{{Ekstr{\"o}m} {et~al.}(2012){Ekstr{\"o}m}, {Georgy}, {Eggenberger},
  {Meynet}, {Mowlavi}, {Wyttenbach}, {Granada}, {Decressin}, {Hirschi},
  {Frischknecht}, {Charbonnel}, \& {Maeder}}]{ekstrom12}
{Ekstr{\"o}m}, S., {Georgy}, C., {Eggenberger}, P., {et~al.} 2012, \aap, 537,
  A146, \dodoi{10.1051/0004-6361/201117751}

\bibitem[{{Eldridge} \& {Stanway}(2022)}]{eldridge22}
{Eldridge}, J.~J., \& {Stanway}, E.~R. 2022, \araa, 60, 455,
  \dodoi{10.1146/annurev-astro-052920-100646}

\bibitem[{{Eldridge} {et~al.}(2017){Eldridge}, {Stanway}, {Xiao}, {McClelland},
  {Taylor}, {Ng}, {Greis}, \& {Bray}}]{eldridge17}
{Eldridge}, J.~J., {Stanway}, E.~R., {Xiao}, L., {et~al.} 2017, \pasa, 34,
  e058, \dodoi{10.1017/pasa.2017.51}

\bibitem[{{Endsley} {et~al.}(2023){Endsley}, {Stark}, {Whitler}, {Topping},
  {Chen}, {Plat}, {Chisholm}, \& {Charlot}}]{endsley23}
{Endsley}, R., {Stark}, D.~P., {Whitler}, L., {et~al.} 2023, \mnras, 524, 2312,
  \dodoi{10.1093/mnras/stad1919}

\bibitem[{{Evans} {et~al.}(2007){Evans}, {Bresolin}, {Urbaneja},
  {Pietrzy{\'n}ski}, {Gieren}, \& {Kudritzki}}]{evans07}
{Evans}, C.~J., {Bresolin}, F., {Urbaneja}, M.~A., {et~al.} 2007, \apj, 659,
  1198, \dodoi{10.1086/511382}

\bibitem[{{Evans} {et~al.}(2019){Evans}, {Castro}, {Gonzalez}, {Garcia},
  {Bastian}, {Cioni}, {Clark}, {Davies}, {Ferguson}, {Kamann}, {Lennon},
  {Patrick}, {Vink}, \& {Weisz}}]{evans19}
{Evans}, C.~J., {Castro}, N., {Gonzalez}, O.~A., {et~al.} 2019, \aap, 622,
  A129, \dodoi{10.1051/0004-6361/201834145}

\bibitem[{{Fan} {et~al.}(2006){Fan}, {Carilli}, \& {Keating}}]{fan06}
{Fan}, X., {Carilli}, C.~L., \& {Keating}, B. 2006, \araa, 44, 415,
  \dodoi{10.1146/annurev.astro.44.051905.092514}

\bibitem[{{Fitzpatrick}(1999)}]{fitzpatrick99}
{Fitzpatrick}, E.~L. 1999, \pasp, 111, 63, \dodoi{10.1086/316293}

\bibitem[{{Garcia} \& {Herrero}(2013)}]{garcia13}
{Garcia}, M., \& {Herrero}, A. 2013, \aap, 551, A74,
  \dodoi{10.1051/0004-6361/201219977}

\bibitem[{{Garcia} {et~al.}(2019){Garcia}, {Herrero}, {Najarro}, {Camacho}, \&
  {Lorenzo}}]{garcia19}
{Garcia}, M., {Herrero}, A., {Najarro}, F., {Camacho}, I., \& {Lorenzo}, M.
  2019, \mnras, 484, 422, \dodoi{10.1093/mnras/sty3503}

\bibitem[{{Garcia} {et~al.}(2014){Garcia}, {Herrero}, {Najarro}, {Lennon}, \&
  {Alejandro Urbaneja}}]{garcia14}
{Garcia}, M., {Herrero}, A., {Najarro}, F., {Lennon}, D.~J., \& {Alejandro
  Urbaneja}, M. 2014, \apj, 788, 64, \dodoi{10.1088/0004-637X/788/1/64}

\bibitem[{{Georgy} {et~al.}(2013){Georgy}, {Ekstr{\"o}m}, {Eggenberger},
  {Meynet}, {Haemmerl{\'e}}, {Maeder}, {Granada}, {Groh}, {Hirschi}, {Mowlavi},
  {Yusof}, {Charbonnel}, {Decressin}, \& {Barblan}}]{georgy13}
{Georgy}, C., {Ekstr{\"o}m}, S., {Eggenberger}, P., {et~al.} 2013, \aap, 558,
  A103, \dodoi{10.1051/0004-6361/201322178}

\bibitem[{{Gordon} {et~al.}(2003){Gordon}, {Clayton}, {Misselt}, {Landolt}, \&
  {Wolff}}]{gordon03}
{Gordon}, K.~D., {Clayton}, G.~C., {Misselt}, K.~A., {Landolt}, A.~U., \&
  {Wolff}, M.~J. 2003, \apj, 594, 279, \dodoi{10.1086/376774}

\bibitem[{{G{\"o}tberg} {et~al.}(2017){G{\"o}tberg}, {de Mink}, \&
  {Groh}}]{gotberg17}
{G{\"o}tberg}, Y., {de Mink}, S.~E., \& {Groh}, J.~H. 2017, \aap, 608, A11,
  \dodoi{10.1051/0004-6361/201730472}

\bibitem[{{Gr{\"a}fener} {et~al.}(2002){Gr{\"a}fener}, {Koesterke}, \&
  {Hamann}}]{grafener02}
{Gr{\"a}fener}, G., {Koesterke}, L., \& {Hamann}, W.~R. 2002, \aap, 387, 244,
  \dodoi{10.1051/0004-6361:20020269}

\bibitem[{{Green} {et~al.}(2015){Green}, {Schlafly}, {Finkbeiner}, {Rix},
  {Martin}, {Burgett}, {Draper}, {Flewelling}, {Hodapp}, {Kaiser}, {Kudritzki},
  {Magnier}, {Metcalfe}, {Price}, {Tonry}, \& {Wainscoat}}]{green15}
{Green}, G.~M., {Schlafly}, E.~F., {Finkbeiner}, D.~P., {et~al.} 2015, \apj,
  810, 25, \dodoi{10.1088/0004-637X/810/1/25}

\bibitem[{{Groh} {et~al.}(2019){Groh}, {Ekstr{\"o}m}, {Georgy}, {Meynet},
  {Choplin}, {Eggenberger}, {Hirschi}, {Maeder}, {Murphy}, {Boian}, \&
  {Farrell}}]{groh19}
{Groh}, J.~H., {Ekstr{\"o}m}, S., {Georgy}, C., {et~al.} 2019, \aap, 627, A24,
  \dodoi{10.1051/0004-6361/201833720}

\bibitem[{{Gull} {et~al.}(2022){Gull}, {Weisz}, {Senchyna}, {Sandford}, {Choi},
  {McLeod}, {El-Badry}, {G{\"o}tberg}, {Gilbert}, {Boyer}, {Dalcanton},
  {GuhaThakurta}, {Goldman}, {Marigo}, {McQuinn}, {Pastorelli}, {Stark},
  {Skillman}, {Ting}, \& {Williams}}]{gull22}
{Gull}, M., {Weisz}, D.~R., {Senchyna}, P., {et~al.} 2022, \apj, 941, 206,
  \dodoi{10.3847/1538-4357/aca295}

\bibitem[{{Hainich} {et~al.}(2019){Hainich}, {Ramachandran}, {Shenar},
  {Sander}, {Todt}, {Gruner}, {Oskinova}, \& {Hamann}}]{hainich19}
{Hainich}, R., {Ramachandran}, V., {Shenar}, T., {et~al.} 2019, \aap, 621, A85,
  \dodoi{10.1051/0004-6361/201833787}

\bibitem[{{Hamann} \& {Gr{\"a}fener}(2003)}]{hamann03}
{Hamann}, W.-R., \& {Gr{\"a}fener}, G. 2003, \aap, 410, 993,
  \dodoi{10.1051/0004-6361:20031308}

\bibitem[{{Hamann} \& {Koesterke}(1998)}]{hamann98}
{Hamann}, W.~R., \& {Koesterke}, L. 1998, \aap, 335, 1003

\bibitem[{{Harris} {et~al.}(2020){Harris}, {Millman}, {van der Walt},
  {Gommers}, {Virtanen}, {Cournapeau}, {Wieser}, {Taylor}, {Berg}, {Smith},
  {Kern}, {Picus}, {Hoyer}, {van Kerkwijk}, {Brett}, {Haldane}, {del R{\'\i}o},
  {Wiebe}, {Peterson}, {G{\'e}rard-Marchant}, {Sheppard}, {Reddy}, {Weckesser},
  {Abbasi}, {Gohlke}, \& {Oliphant}}]{numpy2}
{Harris}, C.~R., {Millman}, K.~J., {van der Walt}, S.~J., {et~al.} 2020, \nat,
  585, 357, \dodoi{10.1038/s41586-020-2649-2}

\bibitem[{{Hawcroft} {et~al.}(2023){Hawcroft}, {Sana}, {Mahy}, {Sundqvist}, {de
  Koter}, {Crowther}, {Bestenlehner}, {Brands}, {David-Uraz}, {Decin}, {Erba},
  {Garcia}, {Hamann}, {Herrero}, {Ignace}, {Kee}, {Kub{\'a}tov{\'a}},
  {Lefever}, {Moffat}, {Najarro}, {Oskinova}, {Pauli}, {Prinja}, {Puls},
  {Sander}, {Shenar}, {St-Louis}, {ud-Doula}, \& {Vink}}]{hawcroft23}
{Hawcroft}, C., {Sana}, H., {Mahy}, L., {et~al.} 2023, arXiv e-prints,
  arXiv:2303.12165, \dodoi{10.48550/arXiv.2303.12165}

\bibitem[{{Heap} {et~al.}(2006){Heap}, {Lanz}, \& {Hubeny}}]{heap06}
{Heap}, S.~R., {Lanz}, T., \& {Hubeny}, I. 2006, \apj, 638, 409,
  \dodoi{10.1086/498635}

\bibitem[{{Herrero} {et~al.}(1992){Herrero}, {Kudritzki}, {Vilchez}, {Kunze},
  {Butler}, \& {Haser}}]{herrero92}
{Herrero}, A., {Kudritzki}, R.~P., {Vilchez}, J.~M., {et~al.} 1992, \aap, 261,
  209

\bibitem[{{Higgins} \& {Vink}(2019)}]{higgins19}
{Higgins}, E.~R., \& {Vink}, J.~S. 2019, \aap, 622, A50,
  \dodoi{10.1051/0004-6361/201834123}

\bibitem[{{Holgado} {et~al.}(2020){Holgado}, {Sim{\'o}n-D{\'\i}az},
  {Haemmerl{\'e}}, {Lennon}, {Barb{\'a}}, {Cervi{\~n}o}, {Castro}, {Herrero},
  {Meynet}, \& {Arias}}]{holgado20}
{Holgado}, G., {Sim{\'o}n-D{\'\i}az}, S., {Haemmerl{\'e}}, L., {et~al.} 2020,
  \aap, 638, A157, \dodoi{10.1051/0004-6361/202037699}

\bibitem[{{Huenemoerder} {et~al.}(2012){Huenemoerder}, {Oskinova}, {Ignace},
  {Waldron}, {Todt}, {Hamaguchi}, \& {Kitamoto}}]{huenemoerder12}
{Huenemoerder}, D.~P., {Oskinova}, L.~M., {Ignace}, R., {et~al.} 2012, \apjl,
  756, L34, \dodoi{10.1088/2041-8205/756/2/L34}

\bibitem[{{Hunter} {et~al.}(2007){Hunter}, {Dufton}, {Smartt}, {Ryans},
  {Evans}, {Lennon}, {Trundle}, {Hubeny}, \& {Lanz}}]{hunter07}
{Hunter}, I., {Dufton}, P.~L., {Smartt}, S.~J., {et~al.} 2007, \aap, 466, 277,
  \dodoi{10.1051/0004-6361:20066148}

\bibitem[{Hunter(2007)}]{matplotlib}
Hunter, J.~D. 2007, Computing In Science \& Engineering, 9, 90

\bibitem[{{Jacobs} {et~al.}(2009){Jacobs}, {Rizzi}, {Tully}, {Shaya},
  {Makarov}, \& {Makarova}}]{jacobs09}
{Jacobs}, B.~A., {Rizzi}, L., {Tully}, R.~B., {et~al.} 2009, \aj, 138, 332,
  \dodoi{10.1088/0004-6256/138/2/332}

\bibitem[{{Joye} \& {Mandel}(2003)}]{joye03}
{Joye}, W.~A., \& {Mandel}, E. 2003, in Astronomical Society of the Pacific
  Conference Series, Vol. 295, Astronomical Data Analysis Software and Systems
  XII, ed. H.~E. {Payne}, R.~I. {Jedrzejewski}, \& R.~N. {Hook}, 489

\bibitem[{{Kniazev} {et~al.}(2005){Kniazev}, {Grebel}, {Pustilnik}, {Pramskij},
  \& {Zucker}}]{kniazev05}
{Kniazev}, A.~Y., {Grebel}, E.~K., {Pustilnik}, S.~A., {Pramskij}, A.~G., \&
  {Zucker}, D.~B. 2005, \aj, 130, 1558, \dodoi{10.1086/432931}

\bibitem[{{Krti{\v{c}}ka} \& {Kub{\'a}t}(2018)}]{krticka18}
{Krti{\v{c}}ka}, J., \& {Kub{\'a}t}, J. 2018, \aap, 612, A20,
  \dodoi{10.1051/0004-6361/201731969}

\bibitem[{{Kudritzki} {et~al.}(1989){Kudritzki}, {Pauldrach}, {Puls}, \&
  {Abbott}}]{kudritzki89}
{Kudritzki}, R.~P., {Pauldrach}, A., {Puls}, J., \& {Abbott}, D.~C. 1989, \aap,
  219, 205

\bibitem[{{Kudritzki} \& {Puls}(2000)}]{kudritzki00}
{Kudritzki}, R.-P., \& {Puls}, J. 2000, \araa, 38, 613,
  \dodoi{10.1146/annurev.astro.38.1.613}

\bibitem[{{Kunth} \& {{\"O}stlin}(2000)}]{kunth00}
{Kunth}, D., \& {{\"O}stlin}, G. 2000, \aapr, 10, 1,
  \dodoi{10.1007/s001590000005}

\bibitem[{{Lagae} {et~al.}(2021){Lagae}, {Driessen}, {Hennicker}, {Kee}, \&
  {Sundqvist}}]{lagae21}
{Lagae}, C., {Driessen}, F.~A., {Hennicker}, L., {Kee}, N.~D., \& {Sundqvist},
  J.~O. 2021, \aap, 648, A94, \dodoi{10.1051/0004-6361/202039972}

\bibitem[{{Lamers} \& {Cassinelli}(1999)}]{lamers99}
{Lamers}, H. J.~G.~L.~M., \& {Cassinelli}, J.~P. 1999, {Introduction to Stellar
  Winds}

\bibitem[{{Lamers} {et~al.}(1995){Lamers}, {Snow}, \& {Lindholm}}]{lamers95}
{Lamers}, H. J.~G.~L.~M., {Snow}, T.~P., \& {Lindholm}, D.~M. 1995, \apj, 455,
  269, \dodoi{10.1086/176575}

\bibitem[{{Lanz} \& {Hubeny}(2003)}]{lanz03}
{Lanz}, T., \& {Hubeny}, I. 2003, \apjs, 146, 417, \dodoi{10.1086/374373}

\bibitem[{{Lee} {et~al.}(2005){Lee}, {Skillman}, \& {Venn}}]{lee05}
{Lee}, H., {Skillman}, E.~D., \& {Venn}, K.~A. 2005, \apj, 620, 223,
  \dodoi{10.1086/427019}

\bibitem[{{Leitherer} {et~al.}(2014){Leitherer}, {Ekstr{\"o}m}, {Meynet},
  {Schaerer}, {Agienko}, \& {Levesque}}]{leitherer14}
{Leitherer}, C., {Ekstr{\"o}m}, S., {Meynet}, G., {et~al.} 2014, \apjs, 212,
  14, \dodoi{10.1088/0067-0049/212/1/14}

\bibitem[{{Leitherer} {et~al.}(1992){Leitherer}, {Robert}, \&
  {Drissen}}]{leitherer92}
{Leitherer}, C., {Robert}, C., \& {Drissen}, L. 1992, \apj, 401, 596,
  \dodoi{10.1086/172089}

\bibitem[{{Leitherer} {et~al.}(1999){Leitherer}, {Schaerer}, {Goldader},
  {Delgado}, {Robert}, {Kune}, {de Mello}, {Devost}, \&
  {Heckman}}]{leitherer99}
{Leitherer}, C., {Schaerer}, D., {Goldader}, J.~D., {et~al.} 1999, \apjs, 123,
  3, \dodoi{10.1086/313233}

\bibitem[{{Lorenzo} {et~al.}(2022){Lorenzo}, {Garcia}, {Najarro}, {Herrero},
  {Cervi{\~n}o}, \& {Castro}}]{lorenzo22}
{Lorenzo}, M., {Garcia}, M., {Najarro}, F., {et~al.} 2022, \mnras, 516, 4164,
  \dodoi{10.1093/mnras/stac2050}

\bibitem[{{Lucy} \& {White}(1980)}]{lucy80}
{Lucy}, L.~B., \& {White}, R.~L. 1980, \apj, 241, 300, \dodoi{10.1086/158342}

\bibitem[{{Maeder}(1987)}]{maeder87}
{Maeder}, A. 1987, \aap, 173, 247

\bibitem[{{Maeder} \& {Meynet}(2000)}]{maeder00}
{Maeder}, A., \& {Meynet}, G. 2000, \araa, 38, 143,
  \dodoi{10.1146/annurev.astro.38.1.143}

\bibitem[{{Maeder} {et~al.}(2009){Maeder}, {Meynet}, {Ekstr{\"o}m}, \&
  {Georgy}}]{maeder09}
{Maeder}, A., {Meynet}, G., {Ekstr{\"o}m}, S., \& {Georgy}, C. 2009,
  Communications in Asteroseismology, 158, 72, \dodoi{10.48550/arXiv.0810.0657}

\bibitem[{{Maeder} {et~al.}(2014){Maeder}, {Przybilla}, {Nieva}, {Georgy},
  {Meynet}, {Ekstr{\"o}m}, \& {Eggenberger}}]{maeder14}
{Maeder}, A., {Przybilla}, N., {Nieva}, M.-F., {et~al.} 2014, \aap, 565, A39,
  \dodoi{10.1051/0004-6361/201220602}

\bibitem[{{Marble} {et~al.}(2010){Marble}, {Engelbracht}, {van Zee}, {Dale},
  {Smith}, {Gordon}, {Wu}, {Lee}, {Kennicutt}, {Skillman}, {Johnson}, {Block},
  {Calzetti}, {Cohen}, {Lee}, \& {Schuster}}]{marble10}
{Marble}, A.~R., {Engelbracht}, C.~W., {van Zee}, L., {et~al.} 2010, \apj, 715,
  506, \dodoi{10.1088/0004-637X/715/1/506}

\bibitem[{{Marcolino} {et~al.}(2022){Marcolino}, {Bouret}, {Rocha-Pinto},
  {Bernini-Peron}, \& {Vink}}]{marcolino22}
{Marcolino}, W.~L.~F., {Bouret}, J.~C., {Rocha-Pinto}, H.~J., {Bernini-Peron},
  M., \& {Vink}, J.~S. 2022, \mnras, 511, 5104, \dodoi{10.1093/mnras/stac452}

\bibitem[{{Markova} {et~al.}(2018){Markova}, {Puls}, \& {Langer}}]{markova18}
{Markova}, N., {Puls}, J., \& {Langer}, N. 2018, \aap, 613, A12,
  \dodoi{10.1051/0004-6361/201731361}

\bibitem[{{Martins}(2018)}]{martins18}
{Martins}, F. 2018, \aap, 616, A135, \dodoi{10.1051/0004-6361/201833050}

\bibitem[{{Martins} \& {Palacios}(2021)}]{martins21}
{Martins}, F., \& {Palacios}, A. 2021, \aap, 645, A67,
  \dodoi{10.1051/0004-6361/202039337}

\bibitem[{{Martins} {et~al.}(2005){Martins}, {Schaerer}, {Hillier},
  {Meynadier}, {Heydari-Malayeri}, \& {Walborn}}]{martins05}
{Martins}, F., {Schaerer}, D., {Hillier}, D.~J., {et~al.} 2005, \aap, 441, 735,
  \dodoi{10.1051/0004-6361:20052927}

\bibitem[{{Martins} {et~al.}(2024){Martins}, {Bouret}, {Hillier}, {Brands},
  {Crowther}, {Herrero}, {Najarro}, {Pauli}, {Puls}, {Ramachandran}, {Sander},
  {Vink}, \& {the XshootU collaboration}}]{martins24}
{Martins}, F., {Bouret}, J.~C., {Hillier}, D.~J., {et~al.} 2024, arXiv
  e-prints, arXiv:2405.01267, \dodoi{10.48550/arXiv.2405.01267}

\bibitem[{{McQuinn} {et~al.}(2015){McQuinn}, {Skillman}, {Dolphin}, {Cannon},
  {Salzer}, {Rhode}, {Adams}, {Berg}, {Giovanelli}, {Girardi}, \&
  {Haynes}}]{mcquinn15}
{McQuinn}, K. B.~W., {Skillman}, E.~D., {Dolphin}, A., {et~al.} 2015, \apj,
  812, 158, \dodoi{10.1088/0004-637X/812/2/158}

\bibitem[{{Meynet} \& {Maeder}(2002)}]{meynet02}
{Meynet}, G., \& {Maeder}, A. 2002, \aap, 390, 561,
  \dodoi{10.1051/0004-6361:20020755}

\bibitem[{{Mokiem} {et~al.}(2007){Mokiem}, {de Koter}, {Vink}, {Puls}, {Evans},
  {Smartt}, {Crowther}, {Herrero}, {Langer}, {Lennon}, {Najarro}, \&
  {Villamariz}}]{mokiem07}
{Mokiem}, M.~R., {de Koter}, A., {Vink}, J.~S., {et~al.} 2007, \aap, 473, 603,
  \dodoi{10.1051/0004-6361:20077545}

\bibitem[{{Morrissey} {et~al.}(2018){Morrissey}, {Matuszewski}, {Martin},
  {Neill}, {Epps}, {Fucik}, {Weber}, {Darvish}, {Adkins}, {Allen}, {Bartos},
  {Belicki}, {Cabak}, {Callahan}, {Cowley}, {Crabill}, {Deich}, {Delecroix},
  {Doppman}, {Hilyard}, {James}, {Kaye}, {Kokorowski}, {Kwok}, {Lanclos},
  {Milner}, {Moore}, {O'Sullivan}, {Parihar}, {Park}, {Phillips}, {Rizzi},
  {Rockosi}, {Rodriguez}, {Salaun}, {Seaman}, {Sheikh}, {Weiss}, \&
  {Zarzaca}}]{morrissey18}
{Morrissey}, P., {Matuszewski}, M., {Martin}, D.~C., {et~al.} 2018, \apj, 864,
  93, \dodoi{10.3847/1538-4357/aad597}

\bibitem[{{Offner} {et~al.}(2023){Offner}, {Moe}, {Kratter}, {Sadavoy},
  {Jensen}, \& {Tobin}}]{offner23}
{Offner}, S.~S.~R., {Moe}, M., {Kratter}, K.~M., {et~al.} 2023, in Astronomical
  Society of the Pacific Conference Series, Vol. 534, Protostars and Planets
  VII, ed. S.~{Inutsuka}, Y.~{Aikawa}, T.~{Muto}, K.~{Tomida}, \& M.~{Tamura},
  275, \dodoi{10.48550/arXiv.2203.10066}

\bibitem[{{O'Sullivan} \& {Chen}(2020)}]{osullivan20}
{O'Sullivan}, D., \& {Chen}, Y. 2020, arXiv e-prints, arXiv:2011.05444,
  \dodoi{10.48550/arXiv.2011.05444}

\bibitem[{{Pauldrach} {et~al.}(1986){Pauldrach}, {Puls}, \&
  {Kudritzki}}]{pauldrach86}
{Pauldrach}, A., {Puls}, J., \& {Kudritzki}, R.~P. 1986, \aap, 164, 86

\bibitem[{{Perez} \& {Granger}(2007)}]{ipython}
{Perez}, F., \& {Granger}, B.~E. 2007, Computing in Science and Engineering, 9,
  21, \dodoi{10.1109/MCSE.2007.53}

\bibitem[{{Prinja} {et~al.}(1990){Prinja}, {Barlow}, \& {Howarth}}]{prinja90}
{Prinja}, R.~K., {Barlow}, M.~J., \& {Howarth}, I.~D. 1990, \apj, 361, 607,
  \dodoi{10.1086/169224}

\bibitem[{{Przybilla} {et~al.}(2010){Przybilla}, {Firnstein}, {Nieva},
  {Meynet}, \& {Maeder}}]{przybilla10}
{Przybilla}, N., {Firnstein}, M., {Nieva}, M.~F., {Meynet}, G., \& {Maeder}, A.
  2010, \aap, 517, A38, \dodoi{10.1051/0004-6361/201014164}

\bibitem[{{Puls} {et~al.}(2020){Puls}, {Najarro}, {Sundqvist}, \&
  {Sen}}]{puls20}
{Puls}, J., {Najarro}, F., {Sundqvist}, J.~O., \& {Sen}, K. 2020, \aap, 642,
  A172, \dodoi{10.1051/0004-6361/202038464}

\bibitem[{{Ramachandran} {et~al.}(2018){Ramachandran}, {Hamann}, {Hainich},
  {Oskinova}, {Shenar}, {Sander}, {Todt}, \& {Gallagher}}]{ramachandran18}
{Ramachandran}, V., {Hamann}, W.~R., {Hainich}, R., {et~al.} 2018, \aap, 615,
  A40, \dodoi{10.1051/0004-6361/201832816}

\bibitem[{{Ramachandran} {et~al.}(2021){Ramachandran}, {Oskinova}, \&
  {Hamann}}]{ramachandran21}
{Ramachandran}, V., {Oskinova}, L.~M., \& {Hamann}, W.~R. 2021, \aap, 646, A16,
  \dodoi{10.1051/0004-6361/202039486}

\bibitem[{{Ramachandran} {et~al.}(2019){Ramachandran}, {Hamann}, {Oskinova},
  {Gallagher}, {Hainich}, {Shenar}, {Sander}, {Todt}, \&
  {Fulmer}}]{ramachandran19}
{Ramachandran}, V., {Hamann}, W.~R., {Oskinova}, L.~M., {et~al.} 2019, \aap,
  625, A104, \dodoi{10.1051/0004-6361/201935365}

\bibitem[{{Rauw} {et~al.}(2015){Rauw}, {Naz{\'e}}, {Wright}, {Drake},
  {Guarcello}, {Prinja}, {Peck}, {Albacete Colombo}, {Herrero}, {Kobulnicky},
  {Sciortino}, \& {Vink}}]{rauw15}
{Rauw}, G., {Naz{\'e}}, Y., {Wright}, N.~J., {et~al.} 2015, \apjs, 221, 1,
  \dodoi{10.1088/0067-0049/221/1/1}

\bibitem[{{Renzo} \& {G{\"o}tberg}(2021)}]{renzo21}
{Renzo}, M., \& {G{\"o}tberg}, Y. 2021, \apj, 923, 277,
  \dodoi{10.3847/1538-4357/ac29c5}

\bibitem[{{Rickard} {et~al.}(2022){Rickard}, {Hainich}, {Hamann}, {Oskinova},
  {Prinja}, {Ramachandran}, {Pauli}, {Todt}, {Sander}, {Shenar}, {Chu}, \&
  {Gallagher}}]{rickard22}
{Rickard}, M.~J., {Hainich}, R., {Hamann}, W.~R., {et~al.} 2022, \aap, 666,
  A189, \dodoi{10.1051/0004-6361/202243281}

\bibitem[{{Rivero Gonz{\'a}lez} {et~al.}(2011){Rivero Gonz{\'a}lez}, {Puls}, \&
  {Najarro}}]{rivero-gonzalez11}
{Rivero Gonz{\'a}lez}, J.~G., {Puls}, J., \& {Najarro}, F. 2011, \aap, 536,
  A58, \dodoi{10.1051/0004-6361/201117101}

\bibitem[{{Roberts-Borsani} {et~al.}(2022){Roberts-Borsani}, {Morishita},
  {Treu}, {Brammer}, {Strait}, {Wang}, {Bradac}, {Acebron}, {Bergamini},
  {Boyett}, {Calabr{\'o}}, {Castellano}, {Fontana}, {Glazebrook}, {Grillo},
  {Henry}, {Jones}, {Malkan}, {Marchesini}, {Mascia}, {Mason}, {Mercurio},
  {Merlin}, {Nanayakkara}, {Pentericci}, {Rosati}, {Santini}, {Scarlata},
  {Trenti}, {Vanzella}, {Vulcani}, \& {Willott}}]{roberts-borsani22}
{Roberts-Borsani}, G., {Morishita}, T., {Treu}, T., {et~al.} 2022, \apjl, 938,
  L13, \dodoi{10.3847/2041-8213/ac8e6e}

\bibitem[{{Roman-Duval} {et~al.}(2020){Roman-Duval}, {Proffitt}, {Taylor},
  {Monroe}, {Fischer}, {Fischer}, {Fullerton}, {Aloisi}, {Britt}, {Busko},
  {Carlberg}, {De Rosa}, {Jedrzejewski}, {Lockwood}, {Frazer}, {Hernandez},
  {James}, {Oliveira}, {Plesha}, {Riedel}, {Riley}, {Sahnow}, {Sankrit},
  {Shaw}, {Smith}, {Sohn}, {Som}, {Ubeda}, \& {Welty}}]{roman-duval20}
{Roman-Duval}, J., {Proffitt}, C.~R., {Taylor}, J.~M., {et~al.} 2020, Research
  Notes of the American Astronomical Society, 4, 205,
  \dodoi{10.3847/2515-5172/abca2f}

\bibitem[{{Rosen}(2022)}]{rosen22}
{Rosen}, A.~L. 2022, \apj, 941, 202, \dodoi{10.3847/1538-4357/ac9f3d}

\bibitem[{{Sana} {et~al.}(2012){Sana}, {de Mink}, {de Koter}, {Langer},
  {Evans}, {Gieles}, {Gosset}, {Izzard}, {Le Bouquin}, \& {Schneider}}]{sana12}
{Sana}, H., {de Mink}, S.~E., {de Koter}, A., {et~al.} 2012, Science, 337, 444,
  \dodoi{10.1126/science.1223344}

\bibitem[{{Sander} {et~al.}(2015){Sander}, {Shenar}, {Hainich},
  {G{\'\i}menez-Garc{\'\i}a}, {Todt}, \& {Hamann}}]{sander15}
{Sander}, A., {Shenar}, T., {Hainich}, R., {et~al.} 2015, \aap, 577, A13,
  \dodoi{10.1051/0004-6361/201425356}

\bibitem[{{Sander} {et~al.}(2017){Sander}, {Hamann}, {Todt}, {Hainich}, \&
  {Shenar}}]{sander17}
{Sander}, A.~A.~C., {Hamann}, W.~R., {Todt}, H., {Hainich}, R., \& {Shenar}, T.
  2017, \aap, 603, A86, \dodoi{10.1051/0004-6361/201730642}

\bibitem[{{Schneider} {et~al.}(2019){Schneider}, {Ohlmann}, {Podsiadlowski},
  {R{\"o}pke}, {Balbus}, {Pakmor}, \& {Springel}}]{schneider19}
{Schneider}, F. R.~N., {Ohlmann}, S.~T., {Podsiadlowski}, P., {et~al.} 2019,
  \nat, 574, 211, \dodoi{10.1038/s41586-019-1621-5}

\bibitem[{{Scott} {et~al.}(2015){Scott}, {Asplund}, {Grevesse}, {Bergemann}, \&
  {Sauval}}]{scott15}
{Scott}, P., {Asplund}, M., {Grevesse}, N., {Bergemann}, M., \& {Sauval}, A.~J.
  2015, \aap, 573, A26, \dodoi{10.1051/0004-6361/201424110}

\bibitem[{{Serenelli} {et~al.}(2021){Serenelli}, {Weiss}, {Aerts}, {Angelou},
  {Baroch}, {Bastian}, {Beck}, {Bergemann}, {Bestenlehner}, {Czekala},
  {Elias-Rosa}, {Escorza}, {Van Eylen}, {Feuillet}, {Gandolfi}, {Gieles},
  {Girardi}, {Lebreton}, {Lodieu}, {Martig}, {Miller Bertolami}, {Mombarg},
  {Morales}, {Moya}, {Nsamba}, {Pavlovski}, {Pedersen}, {Ribas}, {Schneider},
  {Silva Aguirre}, {Stassun}, {Tolstoy}, {Tremblay}, \& {Zwintz}}]{serenelli21}
{Serenelli}, A., {Weiss}, A., {Aerts}, C., {et~al.} 2021, \aapr, 29, 4,
  \dodoi{10.1007/s00159-021-00132-9}

\bibitem[{{Shenar} {et~al.}(2015){Shenar}, {Oskinova}, {Hamann}, {Corcoran},
  {Moffat}, {Pablo}, {Richardson}, {Waldron}, {Huenemoerder}, {Ma{\'\i}z
  Apell{\'a}niz}, {Nichols}, {Todt}, {Naz{\'e}}, {Hoffman}, {Pollock}, \&
  {Negueruela}}]{shenar15}
{Shenar}, T., {Oskinova}, L., {Hamann}, W.~R., {et~al.} 2015, \apj, 809, 135,
  \dodoi{10.1088/0004-637X/809/2/135}

\bibitem[{{Sim{\'o}n-D{\'\i}az}(2020)}]{simon-diaz20}
{Sim{\'o}n-D{\'\i}az}, S. 2020, in Reviews in Frontiers of Modern Astrophysics;
  From Space Debris to Cosmology, 155--187, \dodoi{10.1007/978-3-030-38509-5_6}

\bibitem[{{Sim{\'o}n-D{\'\i}az} \& {Herrero}(2014)}]{simon-diaz14}
{Sim{\'o}n-D{\'\i}az}, S., \& {Herrero}, A. 2014, \aap, 562, A135,
  \dodoi{10.1051/0004-6361/201322758}

\bibitem[{{Skillman} {et~al.}(1989){Skillman}, {Kennicutt}, \&
  {Hodge}}]{skillman89}
{Skillman}, E.~D., {Kennicutt}, R.~C., \& {Hodge}, P.~W. 1989, \apj, 347, 875,
  \dodoi{10.1086/168178}

\bibitem[{{Skillman} {et~al.}(2013){Skillman}, {Salzer}, {Berg}, {Pogge},
  {Haurberg}, {Cannon}, {Aver}, {Olive}, {Giovanelli}, {Haynes}, {Adams},
  {McQuinn}, \& {Rhode}}]{skillman13}
{Skillman}, E.~D., {Salzer}, J.~J., {Berg}, D.~A., {et~al.} 2013, \aj, 146, 3,
  \dodoi{10.1088/0004-6256/146/1/3}

\bibitem[{{Sukhbold} {et~al.}(2016){Sukhbold}, {Ertl}, {Woosley}, {Brown}, \&
  {Janka}}]{sukhbold16}
{Sukhbold}, T., {Ertl}, T., {Woosley}, S.~E., {Brown}, J.~M., \& {Janka}, H.~T.
  2016, \apj, 821, 38, \dodoi{10.3847/0004-637X/821/1/38}

\bibitem[{{Sz{\'e}csi} {et~al.}(2022){Sz{\'e}csi}, {Agrawal}, {W{\"u}nsch}, \&
  {Langer}}]{szecsi22}
{Sz{\'e}csi}, D., {Agrawal}, P., {W{\"u}nsch}, R., \& {Langer}, N. 2022, \aap,
  658, A125, \dodoi{10.1051/0004-6361/202141536}

\bibitem[{{Tang} {et~al.}(2014){Tang}, {Bressan}, {Rosenfield}, {Slemer},
  {Marigo}, {Girardi}, \& {Bianchi}}]{tang14}
{Tang}, J., {Bressan}, A., {Rosenfield}, P., {et~al.} 2014, \mnras, 445, 4287,
  \dodoi{10.1093/mnras/stu2029}

\bibitem[{{Telford} {et~al.}(2021){Telford}, {Chisholm}, {McQuinn}, \&
  {Berg}}]{telford21}
{Telford}, O.~G., {Chisholm}, J., {McQuinn}, K. B.~W., \& {Berg}, D.~A. 2021,
  \apj, 922, 191, \dodoi{10.3847/1538-4357/ac1ce2}

\bibitem[{{Telford} {et~al.}(2023){Telford}, {McQuinn}, {Chisholm}, \&
  {Berg}}]{telford23}
{Telford}, O.~G., {McQuinn}, K. B.~W., {Chisholm}, J., \& {Berg}, D.~A. 2023,
  \apj, 943, 65, \dodoi{10.3847/1538-4357/aca896}

\bibitem[{{Tinsley}(1980)}]{tinsley80}
{Tinsley}, B.~M. 1980, \fcp, 5, 287, \dodoi{10.48550/arXiv.2203.02041}

\bibitem[{{Topping} {et~al.}(2024){Topping}, {Stark}, {Endsley}, {Whitler},
  {Hainline}, {Johnson}, {Robertson}, {Tacchella}, {Chen}, {Alberts}, {Baker},
  {Bunker}, {Carniani}, {Charlot}, {Chevallard}, {Curtis-Lake}, {DeCoursey},
  {Egami}, {Eisenstein}, {Ji}, {Maiolino}, {Williams}, {Willmer}, {Willott}, \&
  {Witstok}}]{topping24}
{Topping}, M.~W., {Stark}, D.~P., {Endsley}, R., {et~al.} 2024, \mnras, 529,
  4087, \dodoi{10.1093/mnras/stae800}

\bibitem[{{Tramper} {et~al.}(2011){Tramper}, {Sana}, {de Koter}, \&
  {Kaper}}]{tramper11}
{Tramper}, F., {Sana}, H., {de Koter}, A., \& {Kaper}, L. 2011, \apjl, 741, L8,
  \dodoi{10.1088/2041-8205/741/1/L8}

\bibitem[{{Tramper} {et~al.}(2014){Tramper}, {Sana}, {de Koter}, {Kaper}, \&
  {Ram{\'\i}rez-Agudelo}}]{tramper14}
{Tramper}, F., {Sana}, H., {de Koter}, A., {Kaper}, L., \&
  {Ram{\'\i}rez-Agudelo}, O.~H. 2014, \aap, 572, A36,
  \dodoi{10.1051/0004-6361/201424312}

\bibitem[{{Trebitsch} {et~al.}(2017){Trebitsch}, {Blaizot}, {Rosdahl},
  {Devriendt}, \& {Slyz}}]{trebitsch17}
{Trebitsch}, M., {Blaizot}, J., {Rosdahl}, J., {Devriendt}, J., \& {Slyz}, A.
  2017, \mnras, 470, 224, \dodoi{10.1093/mnras/stx1060}

\bibitem[{{Trundle} {et~al.}(2007){Trundle}, {Dufton}, {Hunter}, {Evans},
  {Lennon}, {Smartt}, \& {Ryans}}]{trundle07}
{Trundle}, C., {Dufton}, P.~L., {Hunter}, I., {et~al.} 2007, \aap, 471, 625,
  \dodoi{10.1051/0004-6361:20077838}

\bibitem[{{van der Walt} {et~al.}(2011){van der Walt}, {Colbert}, \&
  {Varoquaux}}]{numpy}
{van der Walt}, S., {Colbert}, S.~C., \& {Varoquaux}, G. 2011, Computing in
  Science and Engineering, 13, 22, \dodoi{10.1109/MCSE.2011.37}

\bibitem[{{Vink}(2022)}]{vink22}
{Vink}, J.~S. 2022, \araa, 60, 203, \dodoi{10.1146/annurev-astro-052920-094949}

\bibitem[{{Vink} {et~al.}(1999){Vink}, {de Koter}, \& {Lamers}}]{vink99}
{Vink}, J.~S., {de Koter}, A., \& {Lamers}, H.~J.~G.~L.~M. 1999, \aap, 350,
  181, \dodoi{10.48550/arXiv.astro-ph/9908196}

\bibitem[{{Vink} {et~al.}(2001){Vink}, {de Koter}, \& {Lamers}}]{vink01}
---. 2001, \aap, 369, 574, \dodoi{10.1051/0004-6361:20010127}

\bibitem[{{Vink} {et~al.}(2023){Vink}, {Mehner}, {Crowther}, {Fullerton},
  {Garcia}, {Martins}, {Morrell}, {Oskinova}, {St-Louis}, {ud-Doula}, {Sander},
  {Sana}, {Bouret}, {Kub{\'a}tov{\'a}}, {Marchant}, {Martins}, {Wofford}, {van
  Loon}, {Grace Telford}, {G{\"o}tberg}, {Bowman}, {Erba}, {Kalari},
  {Abdul-Masih}, {Alkousa}, {Backs}, {Barbosa}, {Berlanas}, {Bernini-Peron},
  {Bestenlehner}, {Blomme}, {Bodensteiner}, {Brands}, {Evans}, {David-Uraz},
  {Driessen}, {Dsilva}, {Geen}, {G{\'o}mez-Gonz{\'a}lez}, {Grassitelli},
  {Hamann}, {Hawcroft}, {Herrero}, {Higgins}, {John Hillier}, {Ignace},
  {Istrate}, {Kaper}, {Kee}, {Kehrig}, {Keszthelyi}, {Klencki}, {de Koter},
  {Kuiper}, {Laplace}, {Larkin}, {Lefever}, {Leitherer}, {Lennon}, {Mahy},
  {Ma{\'\i}z Apell{\'a}niz}, {Maravelias}, {Marcolino}, {McLeod}, {de Mink},
  {Najarro}, {Oey}, {Parsons}, {Pauli}, {Pedersen}, {Prinja}, {Ramachandran},
  {Ram{\'\i}rez-Tannus}, {Sabhahit}, {Schootemeijer}, {Reyero Serantes},
  {Shenar}, {Stringfellow}, {Sudnik}, {Tramper}, \& {Wang}}]{vink23}
{Vink}, J.~S., {Mehner}, A., {Crowther}, P.~A., {et~al.} 2023, \aap, 675, A154,
  \dodoi{10.1051/0004-6361/202245650}

\bibitem[{{Virtanen} {et~al.}(2020){Virtanen}, {Gommers}, {Oliphant},
  {Haberland}, {Reddy}, {Cournapeau}, {Burovski}, {Peterson}, {Weckesser},
  {Bright}, {van der Walt}, {Brett}, {Wilson}, {Millman}, {Mayorov}, {Nelson},
  {Jones}, {Kern}, {Larson}, {Carey}, {Polat}, {Feng}, {Moore}, {VanderPlas},
  {Laxalde}, {Perktold}, {Cimrman}, {Henriksen}, {Quintero}, {Harris},
  {Archibald}, {Ribeiro}, {Pedregosa}, {van Mulbregt}, \& {SciPy 1. 0
  Contributors}}]{scipy2}
{Virtanen}, P., {Gommers}, R., {Oliphant}, T.~E., {et~al.} 2020, Nature
  Methods, 17, 261, \dodoi{10.1038/s41592-019-0686-2}

\bibitem[{{Weaver} {et~al.}(1977){Weaver}, {McCray}, {Castor}, {Shapiro}, \&
  {Moore}}]{weaver77}
{Weaver}, R., {McCray}, R., {Castor}, J., {Shapiro}, P., \& {Moore}, R. 1977,
  \apj, 218, 377, \dodoi{10.1086/155692}

\end{thebibliography}
\end{document}